\documentclass[]{pasj02}

\usepackage{amsmath}
\usepackage{txfonts}
\usepackage{natbib,bm}

\usepackage{graphicx,color}  

\newcommand{\So}{\bm{S}_\mathrm{o}}

\jyear{2026}

\begin{document}

\title{Probing disk dynamics and dust evolution through shadows in protoplanetary disks: A case study of the HD~142527 disk}

\author{
 Yuya \textsc{Fukuhara},\altaffilmark{1}\altemailmark$^{,\dag}$\orcid{0000-0002-9660-8947} \email{yfukuhara@asiaa.sinica.edu.tw} 
 Ryuta \textsc{Orihara},\altaffilmark{2,3}\altemailmark$^{,\dag}$\orcid{0000-0003-4039-8933}\email{orihara.r.2163@m.isct.ac.jp} 
 Satoshi \textsc{Okuzumi},\altaffilmark{2}\orcid{0000-0002-1886-0880}
 and 
 Takayuki \textsc{Muto}\altaffilmark{4}
}

\altaffiltext{1}{Institute of Astronomy and Astrophysics, Academia Sinica, 11F of Astronomy-Mathematics Building, No.1, Sec. 4, Roosevelt Rd, Taipei 106319, Taiwan, R.O.C.}
\altaffiltext{2}{Department of Earth and Planetary Sciences, Institute of Science Tokyo, Meguro, Tokyo 152-8551, Japan}
\altaffiltext{3}{Department of Astronomy, Graduate School of Science, The University of Tokyo, 7-3-1 Hongo, Bunkyo-ku, Tokyo 113-0033, Japan}
\altaffiltext{4}{Division of Liberal Arts, Kogakuin University, Tokyo, Japan}

\altaffiltext{$^\dag$}{\footnotesize These authors contributed equally to this work. }

\KeyWords{protoplanetary disks -- methods: observational --- methods: analytical}

\maketitle

\begin{abstract}
Planet formation begins with dust growth and planetesimal formation within protoplanetary disks surrounding young stars. 
To understand these processes, it is essential to estimate dust grain sizes from disk observations. 
In this study, we develop a new method to constrain grain size based on the estimation of cooling timescales. 
Our approach applies to transitional disks that possess an inclined inner disk casting shadows on the outer disk, whose temperature variations serve as a tracer of dust properties. 
By constructing a three-dimensional model of the disk surface using near-infrared scattering light images and comparing it with submillimeter dust continuum maps, we estimate the spatial offset between the irradiated and shadowed regions to derive the cooling timescale. 
We then build an analytic model that calculates the cooling timescale at the dust thermal emission height with an assumed turbulent diffusion intensity to infer the dust surface density and dust grain size. 
Applying this method to the protoplanetary disk around HD~142527, we find that the disk's northern shadowed region cools on a timescale of a few percent of the orbital period and that the maximum grain size consistent with the observations is approximately 0.1-1 mm. 
We also find that the conditions required for the vertical shear instability, which needs a short cooling timescale, are satisfied, allowing turbulence with an intensity consistent with near-infrared observations. 
This study demonstrates that estimating cooling timescales is an effective tool for constraining dust grain size. 
Our approach can be generally applied to other transition disks with inner-disk-induced shadows.
\end{abstract}

\section{Introduction}\label{sec:intro}

Planet formation begins with the growth of micrometer-sized dust particles into kilometer-sized planetesimals within protoplanetary disks surrounding young stars (for a review, see \citealt{DrazkowskaBitsch+:2023aa}).
This dust growth and subsequent planetesimal formation depend on gas turbulence and the spatial distribution of dust particles.
Gas turbulence induces relative velocities between dust grains (e.g., \citealt{OrmelCuzzi2007}), leading to their fragmentation (e.g., \citealt{Brauer:2008aa,Birnstiel:2012aa}), while also governing the spatial distribution of dust via turbulent diffusion (e.g., \citealt{Dubrulle+1995}).
If dust grains concentrate toward the midplane due to low turbulence (e.g., \citealt{DullemondDominik:2005vy}) or toward local maxima of gas pressure (e.g., \citealt{Whipple:1972vv,PinillaBirnstiel+:2012vz}), streaming and gravitational instabilities can set in (e.g., \citealt{Goldreich:1973aa,YoudinGoodman:2005aa,LimSimon+:2025aa,LimSimon+:2026aa}), triggering planetesimal formation.

Throughout these stages, gas cooling plays a critical role in disk evolution and dust growth.
Specifically, the cooling timescale and its spatial distribution control the onset of various (thermo-)hydrodynamic instabilities and the intensity of turbulence driven by them (for a review, see \citealt{LesurFlock+:2023aa}).
For instance, a short cooling timescale (i.e., efficient cooling) is required to sustain turbulence driven by the vertical shear instability (VSI; e.g., \citealt{Urpin2003,NelsonGresselUmurhan2013,LinYoudin2015}), which is a candidate mechanism for determining dust spatial distribution in the outer disk regions (e.g., \citealt{PfeilBirnstiel+:2023aa,FukuharaFlock+:2025aa}). 
Efficient cooling also dictates the strength of turbulence driven by the gravitational instability (GI; e.g., \citealt{Gammie:2001aa,MejiaDurisen+:2005aa,ShiChiang:2014aa}), and spiral structures induced by GI depend on the cooling timescale (e.g., \citealt{CossinsLodato+:2009aa,LongariniLodato+:2021aa,BaehrZhu:2021aa,SuWei:2025aa}; for reviews, see \citealt{KratterLodato:2016aa,BaeIsella+:2023aa}).
In GI-dominated disks, efficient cooling can prevent dust growth through strong diffusion and high relative collision velocities of dust particles (e.g., \citealt{BoothClarke:2016aa,ShiZhu+:2016aa,BaehrZhu:2021ab}), while at the same time promoting disk fragmentation (e.g., \citealt{Gammie:2001aa,BaehrKlahr:2015aa,TakahashiTsukamoto+:2016aa,LeedhamBooth+:2025aa}), which can lead to planet formation.
Furthermore, local cooling influences the formation of substructures, such as gas gaps, spirals, and vortices, induced by embedded planets (e.g., \citealt{RaettigLyra+:2013aa,LesLin:2015aa}).
A moderate cooling timescale, which is comparable to the local dynamical timescale, can suppress spiral arms (e.g., \citealt{ZhangZhu:2020aa}), narrow and deepen planetary gaps \citep{MirandaRafikov:2020aa,MirandaRafikov:2020ab,BaeTeague+:2021aa,ZhangHuang+:2024aa}, and accelerate the decay of vortices \citep{Tarczay-NehezRegaly+:2020aa,RometschZiampras+:2021aa,FungOno:2021aa}.

Dust particles, in turn, can determine the cooling efficiency of protoplanetary disks, since gas molecules are less efficient radiators than dust (e.g., \citealt{Malygin+2017,BarrancoPei+:2018kc}).
Consequently, the cooling timescale and its spatial distribution depend on the size, size distribution, and abundance of dust grains (e.g., \citealt{BaeTeague+:2021aa,FukuharaOkuzumi+:2021ca}).
Therefore, one can expect a strong coupling and coevolution between gas dynamics (turbulence/substructures) and dust properties via cooling processes.

Given this importance, it is desirable to constrain the cooling timescale from observations of protoplanetary disks.
A potential candidate target for this purpose is a transition disk, characterized by a dust-depleted inner cavity.
In these transition disks, light from a central star reaches the outer disk regions directly through the inner cavity, allowing the heating and cooling processes to be observed in a spatially resolved manner.
Moreover, if there is a misaligned inner disk with respect to the outer disk, gas and dust in the outer disk would undergo repeated heating and cooling as they rotate into and out of the shadow cast by the inner disk.
As disk material enters and exits the shadow, its temperature relaxes toward the local radiative equilibrium value on a timescale set by the cooling timescale, so the resulting temperature variation with orbital phase carries direct information on this timescale.
Recent radio interferometric observations with the Atacama Large Millimeter--submillimeter Array (ALMA) have characterized properties of transition disks and identified compact dust disks within the inner cavities of several systems (e.g., \citealt{van-der-Marel:2023aa}).
In some cases, these compact inner disks are misaligned with their outer disks, casting shadows across them.
Such shadows have been observed in near-infrared scattered light in several systems (e.g., HD~142527; \citealt{MarinoPerez+:2015aa}, HD~100453; \citealt{BenistyStolker+:2017aa}, HD~143006; \citealt{BenistyJuhasz+:2018aa}, ZZ~Tau~IRS; \citealt{HashimotoDong+:2024aa}).
Various mechanisms have been proposed to explain this misalignment, including gravitational perturbations from planet--disk or binary--disk interactions (e.g., \citealt{OwenLai:2017aa,MatsakosKonigl:2017aa,FacchiniJuhasz+:2018aa,NealonDipierro+:2018aa,Zhu:2019aa}), interaction with a misaligned stellar magnetic field \citep{BouvierChelli+:1999aa}, anisotropic accretion/infall of material onto the disk \citep{Bate:2018aa,KuffmeierDullemond+:2021aa}, and tidal perturbation from a stellar flyby/encounter \citep{NealonCuello+:2020aa}.

In this paper, we propose a novel method to estimate the local cooling timescale in transition disks with shadows cast by inclined inner disks and to use this cooling timescale to constrain the dust grain size.
Specifically, we use near-infrared and submillimeter observations to estimate the cooling timescale of a disk, and then determine the dust grain size consistent with the estimated cooling timescale.
By applying this approach to the protoplanetary disk around HD~142527, we infer dust evolution in this disk.

This paper is organized as follows.
In section \ref{sec:method}, we describe the method for estimating the cooling timescale using the shadow induced by a misaligned inner disk and the analytic model for constraining the dust grain size.
We present the observation data for the disk around HD~142527 and parameter choices in section \ref{sec:dataset}, to which we apply our methods.
Section \ref{sec:results} presents the main results for estimating the cooling timescale and constraining dust grain size.
In section \ref{sec:VSI_application}, we examine the self-consistency of our model, assuming the vertical shear instability operates as the turbulence-driving mechanism.
We discuss the implications and limitations of our study in section \ref{sec:discussion} and summarize our study in section \ref{sec:conclusions}.

\section{Method}\label{sec:method}

\begin{figure*}[t]
    \begin{center}
    \includegraphics[width=\hsize,bb = 0 0 1136 923]{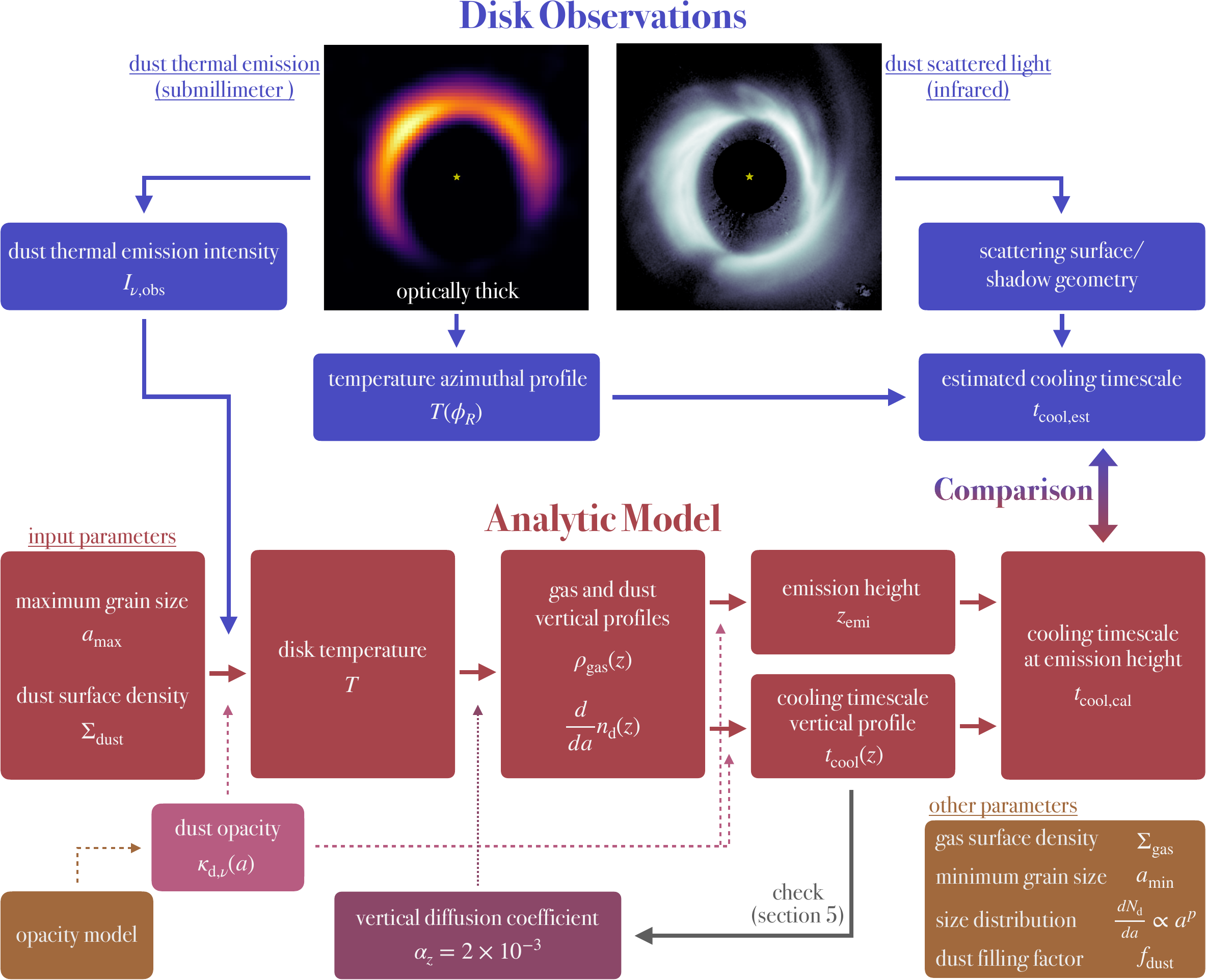}
    \end{center}
    \caption{Schematic overview of our methodology for estimating the cooling timescale and constraining the dust grain size by combining multi-wavelength observations with an analytic model. High-resolution dust thermal emission observations at submillimeter wavelengths provide the emission intensity and azimuthal temperature profile. From near-infrared scattered-light observations, we determine the scattering surface and shadow geometry using the method proposed by \cite[][subsection \ref{subsubsec:shadow_reconstruction}]{OriharaMomose:2025aa}. By measuring the spatial offset between the azimuthal temperature variation and the shadow boundary, we derive the local cooling timescale (subsection \ref{subsubsec:estimate_cooling}). In our analytic model, we consider a protoplanetary disk composed of gas and dust. For a given dust grain size and size distribution (subsection \ref{subsubsec:method_dust_size}), we compute the disk temperature using the observed thermal emission intensity and dust extinction opacity (subsection \ref{subsubsec:method_temperature}). This temperature determines the vertical gas density profile, which, in turn, dictates the vertical dust distribution for an assumed vertical diffusion coefficient (subsection \ref{subsubsec:method_vertical_profile}). Using the adopted dust opacities (subsection \ref{subsubsec:method_opacity}), we calculate the optical depth to estimate the emission heights at a submillimeter wavelength and scattering surface height at a near-infrared wavelength (subsection \ref{subsubsec:method_height}). We then compute the vertical profile of the cooling timescale to extract its theoretical value at this emission height (subsection \ref{subsubsec:method_cooling}). Finally, we constrain the dust grain size by comparing the theoretically predicted cooling timescale at the emission height with the observationally derived one (subsection \ref{subsec:method_procedure}). Additionally, we use the vertical profile of the cooling timescale to verify the validity of the assumed vertical diffusion coefficient (section \ref{sec:VSI_application}). The primary parameters in our model are the maximum grain size and dust surface density, while other parameters include the gas surface density, minimum grain size, size distribution power-law index, and dust filling factor (section \ref{subsec:parameter_choices}). The top left and right panels display the ALMA Band 9 dust continuum emission and VLT/SPHERE scattered-light images of the disk around HD~142527, respectively. }
    \label{fig:overview}
\end{figure*}

In this section, we describe our methodology for estimating the cooling timescale from multi-wavelength observations of the protoplanetary disk and for constraining the dust grain size using an analytic model (see figure \ref{fig:overview} for an overview). 
Our approach relies on the spatial offset between shadow signatures observed at different wavelengths.
First, we construct a three-dimensional model of the outer disk's scattering surface using near-infrared images, based on the method developed by \citet{OriharaMomose:2025aa}.
We then compare the location of this shadowed region in this model with the azimuthal temperature profile derived from dust continuum observations. 
The resulting spatial offset between the irradiated and shadowed regions allows us to estimate the local cooling timescale.
This offset arises because the disk temperature relaxes toward its irradiated or shadowed equilibrium value over a finite cooling timescale.
Therefore, the location of maximum or minimum temperature lags the corresponding change in illumination.
While our strategy shares similarities with that of \citet{TeagueBae+:2022aa}, our method utilizes near-infrared scattered light to trace the disk surface, rather than relying on molecular gas emission.
Separately, we construct an analytic model that calculates the theoretical cooling timescale at the dust thermal emission height for an assumed turbulent diffusion intensity.
By comparing this theoretical cooling timescale with the observationally derived value, we infer the dust grain size.

In the following subsections, we describe our methods step by step.
For the analytic model described in section \ref{subsec:method_ananlytic_model}, we first parameterize the dust grain size and its distribution (subsection \ref{subsubsec:method_dust_size}) and estimate the disk temperature based on the observed dust continuum intensity (subsection \ref{subsubsec:method_temperature}). 
Next, we determine the vertical density profiles for the gas and dust (subsection \ref{subsubsec:method_vertical_profile}). 
By adopting specific dust opacities and porosities (subsection \ref{subsubsec:method_opacity}), we calculate the optical depth to estimate the emission height at a submillimeter wavelength (subsection \ref{subsubsec:method_height}). 
Finally, we compute the vertical profile of the cooling timescale to extract its theoretical value at this emission height (subsection \ref{subsubsec:method_cooling}).

On the observational side described in section \ref{subsec:method_estimate_cooling}, we reconstruct the three-dimensional scattering surface and shadow geometry from infrared scattered-light images (subsection \ref{subsubsec:shadow_reconstruction}).
By comparing this geometry with the azimuthal brightness temperature profiles derived from submillimeter continuum maps, we measure the spatial offsets between the irradiated and shadowed regions. 
These offsets allow us to derive the local cooling timescale within the shadowed regions (subsection \ref{subsubsec:estimate_cooling}).
Finally, by comparing the theoretically predicted cooling timescale at the dust thermal emission height with the observationally derived value, we constrain the dust grain size following the procedure described in section \ref{subsec:method_procedure}.

\subsection{Analytic model for the cooling timescale}\label{subsec:method_ananlytic_model}

    \subsubsection{Dust size distribution models}\label{subsubsec:method_dust_size}
    
    Following \citet{FukuharaOkuzumi:2024aa}, we consider two different types of grain size distribution. 
    The first model is the power-law model, where the grain size distribution is given by
    \begin{equation}\label{eq:Sigmad_a}
        \frac{dN_{\rm d}(a)}{da} =
        \left\{\begin{array}{ll}
        {\displaystyle \frac{(12+3p)\Sigma_{\rm dust}}{4\pi \rho_{\rm int}\left(a_{\rm max}^{4+p}-a_{\rm min}^{4+p}\right)}a^{p},} & a_{\rm min} < a < a_{\rm max}, \\
            0, & {\rm otherwise}, 
        \end{array}\right. 
    \end{equation}
    where $dN_{\rm d}(a)/da$ is the number surface density per unit particle size $a$, $\Sigma_{\rm dust}$ is the total dust mass surface density, $\rho_{\rm int}$ is the grains' internal density, $p~(\neq -4)$ is the slope of the size distribution, and $a_{\rm min}$ and $a_{\rm max}$ are the minimum and maximum particle sizes, respectively.
    The dust surface density, the slope of the size distribution, and the minimum and maximum grain sizes serve as free parameters in this study (see also section \ref{subsec:parameter_choices}).
    Equation~\eqref{eq:Sigmad_a} satisfies the normalization
    \begin{equation}\label{eq:Sigmad}
        \Sigma_{\rm dust} = \int m_{\rm d}(a)\frac{dN_{\rm d}(a)}{da} da,
    \end{equation}    
    where $m_{\rm d} = (4\pi/3)\rho_{\rm int}a^3$ is the particle mass.

    The second model is the single-sized model where all dust particles are assumed to have equal size $a_1$. 
    Formally, the size distribution for the single-sized model can be written as 
    \begin{equation} \label{eq:Sigmad_1}
        \frac{dN_{\rm d}(a)}{da} = \frac{3\Sigma_{\rm dust}}{4\pi\rho_{\rm int}a_1^3} \delta(a-a_1)
    \end{equation}
    where $\delta$ is the delta function.
    When we adopt the single-sized model, we treat $a_1$ as a free parameter.
    Equation~\eqref{eq:Sigmad_1} also fulfills equation~\eqref{eq:Sigmad}.

    \subsubsection{Disk temperature}\label{subsubsec:method_temperature}
    We estimate the disk temperature using the observed intensity of the dust thermal emission.
    Assuming a vertical isothermal disk, the intensity at a frequency $\nu$ can be expressed as (e.g., \citealt{Carrasco-Gonzalez:2019aa,SierraLizano:2020aa,SierraPerez+:2024ab})
    \begin{equation}\label{eq:Iobs}
        I_{\nu,\rm obs} = B_\nu (T) \left[1-\exp{\left(-\frac{\tau_{\rm all}}{\mu}\right)}+\langle\omega_\nu\rangle F(\tau_{\rm all},~\langle\omega_\nu\rangle) \right],
    \end{equation}
    where $T$ is the disk temperature, $B_\nu (T)$ is the Planck function, $\tau_{\rm all}$ is the disk's vertical extinction optical thickness, $\mu \equiv \cos i$ is the cosine of the disk inclination $i$, $\langle\omega_\nu\rangle$ is an averaged dust albedo defined in equation~\eqref{eq:omega_avr}, and $F(\tau_{\rm all},~\langle\omega_\nu\rangle)$ is a function representing the effect of dust scattering on the emergent intensity. 
    Here, $\tau_{\rm all}$ is defined by
    \begin{equation}\label{eq:tau_all}
        \tau_{\rm all} = \int m_{\rm d}(a)\kappa_{{\rm d},\nu}(a)\frac{dN_{\rm d}(a)}{da} da,
    \end{equation}
    where $\kappa_{{\rm d},\nu}(a)$ is the opacity per dust mass at a frequency of $\nu$.
    The averaged dust albedo $\langle\omega_\nu\rangle$ depends on the size distribution and dust extinction opacity, which are described in subsection \ref{subsubsec:method_opacity}.
    The expression of $F(\tau_{\rm all},~\langle\omega_\nu\rangle)$ is given by
    \begin{equation}
        \begin{array}{@{}l@{}l}
            F(\tau_{\rm all}&,~\langle\omega_\nu\rangle) = \\ &\displaystyle\frac{F_1(\tau_{\rm all},~\langle\omega_\nu\rangle) + F_2(\tau_{\rm all},~\langle\omega_\nu\rangle)}{\exp{\left[-\sqrt{3\left(1-\langle\omega_\nu\rangle\right)}\tau_{\rm all}\right]\left(\sqrt{1-\langle\omega_\nu\rangle}-1\right)-\left(\sqrt{1-\langle\omega_\nu\rangle}+1\right)}},
        \end{array}
    \end{equation}
    where
    \begin{equation}
        F_1(\tau_{\rm all},~\langle\omega_\nu\rangle) = \frac{1-\exp{\left[-\sqrt{3\left(1-\langle\omega_\nu\rangle\right)}\tau_{\rm all}-\tau_{\rm all}/\mu\right]}}{\sqrt{3\left(1-\langle\omega_\nu\rangle\right)}\mu+1}
    \end{equation}
    and 
    \begin{equation}
        F_2(\tau_{\rm all},~\langle\omega_\nu\rangle) = \frac{\exp{\left(-\tau_{\rm all}/\mu\right)}-\exp{\left[-\sqrt{3\left(1-\langle\omega_\nu\rangle\right)}\tau_{\rm all}\right]}}{\sqrt{3\left(1-\langle\omega_\nu\rangle\right)}\mu-1}.
    \end{equation}
    For a given dust thermal emission intensity $I_{\nu,\rm obs}$, dust grain size $a$, and dust size distribution $dN_{\rm d}(a)/da$, we estimate the disk temperature at each disk position, $T$, using equation \eqref{eq:Iobs}.

    \subsubsection{Gas and dust vertical distributions}\label{subsubsec:method_vertical_profile}
    We adopt the cylindrical coordinate system $(R,~z)$, where $R$ and $z$ are the distance from the central star and the height from the midplane.

    From vertical hydrostatic equilibrium, the gas density is given by
    \begin{equation}\label{eq:gas_density}
        \rho_{\rm gas} = \frac{\Sigma_{\rm gas}}{\sqrt{2\pi}H_{\rm gas}}\exp{\left(-\frac{z^2}{2H_{\rm gas}^2}\right)},
    \end{equation}
    where $\Sigma_{\rm gas}$ is the gas surface density and $H_{\rm gas}=c_{\rm s}/\Omega_{\rm K}$ is the gas scale height with $c_{\rm s}$ and $\Omega_{\rm K}$ being the isothermal sound speed and Keplerian frequency, respectively.
    The isothermal sound speed is given by $c_{\rm s}=\sqrt{k_{\rm B}T/m_{\rm g}}$, where $k_{\rm B}$ is the Boltzmann constant and $m_{\rm g}$ is the mean molecular mass of the gas.
    The Keplerian frequency is given by $\Omega_{\rm K} = \sqrt{GM_*/R^3}$, where $G$ is the gravitational constant and $M_*$ is the mass of the central star.
    We set $m_{\rm g} = 2.3m_{\rm p}$, where $m_{\rm p}$ is the proton mass.
    In this study, $\Sigma_{\rm gas}$ serves as a free parameter.

    Assuming the balance between settling and diffusion, the vertical distribution of the dust particles can be written as \citep{TakeuchiLin2002}
    \begin{equation}\label{eq:rhoda}
        \frac{dn_{\rm d}(a,z)}{da} = C_{\rm d}(a) \exp{\left[-\frac{z^2}{2H_{\rm gas}^2}-\frac{{\rm St}_{\rm mid}(a)}{\alpha_z}\left(\exp\frac{z^2}{2H_{\rm gas}^2}-1\right) \right]},
    \end{equation}
    where $dn_{\rm d}(a,z)/da$ is the particle number density per unit radius at height $z$, $\mathrm{St}_{{\rm mid}}(a)$ is the Stokes number of the particles at the midplane, $\alpha_z$ is a dimensionless parameter that characterizes the level of the dust vertical diffusion caused by turbulence, and $C_{\rm d}(a)$ is the normalized constant determined by the condition $dN_{\rm d}(a)/da = \int (dn_{\rm d}(a,z)/da)dz$.
    Because most dust particles lie at the region of $z \ll H_{\rm gas}$, the exponential factor in equation \eqref{eq:rhoda} can be approximated as $\exp[-z^2/(2H_{\rm d}^2)]$, yielding \citep{FukuharaOkuzumi+:2021ca}
    \begin{equation}\label{eq:C_d2}
       C_{\rm d}(a) = \frac{1}{\sqrt{2\pi}H_{\rm d}(a,\alpha_z)}\frac{dN_{\rm d}(a)}{da}.
    \end{equation}
    Here, $H_{\rm d}(a,\alpha_z)$ is the scale height of particles with size $a$ given by \citep{Dubrulle+1995,YoudinLithwick2007}
    \begin{equation}\label{eq:Hd}
        H_{\rm d}(a,\alpha_z) = \left[1+\frac{\mathrm{St}_{{\rm mid}}(a)}{\alpha_z} \right]^{-1/2}H_{\rm gas}.
    \end{equation}
    The Stokes number is the product of the stopping time and Keplerian frequency. 
    Assuming that the particle radius is smaller than the mean free path of the gas molecules, gas drag onto the particles follows Epstein's law, which gives \citep[see, e.g., ][]{BirnstielDullemond+:2010aa}
    \begin{equation}\label{eq:St}
        \mathrm{St}_{{\rm mid}}(a) = \frac{\pi \rho_{\rm int} a}{2\Sigma_{\rm gas}}.
    \end{equation}

    The vertical--size distribution $dn_{\rm d}(a,z)/da$ gives the collisional heat transfer and opacity, which control the cooling timescale, as a function of $z$. The mean travel length of gas molecules colliding with dust particles $\ell_{\rm gd}$ is given by 
    \begin{equation}\label{eq:l_dg}
        \ell_{\rm gd} = \left(\int\pi a^2\frac{dn_{\rm d}}{da}  da\right)^{-1}.
    \end{equation}
    This determines the timescale of collisional heat transfer [see equation \eqref{eq:t_coll} in subsection \ref{subsubsec:method_cooling}].

    \subsubsection{Dust opacity and porosity}\label{subsubsec:method_opacity}

    To calculate the disk temperature, emission height, and cooling timescale, we use the frequency-dependent, effective dust extinction opacity per dust mass, denoted as $\kappa_{{\rm d},\nu}(a)$.
    For a given dust grain size $a$ and a a given frequency $\nu$, we evaluate $\kappa_{{\rm d},\nu}(a)$ as
    \begin{equation}\label{eq:kappa_dust}
        \kappa_{\rm d,\nu}(a) = \kappa_{\rm abs,\nu}(a) + \kappa_{\rm sca,eff,\nu}(a), 
    \end{equation}
    where $\kappa_{\rm abs,\nu}$ and $\kappa_{\rm sca,eff,\nu}$ are the dust absorption and effective dust scattering opacities per unit dust mass at different frequencies $\nu$.
    Here, the effective dust scattering opacity is given by
    \begin{equation}
        \kappa_{\rm sca,eff,\nu}(a) = \left[1-g_\nu(a)\right]\kappa_{\rm sca,\nu}(a),
    \end{equation}
    where $\kappa_{\rm sca,\nu}$ is the scattering opacity per unit dust mass and $g_\nu$ is the asymmetry parameter at different frequencies $\nu$.
    Because the second term in equation \eqref{eq:kappa_dust} represents the scattering opacity for isotropic scattering, $\kappa_{\rm d,\nu}(a)$ corresponds to the effective extinction opacity.
    We also define the dust albedo as
    \begin{equation}
        \omega_\nu(a) \equiv \frac{\kappa_{\rm sca,eff,\nu}(a)}{\kappa_{\rm d,\nu}(a)}.
    \end{equation}
    
    To calculate the disk temperature in section \ref{subsubsec:method_temperature}, we use the average of the dust albedo weighted by the dust surface density at a frequency of $\nu$, $\langle\omega_\nu\rangle$, given by
    \begin{equation}
        \langle\omega_\nu\rangle = \frac{\langle\kappa_{\rm sca,eff,\nu}\rangle}{\langle\kappa_{\rm d,\nu}\rangle}.
        \label{eq:omega_avr}
    \end{equation}
    Here, $\langle\kappa_{\rm sca,eff,\nu}\rangle$ and $\langle\kappa_{\rm d,\nu}\rangle$ are the averages of the scattering and extinction dust opacities weighted by the dust surface density at a frequency of $\nu$, defined as    
    \begin{equation}
        \langle\kappa_{\rm sca,eff,\nu}\rangle \equiv {\displaystyle \frac{\int \kappa_{\rm sca,eff,\nu}(a) m_{\rm d}(a)\frac{dN_{\rm d}(a)}{da} da}{\Sigma_{\rm dust}}},
    \end{equation}
    and 
    \begin{equation}
        \langle\kappa_{\rm d,\nu}\rangle\equiv {\displaystyle \frac{\int \kappa_{\rm d,\nu}(a) m_{\rm d}(a)\frac{dN_{\rm d}(a)}{da} da}{\Sigma_{\rm dust}}},
    \end{equation}
    respectively.

    For the fiducial model, we use the opacity model based on \citet{RicciTesti+:2010aa} to compute $\kappa_{\rm d,\nu}$.
    Following \citet{ZormpasBirnstiel+:2022aa} and \citet{DelussuBirnstiel+:2024aa}, we assume spherical dust grains composed of a mixture of astronomical silicates \citep{Draine:2003xf}, amorphous carbon \citep{ZubkoMennella+:1996aa}, and water ice \citep{WarrenBrandt:2008aa} with a volume mixing ratio of 1:2:3 \citep{RicciTesti+:2010aa}.
    The material density of the dust grains, $\rho_{\rm m}$, is set to $1.6~{\rm g~cm^{-3}}$, assuming intrinsic material densities of $3.3$, $1.8$, and $0.92~{\rm g~cm^{-3}}$ for astronomical silicates, amorphous carbon, and water ice, respectively \citep{DominikMin+:2021aa}.

    Alternative opacity models that assume different dust compositions from our fiducial model lead to changes in $\kappa_{\rm d,\nu}$ and thereby affect the optical depth and dust albedo.
    In appendix \ref{appendix:opacity}, we describe these alternative opacity models and investigate their impacts on our results.
    However, within our analytic model, the choice of opacity model has only a minor effect on the estimated dust grain size (see appendix \ref{appendix:result_opacity}).

    Furthermore, we incorporate dust porosity, which alters both the dust extinction opacities and the internal density of the grains.
    By introducing porosity, the dust internal density $\rho_{\rm int}$ can be expressed as
    \begin{equation}\label{eq:rho_int}
        \rho_{\rm int} = f_{\rm dust}\rho_{\rm m},
    \end{equation}
    where $f_{\rm dust}$ is the dust filling factor.
    For completely compact grains, $f_{\rm dust} = 1$, whereas for highly porous grains, $f_{\rm dust}$ approaches zero.

    To calculate $\kappa_{\rm abs, \nu}$, $\kappa_{\rm sca, \nu}$, and $g_\nu$ for a given dust grain size $a$ and filling factor $f_{\rm dust}$ under a specific opacity model, we utilize the \texttt{optool} package \citep{DominikMin+:2021aa}.
    The profiles of $\kappa_{\rm abs, \nu}$, $\kappa_{\rm sca,eff, \nu}$, and $g_\nu$ as functions of the dust grain size, for different opacity models and filling factors, are presented in figure \ref{fig:opacity_Ricci_DIANA_DSHARP} in appendix \ref{appendix:opacity}.

    \subsubsection{Emission/scattering surface height}\label{subsubsec:method_height}
    Using the dust vertical distribution and opacity models, we calculate the emission and scattering surface heights, $z_{\rm emi}$ and $z_{\rm sca}$, respectively, probed by observations at a given frequency $\nu$.
    The vertical optical depth measured from infinity down to a height $z$ is given by
    \begin{equation}\label{eq:tau_z}
        \tau_\nu (z) = \int_z^{\infty} dz^\prime \int da ~ m_{\rm d}(a)\kappa_{{\rm d},\nu}(a)\frac{dn_{\rm d}(a,~z^\prime)}{da}.
    \end{equation}
    Both $z_{\rm emi}$ and $z_{\rm sca}$ are defined analogously, as the height $z$ that satisfies 
    \begin{equation}\label{eq:z_tau_1}
        \frac{\tau_\nu (z)}{\mu'} = 1,
    \end{equation}
    where $\mu'$ denotes the sine of the grazing angle between the radiation and the local disk surface, and takes a different value for each case, as described below.
    This height approximates the effective height of the emitting or scattering layer.
    If the disk is optically so thin that $\tau(z=0)<\mu'$, we set the emission/scattering height to the disk midplane ($z_{\rm emi}=0$ and $z_{\rm sca}=0$).

    For the emission surface height at submillimeter wavelengths, $z_{\rm emi}$, the relevant radiation is that received along the line of sight to the observer, so we set $\mu'$ to the disk's inclination $\mu$.
    For the scattering surface height at near-infrared wavelengths, $z_{\rm sca}$, the relevant radiation is starlight incident on the disk surface.
    Therefore, $\mu'$ corresponds to the sine of the grazing angle at which this starlight illuminates the surface.
    Under the small-angle approximation, this grazing angle can be approximated by the aspect ratio of the scattering surface, $z_{\rm sca}/R$.
    We accordingly adopt $\mu'=0.1$, which is consistent with $z_{\rm sca}/R\approx0.1$ derived for the disk analyzed in this study (see subsection \ref{subsubsec:AnalyticModel_results_fiducial}).

    \subsubsection{Cooling model}\label{subsubsec:method_cooling}
    
    The spatial profile of the gas disk cooling rate depends on the size and spatial distribution of the dust particles \citep{Malygin+2017,BarrancoPei+:2018kc,FukuharaOkuzumi+:2021ca,BaeTeague+:2021aa}.
    Based on our dust model, we approximate the local cooling (thermal relaxation) timescale $t_{\rm cool}$ as \citep{Malygin+2017,PfeilKlahr2019,FukuharaOkuzumi:2024aa}
    \begin{equation}\label{eq:taurelax}
        t_{\rm cool}(z) = \max \{t_{\rm diff}(z),~t_{\rm coll}(z),~t_{\rm emit}(z)\},
    \end{equation}
    where $t_{\rm diff}$, $t_{\rm coll}$, and $t_{\rm emit}$ are the timescales of radiative diffusion, collisional heat transfer, and radiative cooling, respectively.
    For $t_{\rm diff}$ and $t_{\rm emit}$, we use equations (17), (18), and (21) of \citet{FukuharaOkuzumi:2024aa}\footnote{Note that $t_{\rm diff}$ and $t_{\rm emit}$ correspond to $\tau_{\rm diff}$ and $\tau_{\rm emit}$ in \citet{FukuharaOkuzumi:2024aa}, respectively.}.

    Because we focus on optically thin outer disk regions, collisional heat transfer dominates the cooling process in most of our cases (see also section \ref{subsec:AnalyticModel_results}).
    The timescale $t_{\rm coll}$ of collisional heat transfer between the gas molecules and dust particles is given by \citep{BurkeHollenbach:1983aa,BarrancoPei+:2018kc,PfeilBirnstiel+:2023aa}
    \begin{equation}\label{eq:t_coll}
        t_{\rm coll} = \frac{\gamma}{\gamma-1}\frac{\ell_{\rm gd}}{v_{\rm th}},
    \end{equation}
    where $v_{\rm th}$ is the mean relative velocity between the gas molecules and dust particles and $\gamma$ is the heat capacity ratio, which we set to $1.4$.
    In this study, we approximate $v_{\rm th}$ as the mean thermal speed of the gas molecules: $v_{\rm th} =\sqrt{8k_{\rm B}T/\pi m_{\rm g}}$.

\subsection{Observational estimation of the cooling timescale}\label{subsec:method_estimate_cooling}

    \subsubsection{Inferring the cooling time from azimuthal relaxation}\label{subsubsec:estimate_cooling}
    \begin{figure*}[t]
        \begin{center}
        \includegraphics[width=\hsize,bb = -50 0 800 380]{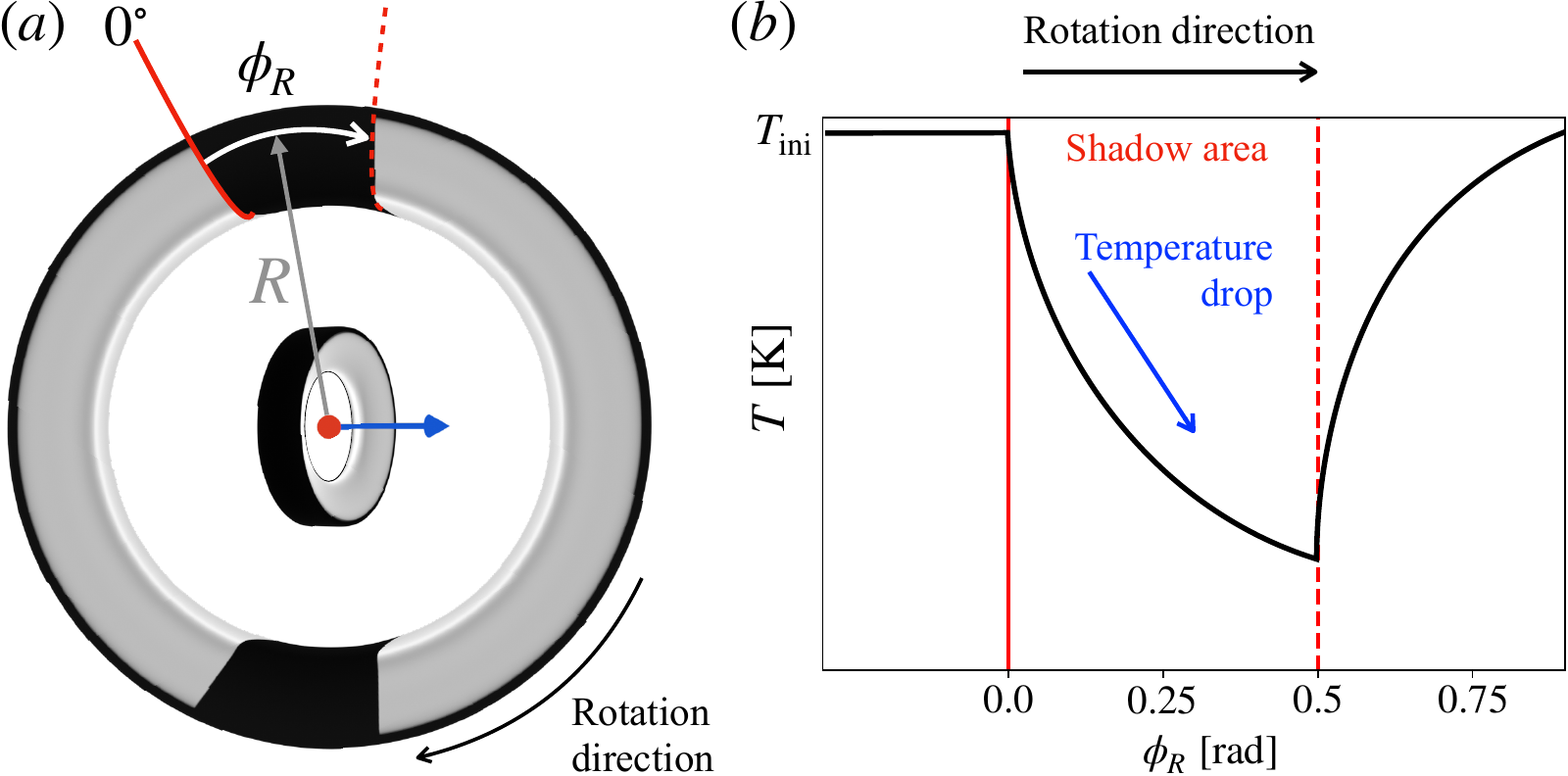}
        \end{center}
        \caption{Schematic illustration of the observational method used to estimate the cooling timescale from the azimuthal temperature variation across a disk shadow. 
        (a) Geometry of the disk at a representative radius $R$. The shadowed sector is shown in black and is bounded by the shadow edges indicated in red. The azimuthal angular distance $\phi_R$ is measured from the upstream shadow boundary along the direction of disk rotation. 
        (b) Schematic temperature profile along the same annulus. The temperature initially equals $T_{\rm ini}$, decreases as the disk material passes through the shadowed region, and subsequently recovers after leaving the shadow as it relaxes toward the unshadowed equilibrium temperature. The red solid and dashed vertical lines mark the entrance and exit boundaries of the shadow, respectively.
    }
        \label{fig:temperature_phi}
    \end{figure*}
    
    In this section, we describe how we infer the radiative cooling timescale directly from observational diagnostics. The basic geometry of the method is illustrated in the left panel of figure~\ref{fig:temperature_phi}. By relating the spatial offset from the observed shadow boundary to the advection time under Keplerian rotation, we convert an observed azimuthal relaxation pattern into an estimate of $t_{\rm cool}$.

    We estimate the cooling timescale from the observed azimuthal relaxation of the disk temperature across a shadow boundary, as schematically shown in the right panel of figure~\ref{fig:temperature_phi}. We assume that a dust advected by Keplerian rotation relaxes toward a local equilibrium, or target, temperature $T_{\rm tar}$ on a characteristic timescale $t_{\rm cool}$. Let $\phi_R$ be the azimuthal angular distance from the observed shadow boundary, measured along the direction of rotation. Under the assumption of first-order cooling, the temperature evolution is written as
    \begin{equation}
    T(\phi_R) = T_{\rm tar} + (T_{\rm ini}-T_{\rm tar})
    \exp\left[-\frac{\phi_R}{t_{\rm cool,est}\Omega_{\rm K}}\right],
    \label{eq:Trelax}
    \end{equation}
    where $T_{\rm ini}$ is the temperature immediately upstream of the boundary and $t_{\rm cool,est}$ is the cooling timescale inferred from observations.
    
    In practice, we first locate the shadow boundary from the scattered-light geometry reconstruction described in section~\ref{subsubsec:shadow_reconstruction}. We then construct an azimuthal brightness-temperature profile, $T_{\rm b}(\phi_R)$, from the continuum data at each radius and use this brightness temperature as $T$ in equation~(\ref{eq:Trelax}). The continuum specific intensity at a frequency $\nu$, $I_\nu$, is converted to $T_{\rm b}$ by inverting the Planck function,
    \begin{equation}
    T_{\rm b}=
    \frac{h\nu}{k_{\rm B}}
    \left[
    \ln\left(
    1+\frac{2h\nu^3}{c^2 I_\nu}
    \right)
    \right]^{-1},
    \label{eq:planck}
    \end{equation}
    where $h$ is the Planck constant and $c$ is the light speed.
    Finally, we fit equation~(\ref{eq:Trelax}) to the observed azimuthal temperature profile to obtain $t_{\rm cool,est}\Omega_{\rm K}$.

    \subsubsection{Determination of the Shadow Boundaries}\label{subsubsec:shadow_reconstruction}
    We reconstruct the three-dimensional scattering-surface geometry of the outer disk using the shadow-based framework developed by \citet{OriharaMomose:2025aa}. This framework jointly fits the horizon curve, the shadow boundaries, and the apparent outer-edge curve in scattered-light images. Simultaneously fitting these features constrains the radial variation of the outer-disk scattering-surface height, together with the orientation and finite thickness of the inner disk.
    We adopt the same coordinate system, geometric definitions, and analytic construction of the model curves as in \citet{OriharaMomose:2025aa}, and summarize below only the parameterization relevant to our analysis.
    
    Following \citet{OriharaMomose:2025aa}, we express the outer-disk scattering surface as a parametric surface $\So(H_{\rm surf},\phi_\mathrm{o})$ in the outer-disk frame, where $H_{\rm surf}$ is the surface height above the midplane and $\phi_\mathrm{o}$ is the azimuthal angle.
    The cylindrical radius is modeled as a function of height, $R(H_{\rm surf})$, parameterized by
    \begin{equation}
    R(H_{\rm surf}) = \left[R_{\rm c}^q + (R_{\rm c}^q - R_{\rm e}^q)\left(\frac{|H_{\rm surf}|}{H_{\rm surf, c}}\right)^q\right]^{1/q},
    \label{eq:surface_height}
    \end{equation}
    where $R_{\rm e}$ is the radius of the inner edge of the outer disk, $R_{\rm c}$ is a characteristic radius, $H_{\rm surf, c}$ is the surface height at $R_c$, and $q$ controls the curvature of the scattering surface (see figure 1 of \citealt{OriharaMomose:2025aa}).
    
    We follow \citet{OriharaMomose:2025aa} to extract and fit the horizon curve, the shadow boundaries, and the apparent outer-edge curve from scattered-light images.
    Specifically, we sample data points from pixels with large image gradients to construct three sets of points corresponding to the horizon, the shadow boundaries, and the apparent outer edge.
    The horizon and apparent outer-edge points are identified from the positive and negative components of the radial image gradients, respectively, whereas the shadow-boundary points are identified from the azimuthal image gradients.
    We then fit these three point sets simultaneously with the corresponding model curves by minimizing the residual distances, thereby estimating the best-fit parameters.

\subsection{Calculation procedure}\label{subsec:method_procedure}

We constrain the maximum dust grain size $a_{\rm max}$ and the dust surface density $\Sigma_{\rm dust}$ by comparing the cooling timescale estimated from observations, $t_{\rm cool, est}$, with that calculated from an analytic model, $t_{\rm cool, cal}$ (see also figure \ref{fig:overview}). 
First, we estimate the local cooling timescale from the near-infrared and submillimeter observations in the following steps.

\begin{enumerate}
    \item Reconstruct the three-dimensional scattering-surface geometry from near-infrared scattered-light observations using equation \eqref{eq:surface_height}, and identify the shadowed regions cast on the outer disk.
    \item Obtain the azimuthal brightness temperature profile in the outer disk region using submillimeter dust thermal emission observations. In this step, we approximate the brightness temperature as the local disk temperature, assuming that the emission is optically thick at the observed wavelength.
    \item Estimate the empirical cooling timescale $t_{\rm cool,est}$ in the shadowed region from the azimuthal temperature profile using equation \eqref{eq:Trelax}. This estimate assumes that the temperature variation reflects the gas's thermal relaxation as it passes through the shadow.
\end{enumerate}

Second, we use an analytic model to calculate the cooling timescale for a given dust grain size and dust surface density, following the steps below.

\begin{enumerate}
\setcounter{enumi}{3}
    \item Determine the dust size distribution for a given set of $a_{\rm max}$ and $\Sigma_{\rm dust}$ using equation \eqref{eq:Sigmad_a} to obtain a theoretically predicted cooling timescale from the analytic model.
    \item Estimate the disk temperature using equation \eqref{eq:Iobs} based on the observed dust thermal emission intensity and the assumed dust size distribution. 
    \item Calculate the vertical density distributions of the gas and dust using equations \eqref{eq:gas_density} and \eqref{eq:rhoda}.
    \item Determine the emission height $z_{\rm emi}$ at the observed submillimeter wavelength using equation \eqref{eq:z_tau_1} based on the adopted dust opacities. Simultaneously, compute the vertical distribution of the cooling timescale using equation \eqref{eq:taurelax}.
    \item Extract the theoretical cooling timescale at this emission height, which we denote hereafter as $t_{\rm cool,cal} \equiv t_{\rm cool}(z_{\rm emi})$.
\end{enumerate}

Finally, we compare the calculated cooling timescale $t_{\rm cool,cal}$ with the observationally derived value, $t_{\rm cool,est}$, to evaluate whether the assumed $a_{\rm max}$ and $\Sigma_{\rm dust}$ successfully reproduce the observationally derived cooling timescale.
We repeat steps 4 through 8 for every set of $a_{\rm max}$ and $\Sigma_{\rm dust}$ within our parameter space (see subsection \ref{subsec:parameter_choices}) to search for the ranges that best match the observational data.
We note that because a single observational constraint (the cooling timescale) is compared against two free parameters ($a_{\rm max}$ and $\Sigma_{\rm dust}$), the best-fit parameter set is not uniquely determined, and the model remains degenerate.

\section{Observation data of the HD~142527 disk and parameter choices}\label{sec:dataset}

We apply the methods described in Section~\ref{sec:method} to the protoplanetary disk around HD~142527. This is a young Herbig Ae/Be system located at a distance of approximately $157~{\rm pc}$ \citep{Gaia-CollaborationVallenari+:2023aa}, with an age of about $3~{\rm Myr}$. The primary star has a mass of $M_\star\simeq2.4M_\odot$ \citep{FukagawaTsukagoshi+:2013aa,ArunMathew+:2019aa}. The disk around HD~142527 is a well-studied transitional disk system characterized by a large dust-depleted cavity and a highly non-axisymmetric outer disk. 
Previous infrared and millimeter observations have shown that the disk contains a prominent cavity extending to $\sim 140~{\rm au}$, while substantial gas remains inside the cavity \citep{PerezCasassus+:2015aa}. 
The outer disk exhibits a strong azimuthal asymmetry in the millimeter continuum emission, often interpreted as a dust trap \citep{FukagawaTsukagoshi+:2013aa,CasassusWright+:2015aa}. 
In scattered-light observations, two localized intensity decrements have been identified along the outer disk and interpreted as shadows cast by a misaligned inner disk \citep{MarinoPerez+:2015aa,AvenhausQuanz+:2017aa}. 
These properties make HD~142527 a particularly suitable target for investigating how stellar irradiation blocked by the inner disk affects the outer disk's thermal structure. 
Although the system hosts a low-mass companion of approximately $0.2M_\odot$, we do not include it in the analytic model because its orbit is too compact to account for the outer disk features considered here \citep{NowakRowther+:2024aa}.

    \subsection{Near-infrared data}\label{subsec:NIR_observation}
    
    To extract the three-dimensional morphology of the shadows projected onto the outer disk surface, we apply the method of \citet{OriharaMomose:2025aa} to a near-infrared scattered-light image of the HD~142527 disk. We use the publicly available H-band polarimetric differential image presented by \citet{HunzikerSchmid+:2021aa}, obtained with the InfraRed Dual-band Imager and Spectrograph (IRDIS) of the Spectro-Polarimetric High-contrast Exoplanet REsearch instrument (SPHERE) at the Very Large Telescope (VLT). The image is a reduced FITS product based on the SPHERE/IRDIS broadband H-band ($1.625~\mu{\rm m}$) observations obtained on May, 31, 2017. The image has a format of $1024 \times 1024$ pixels, a pixel scale of $12.27\,{\rm mas\,pixel}^{-1}$, and an angular resolution of $44\,{\rm mas}$.\footnote{The IRDIS H-band FITS product used here is available from the CDS/VizieR catalogue J/A+A/648/A110, \textit{HD142527 SPHERE polarimetric images}. The corresponding ESO programme ID is 099.C-0601(A).}
    
    We use the H-band $Q_\phi$ image shown in figure~\ref{fig:HD142527_surface}(a) of section \ref{subsec:shadow_result} for the shadow analysis, adopting the same outer-disk geometry as in the temperature analysis below, namely the inclination and position angle of the outer disk. The $Q_\phi$ signal is obtained by transforming the Stokes $Q$ and $U$ images into the azimuthal polarization basis centered on the star, and mainly traces the linearly polarized component of stellar light scattered by the disk surface. It is therefore suitable for investigating both the scattered-light distribution and the non-axisymmetric illumination pattern on the outer-disk surface. In this work, we interpret the azimuthally localized dark regions in the $Q_\phi$ image as shadows cast onto the outer-disk surface by the misaligned inner disk or by material close to the central star.

    To assess the significance of the scattered-light structures, we construct a signal-to-noise ratio (S/N) map using the $U_\phi$ image shown in figure~\ref{fig:images}(b) of appendix \ref{appendix:images} as an empirical estimate of the residual noise. For single scattering under centro-symmetric illumination, polarized scattered light is expected to be contained predominantly in $Q_\phi$, whereas $U_\phi$ should contain little astrophysical signal and is therefore often used as a practical noise estimate in polarized images \citep[e.g.,][]{CanovasMenard+:2015aa}. However, non-azimuthal polarization signals can appear in $U_\phi$ owing to effects such as multiple scattering. Therefore, the S/N map used here should be regarded not as a strict statistical significance map, but as an empirical measure of the scattered-light structures relative to the local residual level. Specifically, we measure the rms fluctuation of the $U_\phi$ image in concentric annuli centered on the star to obtain a radial noise profile, $U_{\phi,\mathrm{rms}}(r)$. The resulting S/N map, shown in figure~\ref{fig:images}(c) of appendix \ref{appendix:images}, is defined as $Q_\phi/U_{\phi,\mathrm{rms}}(r)$ and is used both to evaluate the local significance of the scattered-light structures and to weight the fitting data in the shadow reconstruction.
    
    For the gradient-based extraction of the geometric features, we first apply an inverse hyperbolic sine stretch to the S/N map and then calculate the image gradients. The asinh transformation enhances faint but significant structures while preserving the sign of the signal and remaining approximately linear around zero. These properties are important because the S/N map can contain zero or negative values owing to noise and residual systematics, for which a logarithmic stretch is not well defined. From the gradients of the asinh-stretched S/N map, we identify candidate data points for the horizon curve, shadow boundaries, and apparent outer-edge curve. Specifically, the shadow-boundary points are sampled from pixels with large azimuthal gradients, $\nabla({\rm S/N})\cdot\hat{\boldsymbol{e}}_\phi$, as shown in figure~\ref{fig:images}(d) of appendix \ref{appendix:images}, because the shadows appear as localized brightness variations along the azimuthal direction. The horizon-curve points are sampled from pixels with large positive radial gradients, $\nabla({\rm S/N})\cdot\hat{\boldsymbol{e}}_r>0$, shown in figure~\ref{fig:images}(e) of appendix \ref{appendix:images}, whereas the apparent outer-edge points are sampled from pixels with large negative radial gradients, $\nabla({\rm S/N})\cdot\hat{\boldsymbol{e}}_r<0$, shown in figure~\ref{fig:images}(f) of appendix \ref{appendix:images}.
    
    We then simultaneously fit the sampled data points with the corresponding model curves for the horizon, shadow boundaries, and apparent outer edge. In the MCMC fitting, each sampled point is weighted by its local S/N value, so that higher-significance features have a larger contribution to the likelihood than lower-significance or noise-dominated points. This weighting scheme allows the model to be constrained primarily by robust scattered-light structures while still using the full set of gradient-selected data points. The posterior distributions of the model parameters are explored with an MCMC sampler implemented with \texttt{emcee} \citep{Foreman-MackeyHogg+:2013aa}.

    \subsection{Submillimeter data}\label{subsec:submm_observation}

    We use archival ALMA Band~9 data from project 2015.1.00614.S. The calibrated measurement sets were provided by the East Asian ALMA Regional Center following a support request. Starting from these calibrated visibility data, we performed our own continuum imaging and self-calibration using the Common Astronomy Software Applications package (CASA; \citealt{CASA-TeamBean+:2022aa}). In the self-calibration procedure, scans and spectral windows were combined to derive the calibration solutions. We first carried out iterative phase-only self-calibration with solution intervals of \texttt{solint=inf}, \texttt{60s}, \texttt{30s}, and \texttt{10s}, followed by one final amplitude-and-phase self-calibration with \texttt{solint=inf}. The continuum emission was deconvolved using multiscale CLEAN with spatial scales of 0, 1, 2, 3, and 5 times the synthesized beam size.

    The self-calibration improved the peak signal-to-noise ratio from 22 to 67, while leaving the integrated continuum flux of the disk essentially unchanged. The final self-calibrated image has a synthesized beam of $0.26'' \times 0.21''$ and an rms noise level of $8\times10^{-3}~{\rm Jy~beam}^{-1}$. We converted the continuum specific intensity to brightness temperature by inverting the Planck function given in equation~\ref{eq:planck}. The resulting brightness-temperature map, shown in figure~\ref{fig:HD142527_temp}(a) of section \ref{subsec:shadow_result}, was used to derive the azimuthal temperature distribution.

    \subsection{Parameter choices}\label{subsec:parameter_choices}
    \begin{table*}[t]
    \tbl{Parameter choices for the analytic model presented in section \ref{sec:method} for all runs. }{
    \begin{tabular}{lcccccc} 
    \hline \hline
    Model & $\Sigma_{\rm gas}$ & $p$ & $a_{\rm min}$ & $f_{\rm dust}$ & opacity model &  \\ \hline 
    Fiducial & $10~{\rm g~cm^{-2}}$ & $-2.5$ & $1~{\rm \mu m}$ & $1.0$ & Ricci opacity & subsection \ref{subsubsec:AnalyticModel_results_fiducial} \\
    Low-Gas & $1~{\rm g~cm^{-2}}$ & $-2.5$ & $1~{\rm \mu m}$ & $1.0$ & Ricci opacity & subsection \ref{subsubsec:AnalyticModel_results_low_SigmaGas} \\
    Single-Size & $10~{\rm g~cm^{-2}}$ & -- & -- & $1.0$ & Ricci opacity & subsection \ref{subsubsec:AnalyticModel_results_SizeDis} \\
    Steep-SizeDis & $10~{\rm g~cm^{-2}}$ & $-3.5$ & $1~{\rm \mu m}$ & $1.0$ & Ricci opacity & subsection \ref{subsubsec:AnalyticModel_results_SizeDis} \\
    Small-amin & $10~{\rm g~cm^{-2}}$ & $-2.5$ & $0.1~{\rm \mu m}$ & $1.0$ & Ricci opacity & subsection \ref{subsubsec:AnalyticModel_results_SizeDis} \\
    Steep-SizeDis-Small-amin & $10~{\rm g~cm^{-2}}$ & $-3.5$ & $0.1~{\rm \mu m}$ & $1.0$ & Ricci opacity & subsection \ref{subsubsec:AnalyticModel_results_SizeDis} \\
    Mid-Fill & $10~{\rm g~cm^{-2}}$ & $-2.5$ & $1~{\rm \mu m}$ & $0.3$ & Ricci opacity & subsection \ref{subsubsec:AnalyticModel_results_porosity}\\
    Low-Fill & $10~{\rm g~cm^{-2}}$ & $-2.5$ & $1~{\rm \mu m}$ & $0.1$ & Ricci opacity & subsection \ref{subsubsec:AnalyticModel_results_porosity} \\
    DIANA & $10~{\rm g~cm^{-2}}$ & $-2.5$ & $1~{\rm \mu m}$ & $1.0$ & DIANA opacity & appendix \ref{appendix:result_opacity} \\
    DSHARP & $10~{\rm g~cm^{-2}}$ & $-2.5$ & $1~{\rm \mu m}$ & $1.0$ & DSHARP opacity & appendix \ref{appendix:result_opacity} \\
    \hline
    \end{tabular}} 
    \label{table:run}
    \end{table*}
    
    Throughout this study, we adopt a fixed outer-disk geometry with an inclination of $i_{\rm o}=27^\circ$ and a position angle of ${\rm PA}_{\rm o}=341^\circ$, following the values reported by \citet{FukagawaTsukagoshi+:2013aa}. The same geometry is used consistently in both the shadow analysis and the continuum temperature analysis.

    In the shadow analysis, all other geometric parameters are treated as free parameters. These include the sky-plane offsets of the outer-disk center, $(c_{{\rm o}x},c_{{\rm o}y})$; the inner-edge radius of the outer disk, $R_{\rm e}$; the surface height, $H_{\rm surf,c}$, at the characteristic radius, $R_{\rm c}$; the power-law index, $q$, of the outer-disk surface profile; the apparent outer-edge radius of the outer disk, $R_{\rm out}$; the inclination and position angle of the inner disk, $(i_{\rm i},{\rm PA}_{\rm i})$; and the maximum aspect ratio of the shadow-casting inner structure, $h_r$. The inner-disk center is assumed to coincide with the stellar position, and the sky-plane offsets $(c_{{\rm i}x},c_{{\rm i}y})$ are therefore fixed to $(0,0)$. 

    To estimate the cooling timescale, we adopted uniform priors of $27$--$30~\mathrm{K}$ and $30$--$32~\mathrm{K}$ for the asymptotic temperature, $T_{\rm tar}$, and the initial temperature, $T_{\rm ini}$, respectively. We then used MCMC sampling to infer the posterior distributions of the model parameters. The dimensionless cooling timescale, $t_{\rm cool}\Omega_{\rm K}$, was constrained using the 68\% Bayesian credible interval of its marginalized posterior distribution.

    For the analytic calculation of the cooling timescale, we use the intensity of dust thermal emission at ALMA Band 9 submillimeter wavelengths.
    We fix the radial distance $R$ to $170~{\rm au}$ and the observed dust thermal emission intensity at this radius, $I_{\nu,\rm obs}$, to $0.4~{\rm Jy~beam}^{-1}$ with the beam of $0.26'' \times 0.21''$ at an observing frequency of $\nu = 697~{\rm GHz}$ (corresponding to a wavelength of $\lambda = 0.43~{\rm mm}$).
    In the analytic model, the mass of the central star and outer disk inclination are set to $M_\star=2.4M_\odot$ and $i=27^\circ$, respectively.
    As main constrained parameters, we consider the maximum size of dust grains $a_{\rm max}$ for the power-law size distribution model or the size of dust grains $a_1$ for the single-sized model, and dust surface density $\Sigma_{\rm dust}$.
    We divide the parameter ranges $10~{\rm \mu m}<a<1~{\rm cm}$ or $10~{\rm \mu m}<a_{\rm max}<1~{\rm cm}$ and $0.1~{\rm g~cm^{-2}}<\Sigma_{\rm dust}<10~{\rm g~cm^{-2}}$ into grids of equal logarithmic intervals of $0.1$.
    In addition to the dust grain size and dust surface density, our model involves five free parameters: the gas surface density $\Sigma_{\rm gas}$, the slope of the size distribution $p$, the minimum dust grain size $a_{\rm min}$, the dust filling factor $f_{\rm dust}$, and opacity models.
    Our parameter sets are summarized in table \ref{table:run}.
    As the fiducial model, we select $\Sigma_{\rm gas}=10~{\rm g~cm^{-2}}$, $p=-2.5$, $a_{\rm min}=1~{\rm \mu m}$, $f_{\rm dust}=1.0$, and Ricci opacity model.

\section{Results}\label{sec:results}

In this section, we present the results of applying our method described in section \ref{sec:method} to the protoplanetary disk around HD~142527.
We begin by determining the shadow geometry and estimating the cooling timescale at the disk using the infrared-wavelength scattered light and millimeter-wavelength dust thermal emission observations in subsection \ref{subsec:shadow_result}.
We then show the constraint for the maximum dust grain size that can reproduce the observationally estimated cooling timescale in subsection \ref{subsec:AnalyticModel_results}.

    \subsection{Estimation of the cooling time}\label{subsec:shadow_result}
    
    We aim to estimate the cooling timescale of the outer disk around HD~142527 by comparing the three-dimensional shadow morphology traced in infrared scattered light with the azimuthal temperature distribution traced by the ALMA Band~9 continuum emission. 

    \subsubsection{Shadow morphology on the disk scattering surface}
    Following the procedure described in section~\ref{subsubsec:shadow_reconstruction}, we extracted high-gradient structures from the S/N map constructed from the H-band $Q_\phi$ image and simultaneously fitted the horizon, shadow-boundary, and apparent outer-edge curves. 

\begin{figure*}[t]
    \begin{center}
    \includegraphics[width=\hsize,bb = 00 0 539 174]{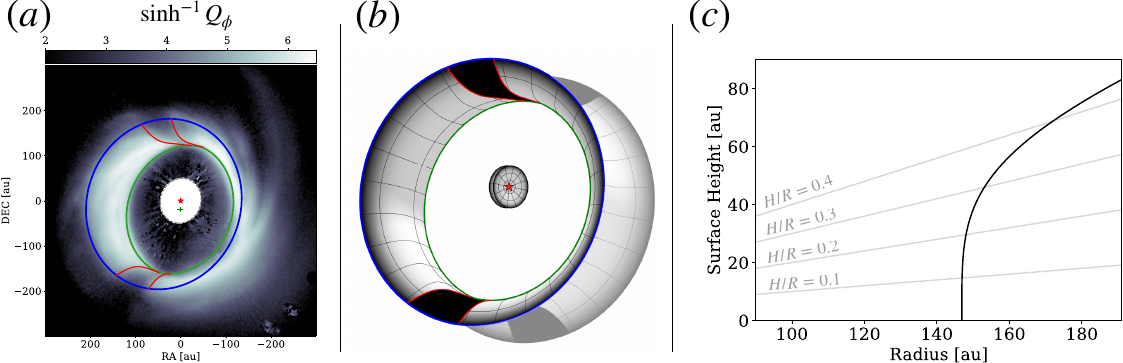}
    \end{center}
    \caption{
    Best-fit scattering-surface model for the HD~142527 disk.
    (a) H-band $Q_\phi$ image overlaid with the best-fit model curves.
    The blue curve indicates the horizon of the outer disk, the red curves show the reconstructed shadow boundaries, and the green curve shows the apparent outer edge of the outer disk.
    The blue cross and red star mark the positions of the outer-disk center and the central star, respectively.
    (b) Three-dimensional view of the best-fit scattering-surface model.
    The black regions indicate the shadowed areas where the inner structure blocks stellar irradiation.
    The colored curves have the same meanings as in panel (a).
    (c) Surface-height profile of the best-fit scattering surface, shown as the black curve, over the radial range where the shadow boundaries are detected.
    The right edge of the profile corresponds to the radius of the apparent outer edge of the outer disk.
    }
    \label{fig:HD142527_surface}
    \end{figure*}

    The western side of the HD~142527 outer disk is not well suited to robust boundary extraction with this procedure because its scattered-light morphology contains spiral-like structures and other non-axisymmetric features (see figure~\ref{fig:HD142527_surface}(a)). 
    We therefore sample boundary points only on the eastern side of the disk, where the apparent outer edge is more clearly traced in the H-band $Q_\phi$ image (see figure \ref{fig:images}).
    
    The center of the inner disk is assumed to coincide with the stellar position. We therefore fix the sky-plane offsets of the inner disk center to $c_{{\rm i},x}=c_{{\rm i},y}=0$, where $c_{{\rm i},x}$ and $c_{{\rm i},y}$ denote the offsets in the declination and R.A. directions, respectively. With this assumption, the fitted inner-disk parameters describe the orientation and vertical extent of the shadow-casting structure, rather than any astrometric displacement of its center.

    Figure~\ref{fig:HD142527_surface} shows the best-fit shadow and scattering-surface model for the HD~142527 disk. Panel (a) shows the H-band $Q_\phi$ image overlaid with the best-fit curves, which trace the outer-disk boundary and the shadow boundaries. Panel (b) presents a three-dimensional view of the reconstructed scattering-surface model, illustrating the geometry of the outer disk surface and the shadowed regions. The model reproduces the locations of the two azimuthally localized dark regions, supporting the interpretation that these features are shadows cast by a misaligned inner disk.

    \begin{table*}[t]
    \caption{Disk Model Parameters for HD~142527}
    \label{tab:disk_params}
    \centering
    \setlength{\tabcolsep}{5pt}
    \renewcommand{\arraystretch}{1.15}
    \begin{tabular}{llllll}
    \hline
    \hline
    Parameter & Description & Search range & Solution & Error & Unit \\
              &             & [min, max]   &          &       &      \\
    \hline
    
    $c_{\mathrm{o}x}$
    & Offset of the outer-disk center in the $x$-direction (Dec)
    & [$-30$, 30]
    & $-19.63$
    & $^{+0.306}_{-0.295}$
    & [au] \\
    
    $c_{\mathrm{o}y}$
    & Offset of the outer-disk center in the $y$-direction (RA)
    & [$-30$, 30]
    & $1.73$
    & $^{+0.255}_{-0.258}$
    & [au] \\
    
    $R_{\rm e}$
    & Radius of the inner edge of the outer disk
    & [120, 160]
    & $146.75$
    & $^{+0.429}_{-0.429}$
    & [au] \\
    
    $H_{\rm surf,c}$
    & Height of the outer-disk surface at $R_{\rm c}=200~\mathrm{au}$
    & [5, 100]
    & $88.18$
    & $^{+2.161}_{-2.067}$
    & [au] \\
    
    $R_{\rm out}$
    & Radius of the apparent outer edge of the outer disk
    & [185, 250]
    & $191.18$
    & $^{+0.731}_{-0.679}$
    & [au] \\
    
    $q$
    & Power index of the outer-disk surface
    & [2, 5]
    & $3.92$
    & $^{+0.205}_{-0.209}$
    & [ ] \\
    
    $c_{\mathrm{i}x}$
    & Offset of the inner-disk center in the $x$-direction (Dec)
    & 0 (fixed)
    & --
    & --
    & [au] \\
    
    $c_{\mathrm{i}y}$
    & Offset of the inner-disk center in the $y$-direction (RA)
    & 0 (fixed)
    & --
    & --
    & [au] \\
    
    $i_{\rm i}$
    & Inclination of the inner disk
    & [0, 90]
    & $34.11$
    & $^{+0.924}_{-0.912}$
    & [$^\circ$] \\
    
    $\mathrm{PA}_{\rm i}$
    & Position angle of the inner disk
    & [0, 360]
    & $179.35$
    & $^{+0.379}_{-0.338}$
    & [$^\circ$] \\
    
    $h_r$
    & Maximum aspect ratio of the inner disk
    & [0.05, 0.2]
    & $0.16$
    & $^{+0.003}_{-0.003}$
    & [ ] \\
    
    \hline
    \end{tabular}
    \end{table*}
    
    The best-fit values are summarized in table~\ref{tab:disk_params}, while the corresponding posterior distributions are shown in the corner plot presented in figure~\ref{fig:corner_plots} of appendix \ref{appendix:corner_plots}.        
    The best-fit inner-disk geometry has an inclination of $i_{\rm i}=34^\circ$ and a position angle of ${\rm PA}_{\rm i}=179^\circ$. Together with the fixed outer-disk geometry, these values imply a misalignment angle of $\Delta\theta_{\rm mis}\sim60^\circ$ between the inner and outer disks.
    The maximum aspect ratio of the optically obscuring inner disk is $h_r=0.16$, representing the maximum height-to-radius ratio of the optically thick dusty component that blocks stellar irradiation and casts shadows on the outer disk surface.

    The inner edge of the outer disk is located at $R_{\rm e}=147~\mathrm{au}$. At the characteristic radius $R_{\rm c}=200~\mathrm{au}$, the scattering surface has a height of $H_{\rm surf,c}=88~\mathrm{au}$ and a power-law index of $q=3.92$. These parameters describe the three-dimensional scattering surface of the outer disk traced by H-band polarized scattered light. 
    The resulting outer-disk surface-height profile is shown in figure~\ref{fig:HD142527_surface}(c). Over the radial range in which the shadow boundaries are detected, from $R_{\rm e}=147~\mathrm{au}$ to $R_{\rm out}=191~\mathrm{au}$, the surface aspect ratio $H_{\rm surf}/R$ ranges from 0.2 to 0.4. This indicates that the H-band scattered light traces a vertically elevated dust surface rather than the disk midplane. We use the inferred surface geometry to deproject the shadow locations onto the three-dimensional outer-disk surface.

    \subsubsection{Cooling timescale of the outer disk}
    
    We next compare the reconstructed shadow geometry with the azimuthal temperature distribution inferred from the ALMA Band~9 continuum image. 
    Because the Band~9 continuum emission is expected to be dominated by optically thick dust thermal emission, the observed brightness temperature can be interpreted as the temperature around the emitting layer defined in equation~(\ref{eq:z_tau_1}). 
    This emitting layer is also shown in the left panel of figure~\ref{fig:tauall_T_Hgas_amax}.

    \begin{figure*}[t]
    \begin{center}
    \includegraphics[width=\hsize,bb = 0 0 452 189]{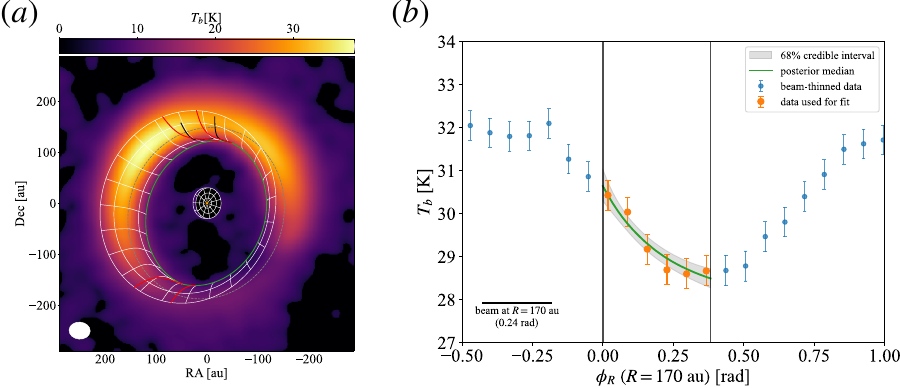}
    \end{center}
    \caption{Comparison between the reconstructed shadow geometry and the azimuthal temperature variation in the HD~142527 outer disk.
    (a) ALMA Band~9 brightness-temperature image overlaid with the surface grid of the best-fit scattering-surface model. The red curves indicate the shadow boundaries reconstructed on the outer-disk surface, while the black curve in the northern region shows the corresponding boundary projected onto the disk midplane. The gray dashed curve marks a radius of $170~\mathrm{au}$ in the midplane. The white grid represents the corresponding coordinates on the disk surface. The synthesized beam is shown in the lower-left corner.
    (b) Azimuthal brightness-temperature profile at $R=170~\mathrm{au}$, shown as a function of the angular distance $\phi_R$ from the upstream shadow boundary in the direction of disk rotation. The vertical lines mark the entrance and exit boundaries of the shadowed region. The blue points show the observed brightness temperatures sampled at intervals of $0.2$ synthesized beams, with error bars derived from the rms noise of the continuum image. The orange points indicate the subset of the sampled data used for the model fitting. The green curve represents the pointwise posterior median of the cooling model, and the gray shaded region indicates the central $68\%$ Bayesian credible interval. The horizontal scale bar indicates the azimuthal extent of one synthesized beam at $R=170~\mathrm{au}$.}
    \label{fig:HD142527_temp}
    \end{figure*}
    
    Figure~\ref{fig:HD142527_temp} compares the reconstructed shadow geometry with the ALMA Band~9 continuum emission and the corresponding azimuthal temperature profile. Panel~(a) shows the ALMA Band~9 continuum image overlaid with the surface grid of the best-fit scattering-surface model. This surface grid is the same as that shown in figure~\ref{fig:HD142527_surface}(b). The red curves indicate the northern shadow boundaries reconstructed on the outer-disk scattering surface from the shadow analysis, whereas the black curves show the corresponding shadow boundaries projected onto the outer-disk midplane. Panel~(b) shows the azimuthal temperature profile at a radius of $170~{\rm au}$ from the outer-disk center. The horizontal axis represents the azimuthal angle measured in the outer-disk midplane, while the vertical axis represents the brightness temperature of the dust continuum emission. The black vertical lines in panel~(b) correspond to the shadow boundaries shown by the black curves in panel~(a) on the disk midplane. Because the disk rotates clockwise, the left line corresponds to the leading boundary, where disk material enters the shadowed region, whereas the right line corresponds to the trailing boundary, where disk material exits the shadowed region.

    A comparison of the shadow-entry boundary with the brightness-temperature data points in figure~\ref{fig:HD142527_temp}(b) suggests that the temperature begins to decrease upstream of the leading boundary. However, this apparent decrease does not necessarily indicate the physical onset of cooling and is likely affected by convolution with the finite synthesized beam. At a radius of $170~{\rm au}$, the synthesized beam corresponds to an azimuthal angular scale of approximately $0.24~{\rm rad}$. Consequently, the temperature structure may be substantially smoothed near the shadow boundary, where the azimuthal temperature gradient is steep. Determining the onset of cooling from the millimeter continuum image alone could therefore lead to a misidentification of its physical location. This further highlights the importance of independently determining the shadow boundaries from scattered-light observations when estimating the cooling timescale.

    In the limit of an instantaneous thermal response, the temperature minimum is expected to coincide with the midplane-projected shadow position. However, if cooling takes a finite time, an angular offset arises between the projected shadow position and the resulting temperature minimum. The observed temperature profile therefore traces the shadow-induced azimuthal temperature variation at $R=170~{\rm au}$ while accounting for the three-dimensional projection of the shadow boundaries. The measured angular offset between the projected shadow position and the temperature minimum is used to estimate the cooling timescale of the outer disk.
    
    We model the shadow-induced temperature decrease using equation~(\ref{eq:Trelax}). In this model, after disk material enters the shadowed region, the dust temperature is assumed to relax toward the equilibrium temperature within the shadow over a finite cooling timescale. As shown in figure~\ref{fig:HD142527_temp}(b), this model is fitted to the azimuthal temperature profile at a radius of $170~{\rm au}$ from the outer-disk center. The gray shaded region indicates the 68\% posterior credible interval of the fitted temperature profile. The marginalized posterior distribution yields a 68\% Bayesian credible interval of $t_{\rm cool,est}\Omega_{\rm K}=0.1$--$0.4$, corresponding to a cooling timescale of approximately 20--90~yr at $R=170~{\rm au}$ for the adopted stellar mass of $M_\star=2.4\,M_\odot$. This result indicates that the thermal response time of the outer disk is shorter than the local Keplerian timescale.
    
    \subsection{Estimation of the dust grain size}\label{subsec:AnalyticModel_results}
    
    In this subsection, we present our constraints on the dust grain size, derived by reproducing the observationally estimated cooling timescale.
    As shown in subsection \ref{subsec:shadow_result}, the observations indicate $0.1\lesssim t_{\rm cool,est}\Omega_{\rm K}\lesssim 0.4$ in the outer region of the HD~142527 disk.
    Hereafter, we refer to this requirement as the ``cooling condition.''
    
    To calculate the vertical distribution of the dust particles, we adopt a vertical dust diffusion coefficient of $\alpha_z=2\times 10^{-3}$, assuming that the dust grains are vertically mixed by gas turbulence.
    This adopted value is consistent with the high scattering surface of the outer disk in HD~142527 presented in subsection \ref{subsec:shadow_result}, as well as with previous near-infrared observations \citep{TazakiMurakawa+:2021aa}.
    We discuss the turbulence-driving mechanisms capable of producing this value of the vertical diffusion coefficient in section \ref{sec:VSI_application}.
    
    Following the steps described in subsection \ref{subsec:method_procedure}, we calculate the theoretical cooling timescale at the emission height, $t_{\rm cool,cal}$, and compare it with the observationally estimated one, $t_{\rm cool,est}$, to constrain the dust grain size.
    In subsection \ref{subsubsec:AnalyticModel_results_fiducial}, we present the constraints on the maximum dust grain size for our fiducial model (see table \ref{table:run} in section \ref{sec:method}).
    Subsequently, in subsections \ref{subsubsec:AnalyticModel_results_low_SigmaGas}, \ref{subsubsec:AnalyticModel_results_SizeDis}, and \ref{subsubsec:AnalyticModel_results_porosity}, we explore how these constraints depend on the gas surface density, dust size distribution, and dust porosity, respectively.

        \subsubsection{Fiducial model}\label{subsubsec:AnalyticModel_results_fiducial}
        
        We find that, in our fiducial model (using the gas surface density of $\Sigma_{\rm gas}=10~{\rm g~cm^{-2}}$, slope of the dust size distribution of $p=-2.5$, minimum grain size of $a_{\rm min}=1~{\rm \mu m}$, dust filling factor of $f_{\rm dust}=1.0$, and Ricci opacity model), the northern region of the HD~142527 disk is predominantly optically thick at the ALMA Band~9 wavelength.
        Specifically, the vertical extinction optical depth 
        at ALMA Band 9, $\tau_{\rm all}$, exceeds unity for the dust surface density $\Sigma_{\rm dust}$ with $\gtrsim 1~{\rm g~cm^{-2}}$, regardless of the maximum grain size $a_{\rm max}$.
        In this optically thick regime, we can determine the local disk temperature $T$ and gas scale height $H_{\rm gas}$.
        For instance, in the case of $\Sigma_{\rm dust}=10~{\rm g~cm^{-2}}$, the temperature and gas scale height become constant for $a_{\rm max}\gtrsim 100~{\rm \mu m}$, yielding $T\approx 35~{\rm K}$ and $H_{\rm gas}\approx 17~{\rm au}$, respectively.
        Across the explored parameter space, depending on $\Sigma_{\rm dust}$ and $a_{\rm max}$, $T$ varies by up to a factor of $\sim 1.5$, causing $H_{\rm gas}$ to vary accordingly by a factor of $\sim 1.1$.
        We provide further details regarding these dependencies in appendix \ref{appendix:temperature}.
    
        \begin{figure}[t]
        \begin{center}
            \includegraphics[width=\hsize,bb = 0 0 420 638]{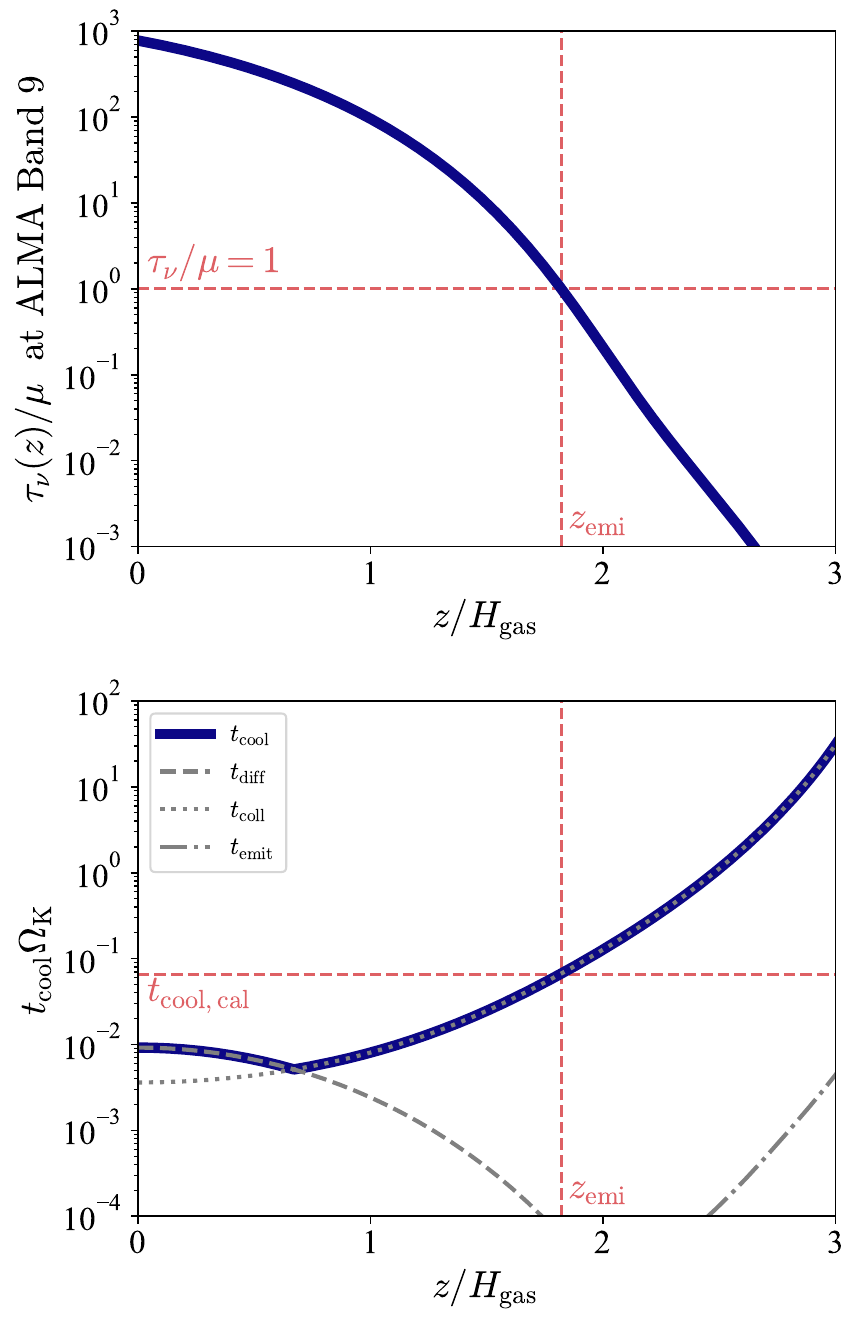}
            \end{center}
            \caption{Calculated vertical optical depth, $\tau_\nu(z)$ at ALMA Band 9 [upper panel; equation \eqref{eq:tau_z}], and cooling timescale, $t_{\rm cool}$ [lower panel; equation \eqref{eq:taurelax}], as functions of height $z$ for $a_{\rm max}=100~{\rm \mu m}$ and $\Sigma_{\rm dust}=10~{\rm g~cm^{-2}}$ in the fiducial model. In the upper panel, the horizontal and vertical dashed lines indicate $\tau_\nu/\mu = 1$ and $z=z_{\rm emi}$, respectively [see equation \eqref{eq:z_tau_1}]. In the lower panel, the horizontal and vertical dashed lines denote $t_{\rm cool}=t_{\rm cool,cal}$ and $z=z_{\rm emi}$, respectively. }
            \label{fig:tau_tcool_z_fiducial}
        \end{figure}
        
        Based on this vertical structure, for each combination of $\Sigma_{\rm dust}$ and $a_{\rm max}$, we determine the ALMA Band 9 emission height $z_{\rm emi}$ and the corresponding calculated cooling timescale, $t_{\rm cool,cal}$, to compare with the observationally estimated value.
        The upper panel of figure \ref{fig:tau_tcool_z_fiducial} illustrates the vertical optical depth profile at ALMA Band 9 for $a_{\rm max}=100~{\rm \mu m}$ and $\Sigma_{\rm dust}=10~{\rm g~cm^{-2}}$ in the fiducial model.
        In this case, the ALMA Band 9 emission height $z_{\rm emi}$, where $\tau_\nu/\mu = 1$, is located at $\approx 1.9 H_{\rm gas}$.
        The lower panel displays the vertical distribution of the cooling timescales for $a_{\rm max}=100~{\rm \mu m}$ and $\Sigma_{\rm dust}=10~{\rm g~cm^{-2}}$ in the fiducial model.
        At $z\gtrsim 0.6H_{\rm gas}$, the cooling timescale is dominated by the collisional heat transfer timescale between gas and dust, which is consistent with previous studies (e.g., \citealt{Malygin+2017,BarrancoPei+:2018kc,FukuharaOkuzumi:2024aa}).
        This figure shows that the cooling timescale at $z_{\rm emi}$, $t_{\rm cool,cal}[=t_{\rm cool}(z_{\rm emi})]$, is $\approx 0.07\Omega_{\rm K}^{-1}$ in this case.
        We note that for all cases in this study, the cooling processes at the emission height are dominated by the collisional heat transfer.

        Our model also predicts that the scattering surface $z_{\rm sca}$ at a wavelength of $1.6~{\rm \mu m}$ lies in a range of $0.2\lesssim z_{\rm sca}/R\lesssim 0.35$, depending on the dust grain size and dust surface density.
        These values are consistent with the height of the scattering surface estimated by the near-infrared observation (see figure \ref{fig:HD142527_surface} in section \ref{subsec:shadow_result}).

        \begin{figure}[t]
            \begin{center}
            \includegraphics[width=\hsize,bb = 0 0 434 349]{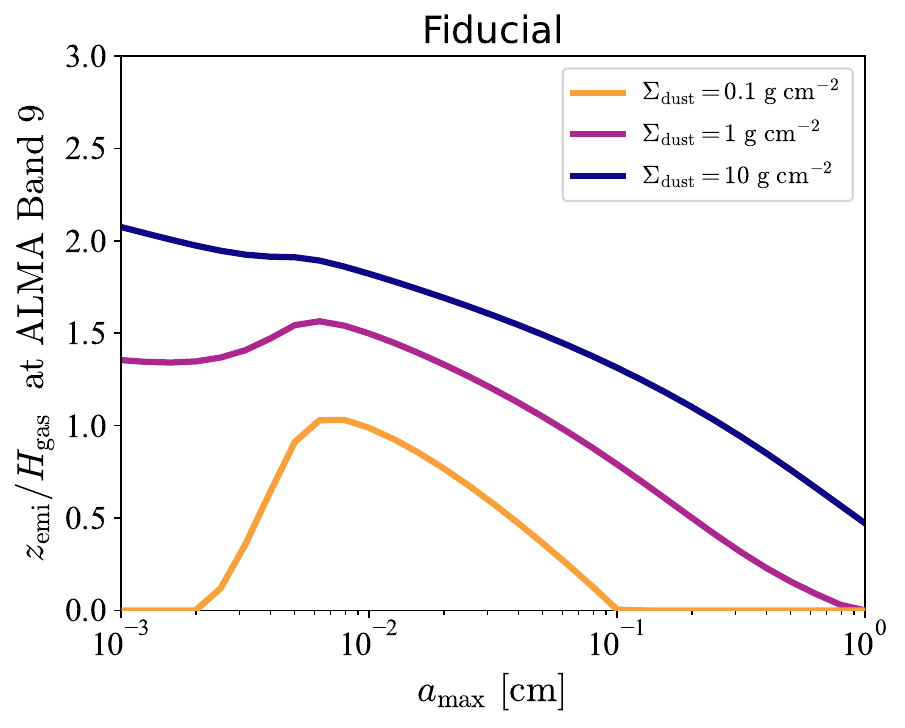}
            \end{center}
            \caption{Calculated ALMA Band 9 emission height $z_{\rm emi}$ normalized by the gas scale height $H_{\rm gas}$ as a function of the maximum dust grain size $a_{\rm max}$ for different values of $\Sigma_{\rm dust}$ in the fiducial model.}
            \label{fig:zobs_amax_fiducial}
        \end{figure}

        Higher dust surface densities yield larger optical depths at ALMA Band 9, resulting in higher emission heights.
        Figure \ref{fig:zobs_amax_fiducial} plots $z_{\rm emi}$ as a function of $a_{\rm max}$ for different values of $\Sigma_{\rm dust}$ in the fiducial model.
        As $\Sigma_{\rm dust}$ decreases from $10~{\rm g~cm^{-2}}$ to $0.1~{\rm g~cm^{-2}}$, $z_{\rm emi}$ drops from $\approx 2.1H_{\rm gas}$ to $0$ (midplane) at $a_{\rm max}=10~{\rm \mu m}$ and from $\approx 1.9H_{\rm gas}$ to $\approx H_{\rm gas}$ at $a_{\rm max}=100~{\rm \mu m}$.
        
        The emission height at ALMA Band 9 also depends on the maximum dust grain size.
        Figure \ref{fig:zobs_amax_fiducial} indicates that for $\Sigma_{\rm dust} \geq 1~{\rm g~cm^{-2}}$, $z_{\rm emi}$ remains nearly constant at $a_{\rm max} \lesssim 100~{\rm \mu m}$.
        This is because the dust extinction opacity at ALMA Band 9, $\kappa_{\rm d,\nu}$, is also nearly constant over this range (see figure \ref{fig:opacity_Ricci_DIANA_DSHARP} in appendix \ref{appendix:opacity}).
        For $\Sigma_{\rm dust} = 0.1~{\rm g~cm^{-2}}$, decreasing $a_{\rm max}$ below $100~{\rm \mu m}$ also lowers $z_{\rm emi}$ because $\kappa_{\rm d,\nu}$ peaks around $a \approx 100~{\rm \mu m}$ at ALMA Band 9.
        Above this peak, at $a_{\rm max} \gtrsim 100~{\rm \mu m}$, $z_{\rm emi}$ falls as $a_{\rm max}$ increases because $\kappa_{\rm d,\nu}$ decreases with increasing grain size.
        When $a_{\rm max}$ reaches $1~{\rm cm}$, these heights drop to $0.5 H_{\rm gas}$ and the midplane ($z_{\rm emi}=0$) for $\Sigma_{\rm dust}=10~{\rm g~cm^{-2}}$ and $\leq 1~{\rm g~cm^{-2}}$, respectively.

        \begin{figure}[t]
            \begin{center}
            \includegraphics[width=\hsize,bb = 0 0 443 349]{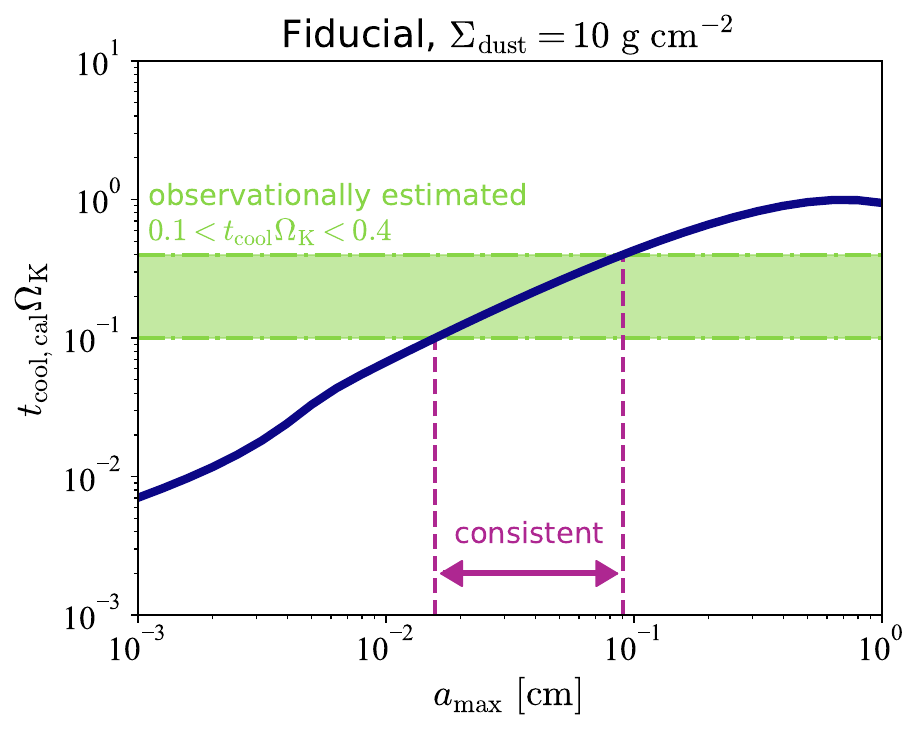}
            \end{center}
            \caption{Calculated cooling timescale at the ALMA Band 9 emission height, $t_{\rm cool,cal}$, as a function of $a_{\rm max}$ for $\Sigma_{\rm dust}=10~{\rm g~cm^{-2}}$ in the fiducial model. The horizontal dash-dotted lines mark the upper ($t_{\rm cool}\Omega_{\rm K}=0.4$) and lower ($0.1$) limits of the observationally estimated range derived in subsection \ref{subsec:shadow_result}. The vertical dashed lines denote the corresponding values of $a_{\rm max}$ that yield these limits. }
            \label{fig:tcool_amax_fiducial}
        \end{figure}

        Figure \ref{fig:tcool_amax_fiducial} displays the calculated cooling timescale at the ALMA Band 9 emission height, $t_{\rm cool,cal}$, as a function of $a_{\rm max}$ for $\Sigma_{\rm dust}=10~{\rm g~cm^{-2}}$ in the fiducial model.
        For $a_{\rm max}\lesssim 100~{\rm \mu m}$, $t_{\rm cool,cal}$ increases with $a_{\rm max}$ because with $z_{\rm obs}$ remaining nearly constant over this range (see also figure \ref{fig:zobs_amax_fiducial}), the total geometric cross section per unit volume and thereby the collisional heat transfer timescale decreases.
        For larger grains, $a_{\rm max}$ affects $t_{\rm cool,cal}$ through two competing effects: the reduction in total geometric cross section, which increases the cooling timescale, and the reduction in the ALMA Band 9 emission height, which decreases it.
        Up to $a_{\rm max}\approx 8~{\rm mm}$, the former effects dominates, and $t_{\rm cool,cal}$ continues to increase with $a_{\rm max}$.
        Beyond this size, the latter effect dominates, and $t_{\rm cool,cal}$ decreases with increasing $a_{\rm max}$, up to $1~{\rm cm}$.

        Comparing $t_{\rm cool,cal}$ with the observationally estimated range of $0.1 \lesssim t_{\rm cool,est}\Omega_{\rm K} \lesssim 0.4$, we find that maximum grain sizes in the range of $150~{\rm \mu m} \lesssim a_{\rm max} \lesssim 900~{\rm \mu m}$ successfully reproduce this cooling condition.
        In contrast, grain sizes smaller than $150~{\rm \mu m}$ or larger than $900~{\rm \mu m}$ yield cooling timescales that are too short or too long, respectively, to match the observations.

        \begin{figure}[t]
            \begin{center}
            \includegraphics[width=\hsize,bb = 0 0 769 595]{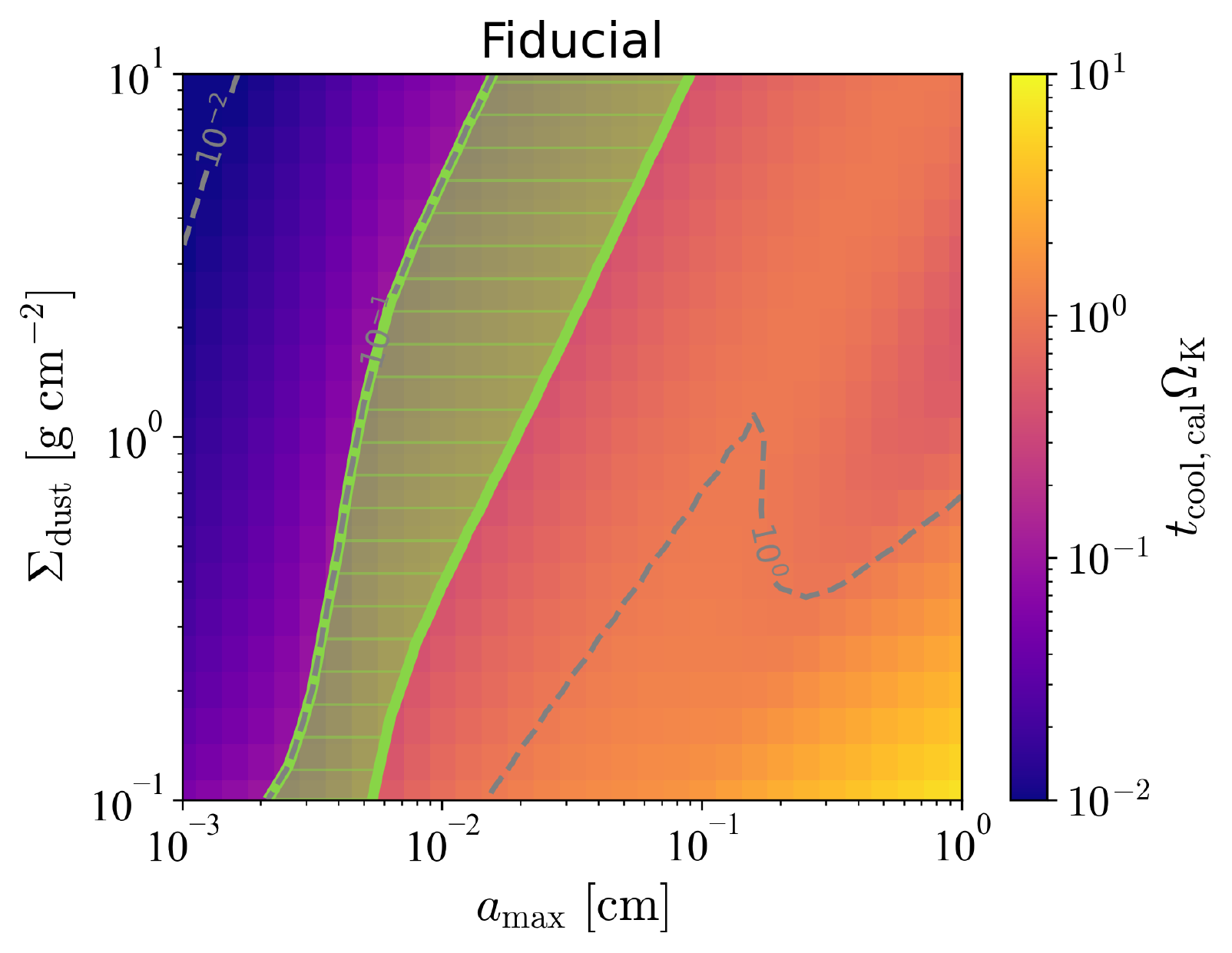}
            \end{center}
            \caption{Calculated cooling timescale at the ALMA Band 9 emission height, $t_{\rm cool,cal}$, across the $a_{\rm max}$--$\Sigma_{\rm dust}$ parameter space for the fiducial model (see table \ref{table:run} for the parameter choice). The dashed contours indicate $t_{\rm cool}\Omega_{\rm K} = 10^{-2},~10^{-1}$, and $1$. The horizontally hatched region represents the parameter space that satisfies the observationally estimated cooling condition ($0.1 \lesssim t_{\rm cool,est}\Omega_{\rm K} \lesssim 0.4$). }
            \label{fig:map_color_fiducial}
        \end{figure}
    
        Next, we examine how the calculated cooling timescale and the resulting constraints on the dust grain size depend on the dust surface density, $\Sigma_{\rm dust}$.
        Figure \ref{fig:map_color_fiducial} maps $t_{\rm cool,cal}$ across the $a_{\rm max}$--$\Sigma_{\rm dust}$ parameter space for the fiducial model.
        The figure shows that $t_{\rm cool,cal}$ decreases as $\Sigma_{\rm dust}$ increases.
        A larger $\Sigma_{\rm dust}$ increases the optical depth and elevates $z_{\rm obs}$ (figure \ref{fig:zobs_amax_fiducial}).
        Because the dust density decreases with height, this elevation reduces the total geometric cross section at the emission height, which by itself would increase the cooling timescale.
        However, a larger $\Sigma_{\rm dust}$ also directly increases the dust density at any given height, and this effect dominates over the reduction caused by the higher emission height, resulting in a decrease in the cooling timescale.

        Overall, as the dust surface density decreases, the grain size required to reproduce the observationally estimated cooling timescale also decreases.
        As shown in figure \ref{fig:map_color_fiducial}, this required dust grain size scales almost linearly with $\Sigma_{\rm dust}$.
        Specifically, the cooling condition is satisfied by $50~{\rm \mu m} \lesssim a_{\rm max} \lesssim 200~{\rm \mu m}$ for $\Sigma_{\rm dust}=1~{\rm g~cm^{-2}}$, and by $20~{\rm \mu m} \lesssim a_{\rm max} \lesssim 50~{\rm \mu m}$ for $\Sigma_{\rm dust}=0.1~{\rm g~cm^{-2}}$.
    
        \subsubsection{Case of low gas surface density}\label{subsubsec:AnalyticModel_results_low_SigmaGas}

        The gas surface density $\Sigma_{\rm gas}$ affects the vertical profile of dust number density $dn_{\rm d}(a,z)/da$, thereby altering the vertical profiles of both the optical depth and the cooling timescale.
        A lower gas surface density leads to increased Stokes numbers for the dust grains [equations \eqref{eq:rhoda}--\eqref{eq:St}], which suppresses their vertical diffusion.
        Consequently, dust grains become more concentrated near the midplane, causing the ALMA Band 9 optical depth $\tau_\nu (z)$ to increase at lower altitudes and decrease at higher altitudes, with the cooling timescale showing the opposite trend (see also section 4.1 of \citealt{FukuharaOkuzumi+:2021ca}).
        Since the emission height at ALMA Band 9, $z_{\rm emi}$, is determined by integrating the optical depth from infinity downward (section \ref{subsubsec:method_height}), this dust settling naturally reduces $z_{\rm emi}$.
        We present the dependence of $z_{\rm emi}$ on $\Sigma_{\rm gas}$ in figure \ref{fig:zobs_parameter} in appendix \ref{appendix:observed_height}.
    
        \begin{figure}[t]
            \begin{center}
            \includegraphics[width=\hsize,bb = 0 0 769 595]{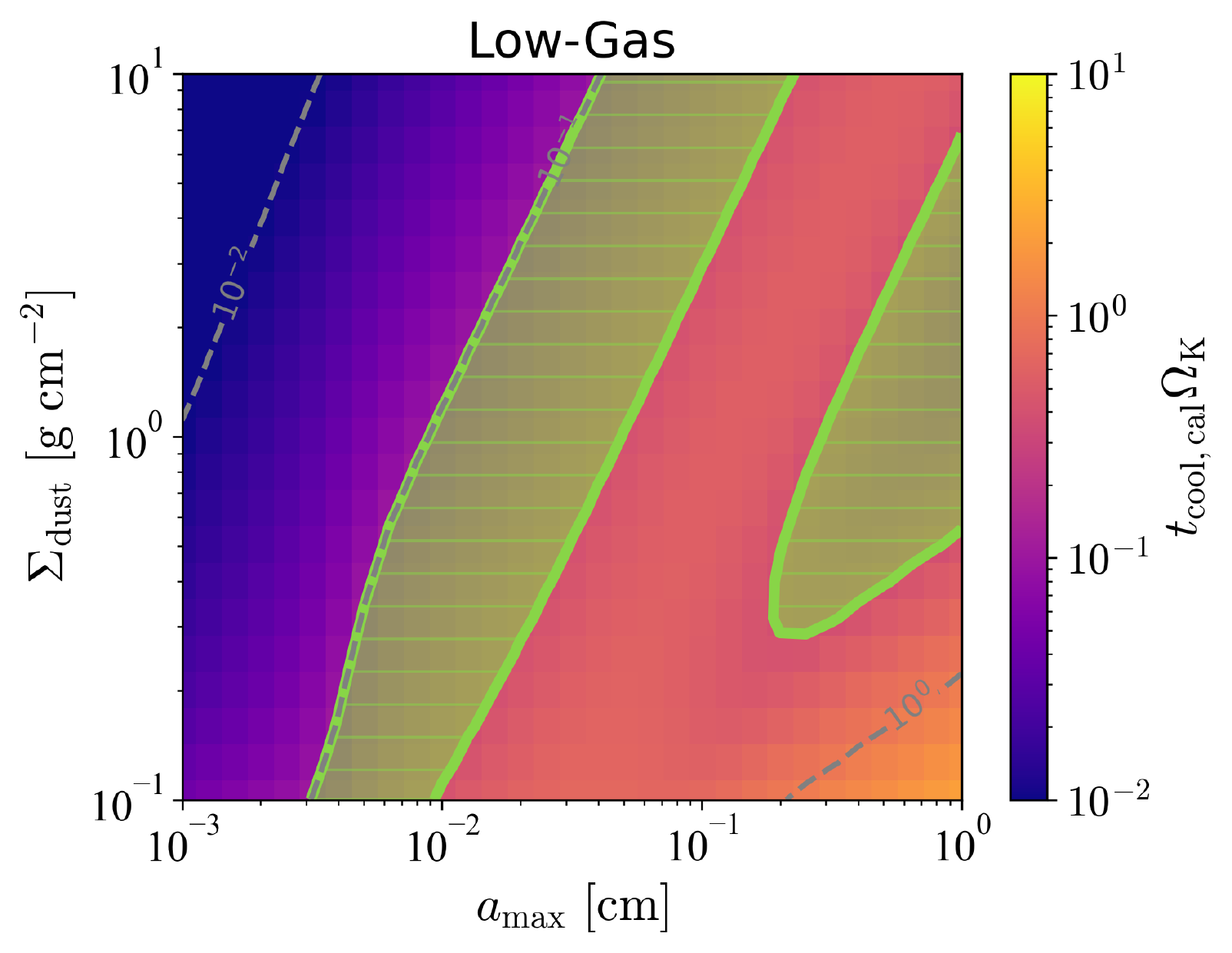}
            \end{center}
            \caption{Same as figure \ref{fig:map_color_fiducial}, but for the Low-Gas model ($\Sigma_{\rm gas}=1.0~{\rm g~cm^{-2}}$, see table \ref{table:run} for other parameter choices). }
            \label{fig:map_color_Sigmag1e0}
        \end{figure}

        This lower emission height, in turn, results in a shorter calculated cooling timescale, $t_{\rm cool,cal}$.
        Figure \ref{fig:map_color_Sigmag1e0} maps $t_{\rm cool,cal}$ and the parameter space satisfying the cooling condition in the $a_{\rm max}$--$\Sigma_{\rm dust}$ plane for the Low-Gas model ($\Sigma_{\rm gas}=1.0~{\rm g~cm^{-2}}$, see table \ref{table:run} for other parameter choices).
        Compared to the fiducial model (figure \ref{fig:map_color_fiducial}), $t_{\rm cool,cal}$ is systematically shorter across the entire parameter space.
        This is because the local cooling timescale decreases closer to the midplane, as seen in the lower panel of figure \ref{fig:tau_tcool_z_fiducial}.
        To compensate for this shorter cooling timescale and successfully reproduce the observations, the required maximum grain size becomes larger.
        For instance, to match $t_{\rm cool,est}$, our model requires a maximum grain size of $390~{\rm \mu m} \lesssim a_{\rm max}\lesssim 2.0~{\rm mm}$ for $\Sigma_{\rm dust}=10~{\rm g~cm^{-2}}$, and $90~{\rm \mu m} \lesssim a_{\rm max}\lesssim 480~{\rm \mu m}$ for $\Sigma_{\rm dust}=1~{\rm g~cm^{-2}}$.
        Additionally, because of the extremely low emission height in this model, large grains ($a_{\rm max}\gtrsim 2~{\rm mm}$) can also satisfy the cooling condition within the range of $0.3~{\rm g~cm^{-2}}\lesssim \Sigma_{\rm dust} \lesssim 7~{\rm g~cm^{-2}}$.
    
        \subsubsection{Dependence on the dust size distribution}\label{subsubsec:AnalyticModel_results_SizeDis}
        
        The slope of the size distribution, $p$, and the minimum grain size, $a_{\rm min}$, decide the overall dust size distribution.
        Altering these parameters affects both the ALMA Band 9 emission height and the local cooling timescale, which in turn shifts the maximum grain size required to match the observations.
        In this subsection, we explore the impact of varying these values, while we set $p=-2.5$ and $a_{\rm min} = 1~{\rm \mu m}$ in the fiducial model.
        
        Together with $a_{\rm max}$, these parameters determine the dust number surface density, $dN_{\rm d}(a)/da$ [equation \eqref{eq:Sigmad_a}], and the particle number density for each grain size at height $z$, $dn_{\rm d}(a,z)/da$ [equation \eqref{eq:rhoda}].
        Consequently, the chosen size distribution alters the total vertical optical depth at ALMA Band 9, $\tau_{\rm all}$, [equation \eqref{eq:tau_all}], the vertical optical depth profile at ALMA Band 9, $\tau_\nu(z)$, and the vertical cooling timescale distribution $t_{\rm cool}(z)$.
        However, variations in $\tau_{\rm all}$ have only a minor effect on our constraints because the disk temperature and gas scale height are relatively insensitive to $\tau_{\rm all}$ (see figure \ref{fig:tauall_T_Hgas_amax} in appendix \ref{appendix:temperature}).
        Instead, the calculated cooling timescale at the ALMA Band 9 emission height, $t_{\rm cool,cal}$, is primarily driven by shifts in the ALMA Band 9 emission height [via $\tau_\nu(z)$] and changes in the local cooling efficiency (via the dust number density).
        We present the emission heights for these different size distribution models in figure \ref{fig:zobs_parameter} in appendix \ref{appendix:observed_height}.
    
        \begin{figure*}[t]
            \begin{center}
            \includegraphics[width=\hsize,bb = 0 0 842 214]{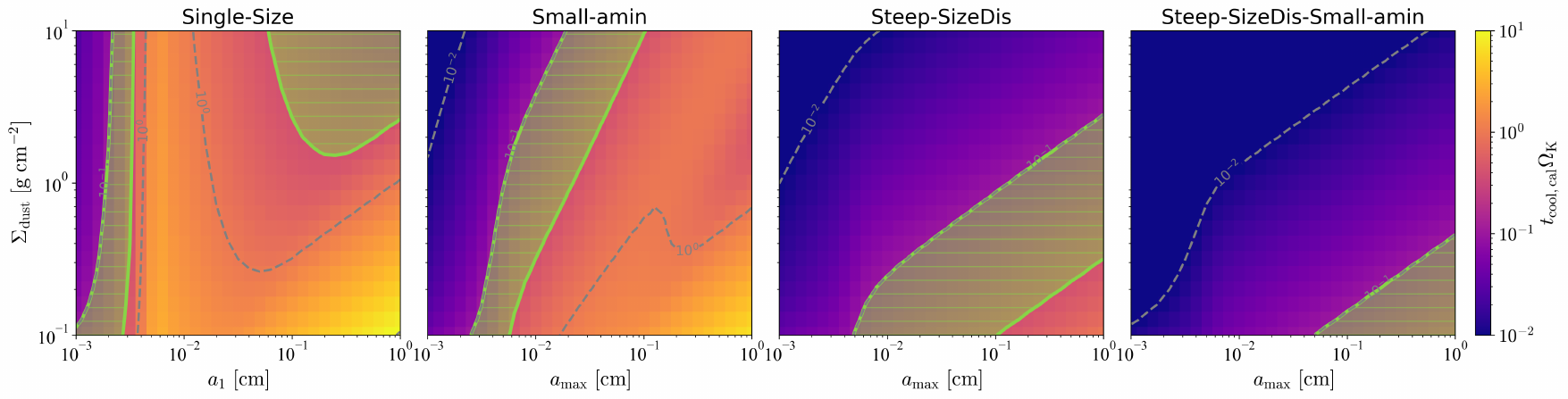}
            \end{center}
            \caption{Same as figure \ref{fig:map_color_fiducial}, but for the Single-Size, Small-amin, Steep-SizeDis, and Steep-SizeDis-Small-amin models from left to right (see table \ref{table:run} for the parameter choice of each model). }
            \label{fig:map_color_SizeDis}
        \end{figure*}

        Figure \ref{fig:map_color_SizeDis} maps $t_{\rm cool,cal}$ and the areas that reproduce $t_{\rm cool,est}$ for four different size distributions. 
        From left to right, the panels display the results for the Single-Size model (plotted against $a_1$), the Small-amin model ($a_{\rm min}=0.1~{\rm \mu m}$), the Steep-SizeDis model ($p=-3.5$), and the Steep-SizeDis-Small-amin model ($a_{\rm min}=0.1~{\rm \mu m}$ and $p=-3.5$).
        In the Single-Size model (left panel), $t_{\rm cool,cal}$ is generally longer than in the fiducial model.
        This is because a continuous size distribution facilitates more efficient collisional heat transfer than a mono-disperse population \citep{FukuharaOkuzumi:2024aa}.
        Consequently, the grain size that satisfies the cooling condition is restricted to $20~{\rm \mu m} \lesssim a_1 \lesssim 30~{\rm \mu m}$ for $\Sigma_{\rm dust} \gtrsim 0.1~{\rm g~cm^{-2}}$, or $a_1 \gtrsim 1~{\rm mm}$ for $\Sigma_{\rm dust} \gtrsim 1.5~{\rm g~cm^{-2}}$.

        For the Small-amin model (second panel), the dependence of $t_{\rm cool,cal}$ on $a_{\rm max}$ and $\Sigma_{\rm dust}$ closely resembles that of the fiducial model (figure \ref{fig:map_color_fiducial}).
        Because the slope of the size distribution is $p=-2.5$ ($> -3$), the largest grains dominate both the dust mass budget and the total geometric cross-section.
        Thus, decreasing the minimum grain size to $0.1~{\rm \mu m}$ has only a negligible effect on the ALMA Band 9 optical depth and collisional heat transfer.
        The grain size required to reproduce the cooling condition changes only slightly, shifting to $180~{\rm \mu m} \lesssim a_{\rm max} \lesssim 1~{\rm m m}$ for $\Sigma_{\rm dust}=10~{\rm g~cm^{-2}}$.

        In contrast, $t_{\rm cool,cal}$ is highly sensitive to the size distribution slope, $p$.
        In the Steep-SizeDis model (third panel), the cooling timescale is systematically shorter than in the fiducial model across the entire parameter space.
        A steeper slope of $p=-3.5$ ($< -3$) implies that the smallest grains dominate the total geometric cross-section, enhancing the collisional heat transfer efficiency, even when large grains are present.
        Due to this rapid cooling, the maximum grain size required must increase, reaching $a_{\rm max} \gtrsim 1.2~{\rm mm}$ for $\Sigma_{\rm dust}=1~{\rm g~cm^{-2}}$.
        For $\Sigma_{\rm dust} \gtrsim 3~{\rm g~cm^{-2}}$, cooling becomes so efficient that no dust population with $a_{\rm max} < 1~{\rm cm}$ can satisfy the observationally estimated condition.

        Finally, combining a steep slope with a smaller minimum grain size (the Steep-SizeDis-Small-amin model; right panel) amplifies this effect, reducing the cooling timescale even further.
        In this scenario, cooling is so efficient that only models with very large maximum grain sizes and low dust surface densities can satisfy the cooling condition.
        Specifically, for $a_{\rm max} \lesssim 1~{\rm mm}$ or $\Sigma_{\rm dust} \gtrsim 0.5~{\rm g~cm^{-2}}$, the calculated cooling timescales are too short to reproduce the observations.

        \subsubsection{Dependence on dust porosity}\label{subsubsec:AnalyticModel_results_porosity}
        So far, we have assumed compact dust grains with a filling factor of $f_{\rm dust}=1.0$.
        Introducing dust porosity (i.e., lowering $f_{\rm dust}$) decreases the internal density of the grains [equation \eqref{eq:rho_int}].
        For a given dust surface density, this lower internal density results in a higher total number of dust grains, thereby increasing the total geometric cross section, which makes the gas-dust collisional heat transfer more efficient and shortens the cooling timescale.
        Porosity also alters the dust opacities (figure \ref{fig:opacity_Ricci_DIANA_DSHARP} in appendix \ref{appendix:opacity}) and, through the resulting change in Stokes number, the vertical distribution of the dust density [equation \eqref{eq:St}].
        Both effects change the emission height at ALMA Band 9 (see figure \ref{fig:zobs_parameter} in appendix \ref{appendix:observed_height}), and the change in the vertical dust distribution further modifies the vertical profile of the cooling timescale, altering $t_{\rm cool, cal}$.

        \begin{figure}[t]
            \begin{center}
            \includegraphics[width=\hsize,bb = 0 0 492 842]{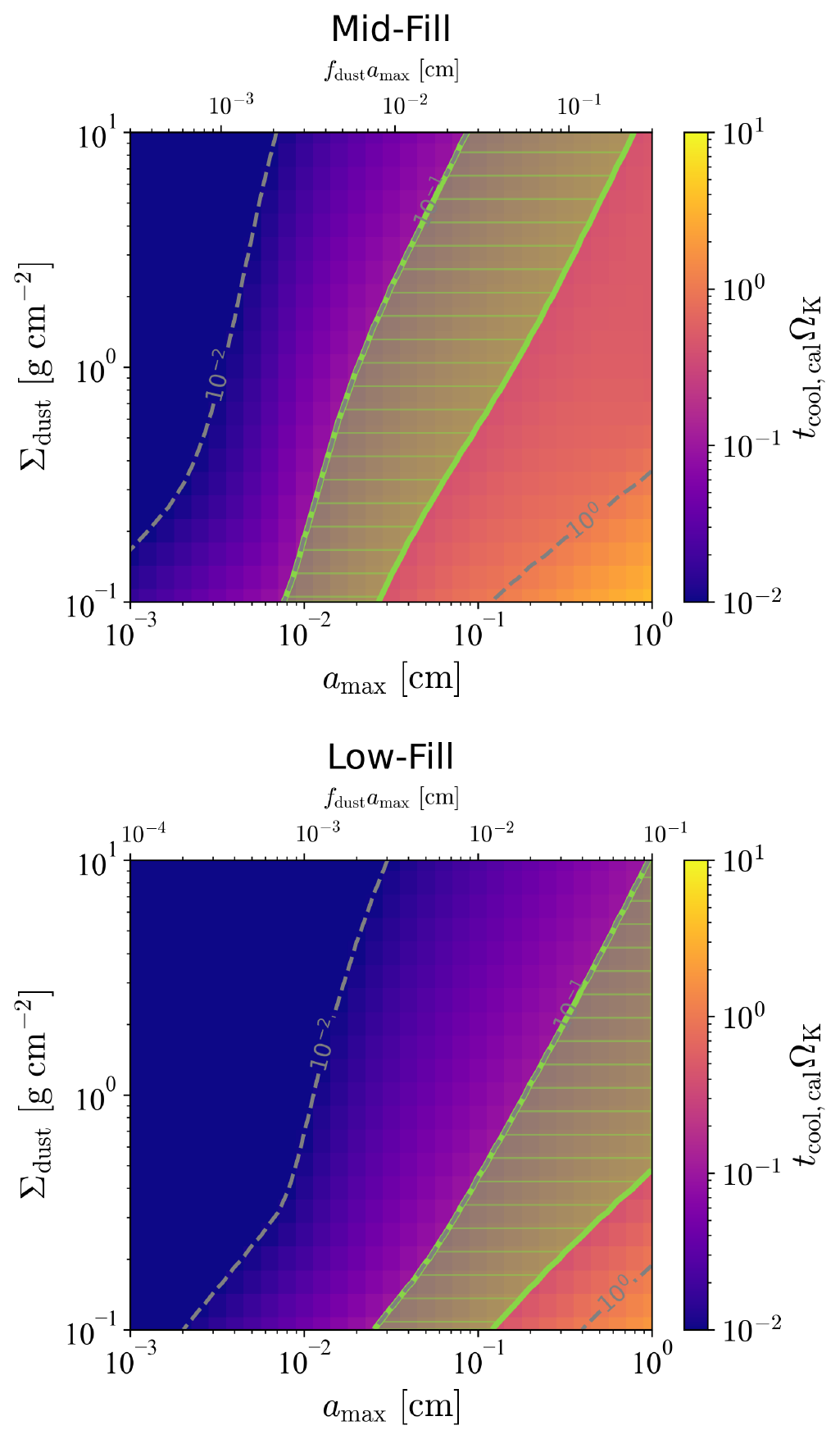}
            \end{center}
            \caption{Same as figure \ref{fig:map_color_fiducial}, but for the Mid-Fill model ($f_{\rm dust}=0.3$; upper panel) and the Low-Fill model ($f_{\rm dust}=0.1$; lower panel). }
            \label{fig:map_color_Poro}
        \end{figure}

        As expected from the enhanced collisional heat transfer, $t_{\rm cool,cal}$ is shorter than in the fiducial model for any given maximum grain size.
        Figure \ref{fig:map_color_Poro} maps $t_{\rm cool,cal}$ across the parameter space for the Mid-Fill model ($f_{\rm dust}=0.3$; upper panel) and Low-Fill model ($f_{\rm dust}=0.1$; lower panel) models.
        To compensate for this efficient cooling and reproduce the observationally estimated timescale, our model requires larger maximum grain sizes.
        Specifically, in the Mid-Fill model, the cooling condition is satisfied by $900~{\rm \mu m} \lesssim a_{\rm max} \lesssim 8~{\rm mm}$ for $\Sigma_{\rm dust}=10~{\rm g~cm^{-2}}$, and by $190~{\rm \mu m} \lesssim a_{\rm max} \lesssim 1.4~{\rm mm}$ for $\Sigma_{\rm dust}=1~{\rm g~cm^{-2}}$.
        In the Low-Fill model, the required grain sizes increase even further, reaching $a_{\rm max} \approx 1~{\rm cm}$ for $\Sigma_{\rm dust}=10~{\rm g~cm^{-2}}$ and $a_{\rm max} \gtrsim 1.7~{\rm mm}$ for $\Sigma_{\rm dust}=1~{\rm g~cm^{-2}}$.

\section{Further constraining dust properties from global VSI condition}\label{sec:VSI_application}

In our analytic model (section \ref{sec:method}), we characterize the vertical distribution of dust particles using the dimensionless vertical diffusion coefficient, $\alpha_z$, which represents the strength of turbulent vertical diffusion.
While $\alpha_z$ is generally a free parameter, we fixed it to $\alpha_z=2\times 10^{-3}$ for the HD~142527 disk in section \ref{subsec:AnalyticModel_results}.
This choice is consistent with the high scattering surface derived in section \ref{subsec:shadow_result} and previous near-infrared observations \citep{TazakiMurakawa+:2021aa}.

From a physical/dynamical perspective, however, the intensity of turbulent diffusion can depend on the local thermodynamic properties.
Specifically, the local cooling timescale dictates the strength of turbulence driven by hydrodynamic instabilities (for a review, \citealt{LesurFlock+:2023aa}).
This expectation motivates us to verify whether the cooling timescales obtained in section \ref{sec:results} can self-consistently sustain turbulence at our assumed level of $\alpha_z=2\times 10^{-3}$.
In this section, we search the $a_{\rm max}$--$\Sigma_{\rm dust}$ parameter space to identify areas that can reproduce this required level of turbulent diffusion. 
A combination of this turbulence condition with the previously established cooling condition gives us a tighter constraint on the dust properties.

In this section, we focus on the vertical shear instability (VSI; e.g., \citealt{NelsonGresselUmurhan2013}) as a primary turbulence-driving mechanism in the outer disk.
Section \ref{subsec:VSI_application_thickness} outlines a criterion for the VSI to operate efficiently.
Section \ref{subsec:VSI_application_results} then presents the constraints on the dust grain sizes that simultaneously satisfy both the observed cooling condition and the required turbulence level.

    \subsection{Condition for VSI-driven turbulence}\label{subsec:VSI_application_thickness}

    Among the hydrodynamic instabilities governed by gas cooling, the VSI (e.g., \citealt{Urpin2003,NelsonGresselUmurhan2013}) is one of the most robust candidates for driving turbulence in the outer regions of protoplanetary disks (for a review, \citealt{LesurFlock+:2023aa}).
    To drive turbulence, the VSI operates in baroclinic disks, relying on a vertical gradient in the gas angular velocity (vertical shear; e.g., \citealt{NelsonGresselUmurhan2013}) and rapid gas thermal relaxation (e.g., \citealt{LinYoudin2015,MangerPfeil+:2021cm}).
    Because VSI-driven turbulence is dominated by vertical gas motions \citep{StollKley2014,ShariffUmurhan:2024aa,LesurLatter+:2025aa}, it is highly efficient at diffusing dust grains in the vertical direction \citep{StollKley:2016vp,Flock:2020aa,DullemondZiampras+:2022aa}.
    Given the rapid cooling controlled by dust, the VSI can dominate in the outer disk regions \citep{Malygin+2017,PfeilKlahr2019}.

    VSI-driven turbulence can also create a geometrically thick dust layer around the midplane \citep{Flock:2020aa,PfeilBirnstiel+:2023aa,PfeilBirnstiel+:2024aa}, which corresponds to the level of $\alpha_z=2\times 10^{-3}$ \citep{FukuharaFlock+:2025aa}. 
    However, this turbulent state is viable only if the cooling efficiency within the diffused dust layer remains sufficient to sustain this turbulence intensity.
    Whether vertical turbulent diffusion can balance dust settling determines whether VSI-driven turbulence can be sustained over the long term \citep{FukuharaOkuzumi:2024aa,FukuharaFlock+:2025aa}.
    Therefore, we can predict the viability of VSI-driven turbulence by assessing the thickness of the rapid cooling regions derived under our assumed turbulent diffusion.

    We obtain the thickness of these VSI-active (rapid cooling) regions from the vertical profile of the cooling timescale calculated in section \ref{subsubsec:method_cooling}.
    The local criterion for the VSI can be written as (e.g., \citealt{Malygin+2017})
    \begin{equation}\label{eq:VSI_criterion}
        t_{\rm cool} \lesssim t_{\rm crit},
    \end{equation}
    where 
    \begin{equation}\label{eq:t_crit}
        t_{\rm crit} = \frac{|q_{\rm temp}|}{\gamma-1}\frac{H_{\rm gas}}{R}\Omega_{\rm K}^{-1}
    \end{equation}
    is the critical cooling time \citep{LinYoudin2015}, with $q_{\rm temp}$ being the radial gradient of the temperature.
    We refer to regions fulfilling this criterion as VSI-active layers\footnote{\citet{FukuharaOkuzumi+:2021ca} and \citet{FukuharaOkuzumi:2024aa} called these VSI zones and VSI-unstable layers, respectively.}.
    Our model shows that the VSI-active region in the HD 142527 disk extends continuously across the midplane, occupying both the upper ($z>0$) and lower ($z<0$) halves of the disk.
    Thus, letting $z_{\rm VSI}$ be the height of the active region's upper boundary, we define the total thickness of the VSI-active layer as $\Delta L_{\rm VSI} = 2z_{\rm VSI}$ \citep{FukuharaOkuzumi+:2023aa}.

    For VSI-driven turbulence to maintain a strong vertical diffusion of $\alpha_z=2\times 10^{-3}$ in a self-consistent manner, the VSI-active layer must be thicker than two gas scale heights \citep{FukuharaOkuzumi:2024aa,FukuharaFlock+:2025aa}:
    \begin{equation}\label{eq:VSIcondition}
        \Delta L_{\rm VSI}\gtrsim 2H_{\rm gas}.
    \end{equation}
    Hereafter, we refer to this requirement as the ``global VSI condition''.

    \subsection{Constraints on dust grain size from the global VSI condition}\label{subsec:VSI_application_results}

    We apply the framework developed in the previous subsection to map the thickness of the VSI-active layer, $\Delta L_{\rm VSI}$, across our parameter space.
    Following the approach in section \ref{subsec:AnalyticModel_results}, we compute $\Delta L_{\rm VSI}$ for every combination of $a_{\rm max}$ and $\Sigma_{\rm dust}$ defined in section \ref{subsec:parameter_choices} to identify the regions that satisfy the global VSI condition [equation \eqref{eq:VSIcondition}].
    When evaluating the critical cooling timescale [equation \eqref{eq:t_crit}], we assume a radial temperature profile with a slope of $|q_{\rm temp}|=0.5$.

    \begin{figure}[t]
        \begin{center}
        \includegraphics[width=\hsize,bb = 0 0 422 349]{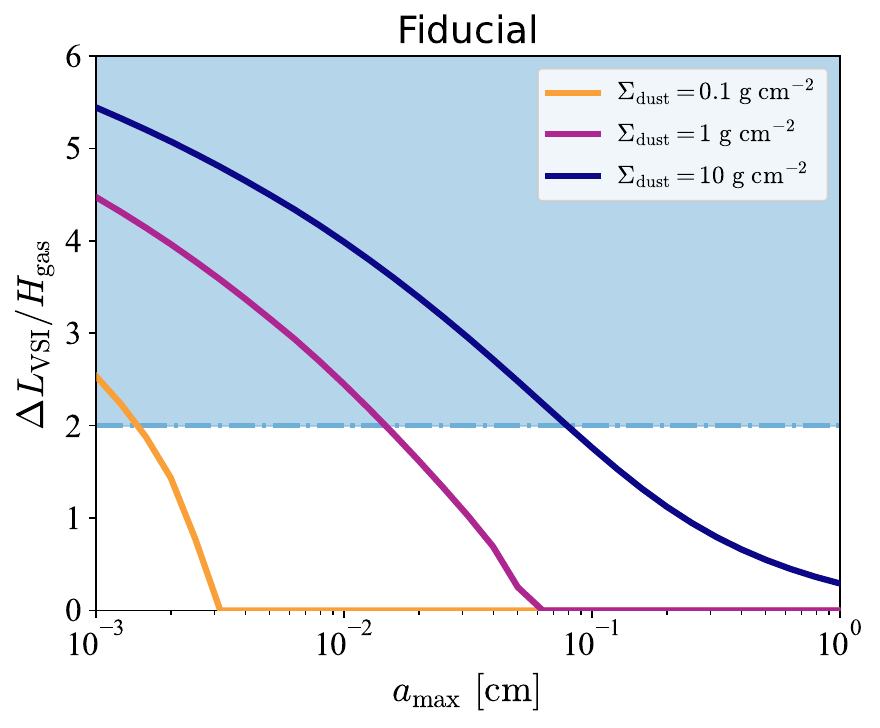}
        \end{center}
        \caption{Thickness of the VSI-active region $\Delta L_{\rm VSI}$ as a function of $a_{\rm max}$ for different values of $\Sigma_{\rm dust}$ in the fiducial model (see table \ref{table:run} for the parameter choice). The horizontal dash-dotted line marks the critical threshold $\Delta L_{\rm VSI}=2H_{\rm gas}$. The shaded area of $\Delta L_{\rm VSI} > 2H_{\rm gas}$ highlights the parameter space that satisfies the global VSI condition, indicating where the VSI can self-consistently drive turbulence at a level of $\alpha_z=2\times 10^{-3}$. }
        \label{fig:LVSI_amax_fiducial}
    \end{figure}

    Figure \ref{fig:LVSI_amax_fiducial} illustrates the thickness of the VSI-active region, $\Delta L_{\rm VSI}$, as a function of the maximum grain size $a_{\rm max}$ for the fiducial model (see table \ref{table:run} for the parameter choice).
    As $a_{\rm max}$ increases, the total geometric cross section of the dust decreases, making cooling via collisional heat transfer less efficient.
    As a result, the VSI-active region shrinks, eventually becoming infinitely thin.
    For the dust surface density of $\Sigma_{\rm dust}=10~{\rm g~cm^{-2}}$, the global VSI condition is satisfied as long as $a_{\rm max} \lesssim 1~{\rm mm}$.
    Decreasing $\Sigma_{\rm dust}$ further reduces the cooling efficiency, requiring even smaller dust grains to sustain VSI-driven turbulence.
    Specifically, the threshold maximum grain size capable of satisfying the global VSI condition decreases to $a_{\rm max} \lesssim 130~{\rm \mu m}$ for $\Sigma_{\rm dust}=1~{\rm g~cm^{-2}}$ and $a_{\rm max} \lesssim 30~{\rm \mu m}$ for $\Sigma_{\rm dust}=0.1~{\rm g~cm^{-2}}$.

    \begin{figure*}[t]
        \begin{center}
        \includegraphics[width=\hsize,bb = 0 0 1370 709]{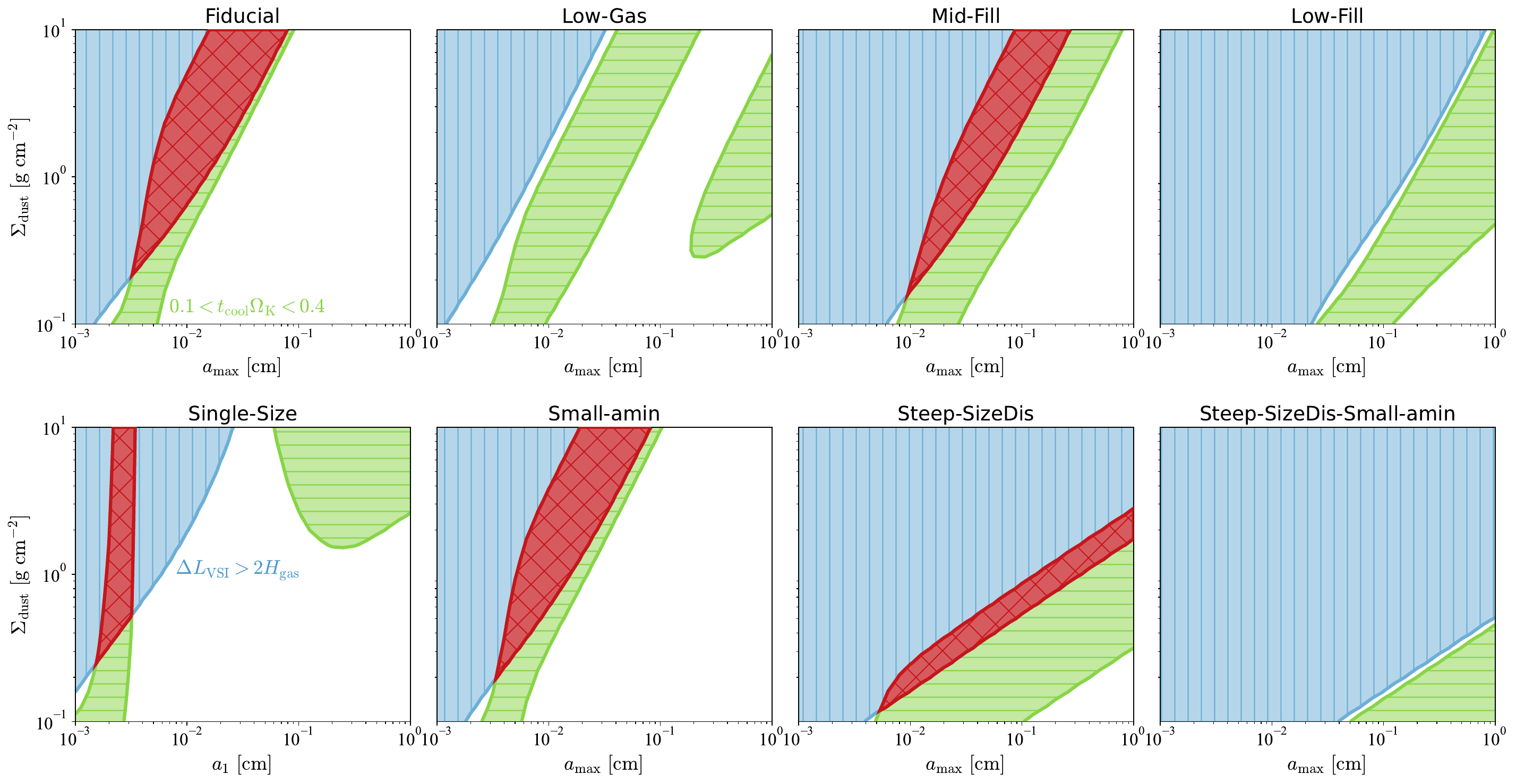}
        \end{center}
        \caption{Area diagram in the $a_{\rm max}$--$\Sigma_{\rm dust}$ plane for the models presented in table \ref{table:run}, excluding the varying opacity models. The vertically and horizontally hatched areas indicate the parameter spaces satisfying the global VSI condition [$\Delta L_{\rm VSI}>2H_{\rm gas}$; equation \eqref{eq:VSIcondition}] and the cooling condition ($0.1<t_{\rm cool,cal}\Omega_{\rm K}< 0.4$), respectively. The cross-hatched regions show the parameter spaces satisfying both conditions. }
        \label{fig:map_VSI_models}
    \end{figure*}

    By comparing the global VSI condition of $\Delta L_{\rm VSI}>2H_{\rm gas}$ with the observational estimated cooling timescale of $0.1\lesssim t_{\rm cool,est}\Omega_{\rm K}\lesssim 0.4$, we identify $a_{\rm max}$ and $\Sigma_{\rm dust}$ that can self-consistently explain both the cooling and the vertical dust distribution in the HD~142527 disk.
    Figure \ref{fig:map_VSI_models} displays the $a_{\rm max}$--$\Sigma_{\rm dust}$ parameter spaces that satisfy both conditions for all models, excluding the varying-opacity models.
    As shown in figure \ref{fig:LVSI_amax_fiducial}, the parameter space satisfying the global VSI condition corresponds to small $a_{\rm max}$ and high $\Sigma_{\rm dust}$.
    For the fiducial model, the maximum grain sizes satisfying both conditions are $150~{\rm \mu m}\lesssim a_{\rm max}\lesssim 800~{\rm \mu m}$ at $\Sigma_{\rm dust}=10~{\rm g~cm^{-2}}$, and $46~{\rm \mu m}\lesssim a_{\rm max}\lesssim 150~{\rm \mu m}$ at $\Sigma_{\rm dust}=1~{\rm g~cm^{-2}}$.
    However, at lower $\Sigma_{\rm dust}$, this overlapping parameter space completely disappears.
    This is because $a_{\rm max}$ reproducing the estimated cooling timescale at low ALMA Band 9 emission heights ($z_{\rm emi}< H_{\rm gas}$; see figure \ref{fig:zobs_amax_fiducial}) allows the condition $t_{\rm cool} < t_{\rm crit}$ to satisfy only in regions with a vertical extent smaller than $H_{\rm gas}$, thereby failing the global VSI condition.

    Because the vertical distribution of the cooling timescale depends on the gas surface density, the $a_{\rm max}$--$\Sigma_{\rm dust}$ parameter spaces satisfying the global VSI condition also depend on it.
    In the Low-Gas model (see table \ref{table:run} for the parameter choice), no $a_{\rm max}$ satisfies both the cooling and global VSI conditions at any $\Sigma_{\rm dust}$.
    Due to enhanced dust settling, the VSI-active region shrinks toward the midplane (see also \citealt{FukuharaOkuzumi+:2021ca}), restricting its thickness to less than two gas scale heights and thus failing to drive strong turbulence.

    Furthermore, the size distribution affects the cooling timescale profile and the resulting VSI-active region.
    The bottom panels of figure \ref{fig:map_VSI_models} show the parameter diagrams for the different size distribution models (see table \ref{table:run} for the parameter choice).
    For the Single-Size model, the global VSI condition [equation \eqref{eq:VSIcondition}] is satisfied only for $a_1\lesssim 200~{\rm \mu m}$ and $\Sigma_{\rm dust} \gtrsim 0.2~{\rm g~cm^{-2}}$.
    Therefore, of the parameter spaces satisfying the cooling condition, the range with larger grains is excluded; only $20~{\rm \mu m}\lesssim a_1\lesssim 30~{\rm \mu m}$ at $\Sigma_{\rm dust}\gtrsim 0.3~{\rm g~cm^{-2}}$ satisfies both conditions.
    For the Small-amin model, the parameter range satisfying both conditions remains similar to that of the fiducial model.
    Conversely, for steep size distributions, $a_{\rm max}$ is further restricted by the global VSI condition.
    In the Steep-SizeDis model, the range satisfying both conditions is $1.2~{\rm mm}\lesssim a_{\rm max}\lesssim 3.7~{\rm mm}$ at $\Sigma_{\rm dust}=1~{\rm g~cm^{-2}}$.
    For the Steep-SizeDis-Small-amin model, the overlapping parameter space disappears entirely.

    Moreover, dust porosity can alter the parameter space satisfying the global VSI condition by changing the cooling timescale distribution.
    As discussed in section \ref{subsubsec:AnalyticModel_results_porosity}, higher porosity shortens the cooling timescale for a given $a_{\rm max}$. 
    Reproducing the cooling condition therefore requires larger grains as porosity increases, and the parameter space satisfying the global VSI condition shifts toward larger grain sizes for the same reason.
    For the Mid-Fill model, both conditions limit the maximum grain size to $900~{\rm \mu m}\lesssim a_{\rm max}\lesssim 2.8~{\rm mm}$ at $\Sigma_{\rm dust}=10~{\rm g~cm^{-2}}$, and $190~{\rm \mu m}\lesssim a_{\rm max}\lesssim 530~{\rm \mu m}$ at $\Sigma_{\rm dust}=1~{\rm g~cm^{-2}}$.
    For the Low-Fill model, the overlapping space almost disappears.
    This trend reflects a difference in how the two conditions depend on $f_{\rm dust}$: the global VSI condition is determined solely by the vertical profile of the cooling timescale, whereas the cooling condition depends additionally on the emission height, making it more sensitive to $f_{\rm dust}$ than the global VSI condition.
    We therefore conclude that the dust reproducing both the short cooling timescale and the high scattering surface height must be relatively compact, requiring a filling factor of $f_{\rm dust} \gtrsim 0.1$.
 
\section{General discussion}\label{sec:discussion}

    \subsection{Dust properties and growth limits in the HD~142527 disk}\label{subsec:_discussion_dust_growth}

    As demonstrated in section \ref{subsec:AnalyticModel_results}, the short cooling timescale observationally derived for the HD~142527 disk (section \ref{subsec:shadow_result}) can be explained by maximum dust grain sizes in the range of $150~{\rm \mu m}\lesssim a_{\rm max}\lesssim 900~{\rm \mu m}$ with $\Sigma_{\rm gas} = \Sigma_{\rm dust}=10~{\rm g~cm^{-2}}$, $p=-2.5$, and $f_{\rm dust}=1.0$.
    In the northern region of this disk, these inferred submillimeter sizes are consistent with estimates from submillimeter polarization observations \citep{KataokaTsukagoshi+:2016ae,OhashiKataoka+:2018aa}.
    However, multi-wavelength dust thermal emission observations indicate low spectral indices \citep{CasassusWright+:2015aa,SoonMomose+:2019aa,YenGu:2020aa}, implying the presence of millimeter-sized dust grains.
    Since smaller values of $f_{\rm dust}$ and $p$ lead to larger inferred maximum grain sizes in our model (see figures \ref{fig:map_color_SizeDis}, \ref{fig:map_color_Poro}, and \ref{fig:map_VSI_models}), dust with a relatively porous structure or a steep size distribution may explain the low spectral indices in addition to the short cooling timescale.
    To further constrain dust properties, incorporating multi-wavelength dust continuum emission into the current model should be addressed in future work.

    Regarding the composition and internal structure, the temperature of the outer disk region is estimated to be $T\approx 35~{\rm K}$ (see figure \ref{fig:tauall_T_Hgas_amax}), which implies that the dust particles would contain abundant volatiles, including H$_2$O and CO$_2$ ices.
    Furthermore, applying the global VSI condition discussed in section \ref{sec:VSI_application} suggests that the dust requires a moderate filling factor of $f_{\rm dust} \gtrsim 0.1$ (figure \ref{fig:map_VSI_models}).
    Future modeling is necessary to verify whether dust exhibiting these specific properties can simultaneously reproduce the full suite of available multi-wavelength observations.
    
    From the perspective of grain sticking, a submillimeter maximum size also implies that dust growth in the disk is limited either by collisional fragmentation with a low threshold velocity (e.g., \citealt{BirnstielOrmel+:2011aa}) or by bouncing between dust particles (e.g., \citealt{KotheBlum+:2013aa,ArakawaOkuzumi+:2023aa,DominikDullemond:2024aa,OshiroTatsuuma+:2025aa}).
    When fragmentation limits dust coagulation and turbulent mixing dominates the relative velocities between particles, the maximum grain size determined by fragmentation can be approximated as (e.g., \citealt{BirnstielDullemond+:2009aa,Birnstiel:2012aa})
    \begin{equation}\label{eq:a_max_vfrag}
        a_{\rm max} \approx \frac{2\Sigma_{\rm gas}}{\pi\rho_{\rm int}}\frac{v_{\rm frag}^2}{c_{\rm s}^2}\frac{1}{\alpha_{\rm mix}},
    \end{equation}
    where $v_{\rm frag}$ is the fragmentation threshold velocity, and $\alpha_{\rm mix}$ is the dimensionless parameter characterizing the turbulent mixing of dust particles.
    Assuming $\alpha_{\rm mix} = \alpha_z$, and substituting the derived range of $150~{\rm \mu m}\lesssim a_{\rm max}\lesssim 900~{\rm \mu m}$ alongside $\alpha_z=2\times 10^{-3}$ and $\Sigma_{\rm gas}=10~{\rm g~cm^{-2}}$ into equation \eqref{eq:a_max_vfrag}, we obtain $1.0~{\rm m~s^{-1}} \lesssim v_{\rm frag} \lesssim 2.4~{\rm m~s^{-1}}$ for compact dust.
    This low fragmentation velocity is consistent with recent experimental results for water ice at low temperatures \citep{GundlachSchmidt+:2018aa,MusiolikWurm:2019aa} and CO$_2$ ice \citep{FritscherTeiser:2021aa}.

    If bouncing limits dust growth, the maximum grain size reaches approximately $100~{\rm \mu m}$ (e.g., \citealt{DominikDullemond:2024aa}), which is also consistent with our inferred sizes.
    To identify the exact mechanism determining the dust grain size in the disk, future work must incorporate numerical dust growth simulations including fragmentation and bouncing, verifying whether they can simultaneously reproduce the observationally estimated cooling timescale and the high scattering surface.

    Moreover, the cooling timescale of $20~{\rm yr}\lesssim t_{\rm cool,est} \lesssim 90~{\rm yr}$ constrained in section \ref{subsec:shadow_result} is longer than the collisional heat transfer timescale between gas molecules and dust particles estimated for the disk around TW~Hya ($t_{\rm coll}\sim 10~{\rm yr}$; \citealt{TeagueBae+:2022aa}).
    Because the collisional timescale dominates the overall cooling timescale in our model (see the lower panel of figure \ref{fig:tau_tcool_z_fiducial}), as is typical for outer disk regions (e.g., \citealt{Malygin+2017,PfeilKlahr2019}), both studies essentially estimate the same physical timescale for different disks.
    This difference between the HD~142527 and TW~Hya disks can be attributed to variations in local and global disk properties, such as disk age, radial location, temperature, gas and dust densities, and dust grain size.
    Detailed comparative modeling is required in future work to investigate the origin of this timescale difference.

    \subsection{Limitation of the presented method}\label{subsec:discuss_limitation}
    The method proposed in this paper is subject to several important limitations that should be addressed in future work.
    First, we assume that the location of the shadow cast by an axisymmetric, misaligned inner disk remains constant over the relevant cooling timescale.
    Very recently, near-infrared observations have detected temporal variations in the locations of the brightness dips and peaks in several protoplanetary disks \citep{MullinLucas+:2026aa}, with one possible mechanism being a moving shadow caused by the precession of a misaligned inner disk.
    Our method remains valid provided that the timescale of the shadow's temporal variation is sufficiently long compared to both the cooling and Keplerian timescales in the outer disk.
    However, a rapid movement of the shadow could alter the thermal response of the outer disk, thereby affecting our estimation of the cooling timescale.
    Quantifying this effect requires detailed modeling of the azimuthal temperature distribution and its time evolution driven by a rapidly moving shadow, which we defer to future work.

    Second, we assume that the gas and dust in the outer disk remain in a steady state on Keplerian timescales.
    Local cooling due to the inner-disk-induced shadow can create azimuthal temperature variations.
    This asymmetric temperature distribution triggers the formation of gas spirals in the outer disk driven by azimuthal pressure gradients \citep{MontesinosPerez+:2016aa,ZhangZhu:2024aa}.
    These spirals can subsequently break up to form ring structures \citep{ZhuZhang+:2025aa} and non-axisymmetric crescents or vortices \citep{ZhangZhu:2024aa}.
    The dominant substructures transition from spirals to rings, and eventually to crescents, as the azimuthal temperature variation increases or the cooling timescale decreases \citep{SuBai:2024aa,ZiamprasDullemond+:2025aa}.
    Furthermore, these gas substructures can trap millimeter-sized dust particles in the pressure maxima of the spirals \citep{CuelloMontesinos+:2019aa}, leading to the formation of dust rings or crescents via aerodynamic drag \citep{ZiamprasDullemond+:2025aa}.
    Moreover, \citet{ZhangZhu+:2025aa} recently found that azimuthal temperature variations in the outer disk, when induced by an inner disk that is neither coplanar nor strictly polar, can develop a warp structure in the outer disk.
    These effects imply that, in the limit of an extremely short cooling timescale, the shadows cast by the inner disk dictate the gas kinematics and the spatial distributions of both gas and dust in the outer disk.
    Therefore, our derived cooling timescale, which corresponds to a few percent of the orbital period, suggests these warp structures would form over a long time in the HD~142527 disk.
    To understand this long-term evolution, future disk modeling must incorporate gas and dust dynamics driven by the shadow-induced azimuthal temperature variations.

    Third, we neglect magnetic fields and self-gravity, both of which can influence the dynamics and distributions of the gas and dust.
    For the protoplanetary disk around HD~142527, which is the primary target of our application in section \ref{sec:results}, \citet{OhashiMuto+:2025aa} recently reported multi-wavelength observations of polarized thermal emission and derived a magnetic field strength of $\sim 0.3~{\rm mG}$ at a radius of $\sim 200~{\rm au}$ in the southern part of the disk.
    If the magnetic field in the northern part of the disk is comparable to that in the southern part, the gas and dust dynamics may be largely governed by magnetohydrodynamic effects.
    Additionally, gas self-gravity can alter the disk dynamics via gravitational instabilities.
    In the HD~142527 disk, polarized scattered-light observations have revealed spiral structures \citep{AvenhausQuanz+:2014aa}, suggesting that these features might be driven by the gravitational instability (GI).
    In our fiducial model, the assumed gas surface density yields a minimum Toomre's $Q_{\rm T}$ parameter of $Q_{\rm T} \approx 2.3$ (see appendix \ref{appendix:temperature}), implying that the disk is currently stable against the GI.
    However, if we were to assume a more massive disk, the GI could dominate the dust dynamics, which would modify our conclusions.

    \subsection{Possible applications to other protoplanetary disks}\label{subsec:discussion_other_disks}
    The framework presented in this work can be applied to other transition disks that exhibit shadow-like non-axisymmetric structures in their outer disks. Recent high-angular-resolution observations suggest that inner--outer disk misalignments, as inferred for the HD~142527 disk, are present in several transition disk systems. In particular, \citet{BohnBenisty+:2022aa} compared inner-disk geometries constrained by VLTI/GRAVITY with outer-disk geometries derived from ALMA molecular-line observations, and found significant misalignments in several transition disks. In addition, \citet{Francisvan-der-Marel:2020aa} reported compact inner dust disks inside the cavities of many transition disks, with a substantial fraction of the resolved inner disks being misaligned with their outer disks. These results indicate that the geometric configuration required for an inner disk to cast shadows onto an outer disk may not be rare. 
    
    Promising targets for applying this method are transition disks in which shadows, spirals, or azimuthal brightness asymmetries have been reported in scattered-light and/or ALMA observations. 
    Examples include HD~100453 \citep{OriharaMomose:2025aa}, J1604--2130 \citep{MayamaHashimoto+:2012aa,PinillaBenisty+:2018aa,MayamaAkiyama+:2018aa}, DoAr~44 \citep{CasassusAvenhaus+:2018aa}, and CQ~Tau \citep{UyamaMuto+:2020aa,WolferFacchini+:2021aa,HammondChristiaens+:2022aa}. 
    Applying our method to these systems would enable a systematic investigation of cooling times across disks with different shadow geometries and disk structures, which we leave for future work.
  
\section{Conclusions}\label{sec:conclusions}
We present a novel method that estimates the cooling timescale using multi-wavelength observations and constrains the dust grain size in protoplanetary disks, utilizing shadows in the outer disk caused by the misaligned inner disk (figure \ref{fig:overview}).
We also apply this method to the protoplanetary disk around HD~142527 as a proof-of-concept.
Our key findings are summarized as follows.
\begin{enumerate}
    \item The reconstructed three-dimensional geometry of the shadows projected onto the outer disk is derived from the VLT H-band $Q_\phi$ scattered-light image.
    The resulting model indicates that the azimuthally localized dark regions in the scattered-light image can be interpreted as shadows cast by an inner disk misaligned by approximately $60^\circ$ with respect to the outer disk. 
    The shadow-casting structure has a maximum aspect ratio of $h_{\rm r}=0.16$, suggesting that the optically thick inner dusty component has a finite vertical extent (figure~\ref{fig:HD142527_surface}).
    
    \item For HD~142527, the reconstructed shadow geometry is compared with the azimuthal temperature distribution derived from the ALMA submillimeter continuum image.
    After accounting for the projection of the three-dimensional shadows onto the disk midplane, the modeled shadow-induced temperature decrease yields a dimensionless cooling time of $t_{\rm cool,est}\Omega_{\rm K}=0.1$--$0.4$ at $R=170~{\rm au}$.
    For a stellar mass of $M_\star=2.4\,M_\odot$, this corresponds to a cooling timescale of $t_{\rm cool,est}=20$--$90~{\rm yr}$. 
    This result suggests that the outer disk of HD~142527 responds thermally to shadowing on a timescale shorter than the local Keplerian timescale (figure~\ref{fig:HD142527_temp}).

    \item We find that submillimeter-sized dust particles can reproduce the observationally estimated cooling timescale. Assuming the compact dust with a size-distribution slope of $-2.5$, the maximum dust grain size $a_{\rm max}$ reproducing the observationally estimated cooling timescale is $150~{\rm \mu m}\lesssim a_{\rm max} \lesssim900~{\rm \mu m}$ at a dust surface density of $\Sigma_{\rm dust}=10~{\rm g~cm^{-2}}$ (figure \ref{fig:tcool_amax_fiducial}). Grain sizes below or above this range yield cooling timescales that are too short or too long, respectively, to match the observations.

    \item As the dust surface density decreases, the constrained maximum grain size decreases. If $\Sigma_{\rm dust}=1.0~{\rm g~cm^{-2}}$ and $0.1~{\rm g~cm^{-2}}$, the range of maximum grain size reproducing the observationally estimated cooling timescale is $50~{\rm \mu m}\lesssim a_{\rm max}\lesssim 200~{\rm \mu m}$ and $20~{\rm \mu m}\lesssim a_{\rm max}\lesssim 50~{\rm \mu m}$, respectively (figure \ref{fig:map_color_fiducial}).
    
    \item The constrained dust grain size also depends on the gas surface density, size distribution, and dust filling factor, as well as the dust surface density. The low gas surface density decreases the cooling timescale at the ALMA Band 9 emission height, $t_{\rm cool,cal}$ (figure \ref{fig:map_color_Sigmag1e0}). The size distribution of dust grains affects $t_{\rm cool,cal}$ through changes in both the abundance of each grain size and the total cross section of the dust population (figure \ref{fig:map_color_SizeDis}). Porous dust yields a shorter cooling timescale than compact dust (figure \ref{fig:map_color_Poro}). Any physical variation that concentrates dust near the midplane (e.g., a lower gas surface density) or increases the total geometric cross section (e.g., a steeper size distribution, a smaller minimum grain size, or higher porosity) enhances the collisional cooling efficiency, and thus requires a larger $a_{\rm max}$ to reproduce $t_{\rm cool,est}$.
    
    \item In addition to the condition reproducing $t_{\rm cool,est}$, we check the condition explaining the high scattering surface. Assuming the VSI as a turbulence-driving mechanism, we calculate the VSI-active layer's thickness $\Delta L_{\rm VSI}$ and search the $a_{\rm max}$--$\Sigma_{\rm dust}$ parameter space that satisfies the $\Delta L_{\rm VSI} > 2H_{\rm gas}$ (global VSI condition), where $H_{\rm gas}$ is the gas scale height. This condition is one under which the VSI can drive turbulence of sufficient intensity to be consistent with the assumed vertical diffusion coefficient. Considering both cooling and global VSI conditions, we find that $\Sigma_{\rm dust}$ and $a_{\rm max}$ can be further constrained (figure \ref{fig:map_VSI_models}). 

\end{enumerate}

This study demonstrates that estimating cooling timescales is an effective tool to constrain dust grain size. 
Our approach can also be generally applied to other transition disks with inner-disk-induced shadows in the outer disks.

\section*{Funding}
This work was supported by JSPS KAKENHI Grant Numbers JP24K00674 and JP26KJ0122. 

\begin{ack}
    We thank Shangjia Zhang, Hsi-Wei Yen, and Jun Hashimoto for a useful discussion.
    Based on observations collected at the European Southern Observatory under ESO programme 099.C-0601(A). The IRDIS H-band FITS product used in this work was obtained from the CDS/VizieR catalogue J/A+A/648/A110.
    This paper makes use of the following ALMA data: ADS/JAO.ALMA\#2015.1.00614.S. ALMA is a partnership of ESO (representing its member states), NSF (USA) and NINS (Japan), together with NRC (Canada), NSTC and ASIAA (Taiwan), and KASI (Republic of Korea), in cooperation with the Republic of Chile. The Joint ALMA Observatory is operated by ESO, AUI/NRAO and NAOJ.
    We thank the staff of the East Asian ALMA Regional Center for preparing and providing the calibrated measurement sets used in this study.
    Data analysis was in part carried out on the Multi-wavelength Data Analysis System operated by the Astronomy Data Center (ADC), National Astronomical Observatory of Japan.
\end{ack}

\bibliographystyle{aasjournal}
\bibliography{FukuharaOrihara+}

\begin{thebibliography}{}
\expandafter\ifx\csname natexlab\endcsname\relax\def\natexlab#1{#1}\fi
\providecommand{\url}[1]{\href{#1}{#1}}
\providecommand{\dodoi}[1]{doi:~\href{http://doi.org/#1}{\nolinkurl{#1}}}
\providecommand{\doeprint}[1]{\href{http://ascl.net/#1}{\nolinkurl{http://ascl.net/#1}}}
\providecommand{\doarXiv}[1]{\href{https://arxiv.org/abs/#1}{\nolinkurl{https://arxiv.org/abs/#1}}}

\bibitem[{{Arakawa} {et~al.}(2023){Arakawa}, {Okuzumi}, {Tatsuuma}, {Tanaka},
  {Kokubo}, {Nishiura}, {Furuichi}, \& {Nakamoto}}]{ArakawaOkuzumi+:2023aa}
{Arakawa}, S., {Okuzumi}, S., {Tatsuuma}, M., {et~al.} 2023, \apjl, 951, L16,
  \dodoi{10.3847/2041-8213/acdb5f}

\bibitem[{{Arun} {et~al.}(2019){Arun}, {Mathew}, {Manoj}, {Ujjwal}, {Kartha},
  {Viswanath}, {Narang}, \& {Paul}}]{ArunMathew+:2019aa}
{Arun}, R., {Mathew}, B., {Manoj}, P., {et~al.} 2019, \aj, 157, 159,
  \dodoi{10.3847/1538-3881/ab0ca1}

\bibitem[{{Avenhaus} {et~al.}(2014){Avenhaus}, {Quanz}, {Schmid}, {Meyer},
  {Garufi}, {Wolf}, \& {Dominik}}]{AvenhausQuanz+:2014aa}
{Avenhaus}, H., {Quanz}, S.~P., {Schmid}, H.~M., {et~al.} 2014, \apj, 781, 87,
  \dodoi{10.1088/0004-637X/781/2/87}

\bibitem[{{Avenhaus} {et~al.}(2017){Avenhaus}, {Quanz}, {Schmid}, {Dominik},
  {Stolker}, {Ginski}, {de Boer}, {Szul^^c3^^a1gyi}, {Garufi}, {Zurlo},
  {Hagelberg}, {Benisty}, {Henning}, {M^^c3^^a9nard}, {Meyer}, {Baruffolo},
  {Bazzon}, {Beuzit}, {Costille}, {Dohlen}, {Girard}, {Gisler}, {Kasper},
  {Mouillet}, {Pragt}, {Roelfsema}, {Salasnich}, \&
  {Sauvage}}]{AvenhausQuanz+:2017aa}
---. 2017, \aj, 154, 33, \dodoi{10.3847/1538-3881/aa7560}

\bibitem[{{Bae} {et~al.}(2023){Bae}, {Isella}, {Zhu}, {Martin}, {Okuzumi}, \&
  {Suriano}}]{BaeIsella+:2023aa}
{Bae}, J., {Isella}, A., {Zhu}, Z., {et~al.} 2023, in Astronomical Society of
  the Pacific Conference Series, Vol. 534, Astronomical Society of the Pacific
  Conference Series, ed. S.~{Inutsuka}, Y.~{Aikawa}, T.~{Muto}, K.~{Tomida}, \&
  M.~{Tamura}, 423

\bibitem[{{Bae} {et~al.}(2021){Bae}, {Teague}, \& {Zhu}}]{BaeTeague+:2021aa}
{Bae}, J., {Teague}, R., \& {Zhu}, Z. 2021, \apj, 912, 56,
  \dodoi{10.3847/1538-4357/abe45e}

\bibitem[{{Baehr} \& {Klahr}(2015)}]{BaehrKlahr:2015aa}
{Baehr}, H., \& {Klahr}, H. 2015, \apj, 814, 155,
  \dodoi{10.1088/0004-637X/814/2/155}

\bibitem[{{Baehr} \& {Zhu}(2021{\natexlab{a}})}]{BaehrZhu:2021aa}
{Baehr}, H., \& {Zhu}, Z. 2021{\natexlab{a}}, \apj, 909, 135,
  \dodoi{10.3847/1538-4357/abddb3}

\bibitem[{{Baehr} \& {Zhu}(2021{\natexlab{b}})}]{BaehrZhu:2021ab}
---. 2021{\natexlab{b}}, \apj, 909, 136, \dodoi{10.3847/1538-4357/abddb4}

\bibitem[{{Barranco} {et~al.}(2018){Barranco}, {Pei}, \&
  {Marcus}}]{BarrancoPei+:2018kc}
{Barranco}, J.~A., {Pei}, S., \& {Marcus}, P.~S. 2018, \apj, 869, 127,
  \dodoi{10.3847/1538-4357/aaec80}

\bibitem[{{Bate}(2018)}]{Bate:2018aa}
{Bate}, M.~R. 2018, \mnras, 475, 5618, \dodoi{10.1093/mnras/sty169}

\bibitem[{{Benisty} {et~al.}(2017){Benisty}, {Stolker}, {Pohl}, {de Boer},
  {Lesur}, {Dominik}, {Dullemond}, {Langlois}, {Min}, {Wagner}, {Henning},
  {Juhasz}, {Pinilla}, {Facchini}, {Apai}, {van Boekel}, {Garufi}, {Ginski},
  {M^^c3^^a9nard}, {Pinte}, {Quanz}, {Zurlo}, {Boccaletti}, {Bonnefoy},
  {Beuzit}, {Chauvin}, {Cudel}, {Desidera}, {Feldt}, {Fontanive}, {Gratton},
  {Kasper}, {Lagrange}, {LeCoroller}, {Mouillet}, {Mesa}, {Sissa}, {Vigan},
  {Antichi}, {Buey}, {Fusco}, {Gisler}, {Llored}, {Magnard}, {Moeller-Nilsson},
  {Pragt}, {Roelfsema}, {Sauvage}, \& {Wildi}}]{BenistyStolker+:2017aa}
{Benisty}, M., {Stolker}, T., {Pohl}, A., {et~al.} 2017, \aap, 597, A42,
  \dodoi{10.1051/0004-6361/201629798}

\bibitem[{{Benisty} {et~al.}(2018){Benisty}, {Juh^^c3^^a1sz}, {Facchini},
  {Pinilla}, {de Boer}, {P^^c3^^a9rez}, {Keppler}, {Muro-Arena}, {Villenave},
  {Andrews}, {Dominik}, {Dullemond}, {Gallenne}, {Garufi}, {Ginski}, \&
  {Isella}}]{BenistyJuhasz+:2018aa}
{Benisty}, M., {Juh^^c3^^a1sz}, A., {Facchini}, S., {et~al.} 2018, \aap, 619,
  A171, \dodoi{10.1051/0004-6361/201833913}

\bibitem[{{Birnstiel} {et~al.}(2009){Birnstiel}, {Dullemond}, \&
  {Brauer}}]{BirnstielDullemond+:2009aa}
{Birnstiel}, T., {Dullemond}, C.~P., \& {Brauer}, F. 2009, \aap, 503, L5,
  \dodoi{10.1051/0004-6361/200912452}

\bibitem[{{Birnstiel} {et~al.}(2010){Birnstiel}, {Dullemond}, \&
  {Brauer}}]{BirnstielDullemond+:2010aa}
---. 2010, \aap, 513, A79, \dodoi{10.1051/0004-6361/200913731}

\bibitem[{{Birnstiel} {et~al.}(2012){Birnstiel}, {Klahr}, \&
  {Ercolano}}]{Birnstiel:2012aa}
{Birnstiel}, T., {Klahr}, H., \& {Ercolano}, B. 2012, \aap, 539, A148,
  \dodoi{10.1051/0004-6361/201118136}

\bibitem[{{Birnstiel} {et~al.}(2011){Birnstiel}, {Ormel}, \&
  {Dullemond}}]{BirnstielOrmel+:2011aa}
{Birnstiel}, T., {Ormel}, C.~W., \& {Dullemond}, C.~P. 2011, \aap, 525, A11,
  \dodoi{10.1051/0004-6361/201015228}

\bibitem[{{Birnstiel} {et~al.}(2018){Birnstiel}, {Dullemond}, {Zhu}, {Andrews},
  {Bai}, {Wilner}, {Carpenter}, {Huang}, {Isella}, {Benisty}, {P^^c3^^a9rez},
  \& {Zhang}}]{BirnstielDullemond+:2018aa}
{Birnstiel}, T., {Dullemond}, C.~P., {Zhu}, Z., {et~al.} 2018, \apjl, 869, L45,
  \dodoi{10.3847/2041-8213/aaf743}

\bibitem[{{Bohn} {et~al.}(2022){Bohn}, {Benisty}, {Perraut}, {van der Marel},
  {W^^c3^^b6lfer}, {van Dishoeck}, {Facchini}, {Manara}, {Teague}, {Francis},
  {Berger}, {Garcia-Lopez}, {Ginski}, {Henning}, {Kenworthy}, {Kraus},
  {M^^c3^^a9nard}, {M^^c3^^a9rand}, \& {P^^c3^^a9rez}}]{BohnBenisty+:2022aa}
{Bohn}, A.~J., {Benisty}, M., {Perraut}, K., {et~al.} 2022, \aap, 658, A183,
  \dodoi{10.1051/0004-6361/202142070}

\bibitem[{{Booth} \& {Clarke}(2016)}]{BoothClarke:2016aa}
{Booth}, R.~A., \& {Clarke}, C.~J. 2016, \mnras, 458, 2676,
  \dodoi{10.1093/mnras/stw488}

\bibitem[{{Bouvier} {et~al.}(1999){Bouvier}, {Chelli}, {Allain}, {Carrasco},
  {Costero}, {Cruz-Gonzalez}, {Dougados}, {Fern^^c3^^a1ndez}, {Mart^^c3^^adn},
  {M^^c3^^a9nard}, {Mennessier}, {Mujica}, {Recillas}, {Salas}, {Schmidt}, \&
  {Wichmann}}]{BouvierChelli+:1999aa}
{Bouvier}, J., {Chelli}, A., {Allain}, S., {et~al.} 1999, \aap, 349, 619

\bibitem[{{Brauer} {et~al.}(2008){Brauer}, {Dullemond}, \&
  {Henning}}]{Brauer:2008aa}
{Brauer}, F., {Dullemond}, C.~P., \& {Henning}, T. 2008, \aap, 480, 859,
  \dodoi{10.1051/0004-6361:20077759}

\bibitem[{{Burke} \& {Hollenbach}(1983)}]{BurkeHollenbach:1983aa}
{Burke}, J.~R., \& {Hollenbach}, D.~J. 1983, \apj, 265, 223,
  \dodoi{10.1086/160667}

\bibitem[{{Canovas} {et~al.}(2015){Canovas}, {M^^c3^^a9nard}, {de Boer},
  {Pinte}, {Avenhaus}, \& {Schreiber}}]{CanovasMenard+:2015aa}
{Canovas}, H., {M^^c3^^a9nard}, F., {de Boer}, J., {et~al.} 2015, \aap, 582,
  L7, \dodoi{10.1051/0004-6361/201527267}

\bibitem[{{Carrasco-Gonz^^c3^^a1lez} {et~al.}(2019){Carrasco-Gonz^^c3^^a1lez},
  {Sierra}, {Flock}, {Zhu}, {Henning}, {Chandler}, {Galv^^c3^^a1n-Madrid},
  {Mac^^c3^^adas}, {Anglada}, {Linz}, {Osorio}, {Rodr^^c3^^adguez}, {Testi},
  {Torrelles}, {P^^c3^^a9rez}, \& {Liu}}]{Carrasco-Gonzalez:2019aa}
{Carrasco-Gonz^^c3^^a1lez}, C., {Sierra}, A., {Flock}, M., {et~al.} 2019, \apj,
  883, 71, \dodoi{10.3847/1538-4357/ab3d33}

\bibitem[{{CASA Team} {et~al.}(2022){CASA Team}, {Bean}, {Bhatnagar}, {Castro},
  {Donovan Meyer}, {Emonts}, {Garcia}, {Garwood}, {Golap}, {Gonzalez Villalba},
  {Harris}, {Hayashi}, {Hoskins}, {Hsieh}, {Jagannathan}, {Kawasaki},
  {Keimpema}, {Kettenis}, {Lopez}, {Marvil}, {Masters}, {McNichols},
  {Mehringer}, {Miel}, {Moellenbrock}, {Montesino}, {Nakazato}, {Ott}, {Petry},
  {Pokorny}, {Raba}, {Rau}, {Schiebel}, {Schweighart}, {Sekhar}, {Shimada},
  {Small}, {Steeb}, {Sugimoto}, {Suoranta}, {Tsutsumi}, {van Bemmel},
  {Verkouter}, {Wells}, {Xiong}, {Szomoru}, {Griffith}, {Glendenning}, \&
  {Kern}}]{CASA-TeamBean+:2022aa}
{CASA Team}, {Bean}, B., {Bhatnagar}, S., {et~al.} 2022, \pasp, 134, 114501,
  \dodoi{10.1088/1538-3873/ac9642}

\bibitem[{{Casassus} {et~al.}(2015){Casassus}, {Wright}, {Marino}, {Maddison},
  {Wootten}, {Roman}, {P^^c3^^a9rez}, {Pinilla}, {Wyatt}, {Moral},
  {M^^c3^^a9nard}, {Christiaens}, {Cieza}, \& {van der
  Plas}}]{CasassusWright+:2015aa}
{Casassus}, S., {Wright}, C.~M., {Marino}, S., {et~al.} 2015, \apj, 812, 126,
  \dodoi{10.1088/0004-637X/812/2/126}

\bibitem[{{Casassus} {et~al.}(2018){Casassus}, {Avenhaus}, {P^^c3^^a9rez},
  {Navarro}, {C^^c3^^a1rcamo}, {Marino}, {Cieza}, {Quanz}, {Alarc^^c3^^b3n},
  {Zurlo}, {Osses}, {Rannou}, {Rom^^c3^^a1n}, \&
  {Barraza}}]{CasassusAvenhaus+:2018aa}
{Casassus}, S., {Avenhaus}, H., {P^^c3^^a9rez}, S., {et~al.} 2018, \mnras, 477,
  5104, \dodoi{10.1093/mnras/sty894}

\bibitem[{{Cossins} {et~al.}(2009){Cossins}, {Lodato}, \&
  {Clarke}}]{CossinsLodato+:2009aa}
{Cossins}, P., {Lodato}, G., \& {Clarke}, C.~J. 2009, \mnras, 393, 1157,
  \dodoi{10.1111/j.1365-2966.2008.14275.x}

\bibitem[{{Cuello} {et~al.}(2019){Cuello}, {Montesinos}, {Stammler}, {Louvet},
  \& {Cuadra}}]{CuelloMontesinos+:2019aa}
{Cuello}, N., {Montesinos}, M., {Stammler}, S.~M., {Louvet}, F., \& {Cuadra},
  J. 2019, \aap, 622, A43, \dodoi{10.1051/0004-6361/201731732}

\bibitem[{{Delussu} {et~al.}(2024){Delussu}, {Birnstiel}, {Miotello},
  {Pinilla}, {Rosotti}, \& {Andrews}}]{DelussuBirnstiel+:2024aa}
{Delussu}, L., {Birnstiel}, T., {Miotello}, A., {et~al.} 2024, \aap, 688, A81,
  \dodoi{10.1051/0004-6361/202450328}

\bibitem[{{Dominik} \& {Dullemond}(2024)}]{DominikDullemond:2024aa}
{Dominik}, C., \& {Dullemond}, C.~P. 2024, \aap, 682, A144,
  \dodoi{10.1051/0004-6361/202347716}

\bibitem[{{Dominik} {et~al.}(2021){Dominik}, {Min}, \&
  {Tazaki}}]{DominikMin+:2021aa}
{Dominik}, C., {Min}, M., \& {Tazaki}, R. 2021, {OpTool: Command-line driven
  tool for creating complex dust opacities}, Astrophysics Source Code Library,
  record ascl:2104.010

\bibitem[{{Draine}(2003)}]{Draine:2003xf}
{Draine}, B.~T. 2003, \apj, 598, 1026, \dodoi{10.1086/379123}

\bibitem[{{Dr{\k{a}}^^c5^^bckowska} {et~al.}(2023){Dr{\k{a}}^^c5^^bckowska},
  {Bitsch}, {Lambrechts}, {Mulders}, {Harsono}, {Vazan}, {Liu}, {Ormel},
  {Kretke}, \& {Morbidelli}}]{DrazkowskaBitsch+:2023aa}
{Dr{\k{a}}^^c5^^bckowska}, J., {Bitsch}, B., {Lambrechts}, M., {et~al.} 2023,
  in Astronomical Society of the Pacific Conference Series, Vol. 534,
  Protostars and Planets VII, ed. S.~{Inutsuka}, Y.~{Aikawa}, T.~{Muto},
  K.~{Tomida}, \& M.~{Tamura}, 717, \dodoi{10.48550/arXiv.2203.09759}

\bibitem[{{Dubrulle} {et~al.}(1995){Dubrulle}, {Morfill}, \&
  {Sterzik}}]{Dubrulle+1995}
{Dubrulle}, B., {Morfill}, G., \& {Sterzik}, M. 1995, \icarus, 114, 237,
  \dodoi{10.1006/icar.1995.1058}

\bibitem[{{Dullemond} \& {Dominik}(2005)}]{DullemondDominik:2005vy}
{Dullemond}, C.~P., \& {Dominik}, C. 2005, \aap, 434, 971,
  \dodoi{10.1051/0004-6361:20042080}

\bibitem[{{Dullemond} {et~al.}(2022){Dullemond}, {Ziampras}, {Ostertag}, \&
  {Dominik}}]{DullemondZiampras+:2022aa}
{Dullemond}, C.~P., {Ziampras}, A., {Ostertag}, D., \& {Dominik}, C. 2022,
  \aap, 668, A105, \dodoi{10.1051/0004-6361/202244218}

\bibitem[{{Facchini} {et~al.}(2018){Facchini}, {Juh^^c3^^a1sz}, \&
  {Lodato}}]{FacchiniJuhasz+:2018aa}
{Facchini}, S., {Juh^^c3^^a1sz}, A., \& {Lodato}, G. 2018, \mnras, 473, 4459,
  \dodoi{10.1093/mnras/stx2523}

\bibitem[{{Flock} {et~al.}(2020){Flock}, {Turner}, {Nelson}, {Lyra}, {Manger},
  \& {Klahr}}]{Flock:2020aa}
{Flock}, M., {Turner}, N.~J., {Nelson}, R.~P., {et~al.} 2020, \apj, 897, 155,
  \dodoi{10.3847/1538-4357/ab9641}

\bibitem[{{Foreman-Mackey} {et~al.}(2013){Foreman-Mackey}, {Hogg}, {Lang}, \&
  {Goodman}}]{Foreman-MackeyHogg+:2013aa}
{Foreman-Mackey}, D., {Hogg}, D.~W., {Lang}, D., \& {Goodman}, J. 2013, \pasp,
  125, 306, \dodoi{10.1086/670067}

\bibitem[{{Francis} \& {van der Marel}(2020)}]{Francisvan-der-Marel:2020aa}
{Francis}, L., \& {van der Marel}, N. 2020, \apj, 892, 111,
  \dodoi{10.3847/1538-4357/ab7b63}

\bibitem[{{Fritscher} \& {Teiser}(2021)}]{FritscherTeiser:2021aa}
{Fritscher}, M., \& {Teiser}, J. 2021, \apj, 923, 134,
  \dodoi{10.3847/1538-4357/ac2df4}

\bibitem[{{Fukagawa} {et~al.}(2013){Fukagawa}, {Tsukagoshi}, {Momose}, {Saigo},
  {Ohashi}, {Kitamura}, {Inutsuka}, {Muto}, {Nomura}, {Takeuchi}, {Kobayashi},
  {Hanawa}, {Akiyama}, {Honda}, {Fujiwara}, {Kataoka}, {Takahashi}, \&
  {Shibai}}]{FukagawaTsukagoshi+:2013aa}
{Fukagawa}, M., {Tsukagoshi}, T., {Momose}, M., {et~al.} 2013, \pasj, 65, L14,
  \dodoi{10.1093/pasj/65.6.L14}

\bibitem[{{Fukuhara} {et~al.}(2025){Fukuhara}, {Flock}, {Okuzumi}, \&
  {Tominaga}}]{FukuharaFlock+:2025aa}
{Fukuhara}, Y., {Flock}, M., {Okuzumi}, S., \& {Tominaga}, R.~T. 2025, \aap,
  701, A72, \dodoi{10.1051/0004-6361/202555624}

\bibitem[{{Fukuhara} \& {Okuzumi}(2024)}]{FukuharaOkuzumi:2024aa}
{Fukuhara}, Y., \& {Okuzumi}, S. 2024, \pasj, 76, 708,
  \dodoi{10.1093/pasj/psae042}

\bibitem[{{Fukuhara} {et~al.}(2021){Fukuhara}, {Okuzumi}, \&
  {Ono}}]{FukuharaOkuzumi+:2021ca}
{Fukuhara}, Y., {Okuzumi}, S., \& {Ono}, T. 2021, \apj, 914, 132,
  \dodoi{10.3847/1538-4357/abfe5c}

\bibitem[{{Fukuhara} {et~al.}(2023){Fukuhara}, {Okuzumi}, \&
  {Ono}}]{FukuharaOkuzumi+:2023aa}
---. 2023, \pasj, 75, 233, \dodoi{10.1093/pasj/psac107}

\bibitem[{{Fung} \& {Ono}(2021)}]{FungOno:2021aa}
{Fung}, J., \& {Ono}, T. 2021, \apj, 922, 13, \dodoi{10.3847/1538-4357/ac1d4e}

\bibitem[{{Gaia Collaboration} {et~al.}(2023){Gaia Collaboration}, {Vallenari},
  {Brown}, {Prusti}, {de Bruijne}, {Arenou}, {Babusiaux}, {Biermann},
  {Creevey}, {Ducourant}, \& et~al.}]{Gaia-CollaborationVallenari+:2023aa}
{Gaia Collaboration}, {Vallenari}, A., {Brown}, A.~G.~A., {et~al.} 2023, \aap,
  674, A1, \dodoi{10.1051/0004-6361/202243940}

\bibitem[{{Gammie}(2001)}]{Gammie:2001aa}
{Gammie}, C.~F. 2001, \apj, 553, 174, \dodoi{10.1086/320631}

\bibitem[{{Goldreich} \& {Ward}(1973)}]{Goldreich:1973aa}
{Goldreich}, P., \& {Ward}, W.~R. 1973, \apj, 183, 1051, \dodoi{10.1086/152291}

\bibitem[{{Gundlach} {et~al.}(2018){Gundlach}, {Schmidt}, {Kreuzig},
  {Bischoff}, {Rezaei}, {Kothe}, {Blum}, {Grzesik}, \&
  {Stoll}}]{GundlachSchmidt+:2018aa}
{Gundlach}, B., {Schmidt}, K.~P., {Kreuzig}, C., {et~al.} 2018, \mnras, 479,
  1273, \dodoi{10.1093/mnras/sty1550}

\bibitem[{{Hammond} {et~al.}(2022){Hammond}, {Christiaens}, {Price},
  {Ubeira-Gabellini}, {Baird}, {Calcino}, {Benisty}, {Lodato}, {Testi},
  {Pinte}, {Toci}, \& {Fedele}}]{HammondChristiaens+:2022aa}
{Hammond}, I., {Christiaens}, V., {Price}, D.~J., {et~al.} 2022, \mnras, 515,
  6109, \dodoi{10.1093/mnras/stac2119}

\bibitem[{{Hashimoto} {et~al.}(2024){Hashimoto}, {Dong}, {Muto}, {Liu}, \&
  {Terada}}]{HashimotoDong+:2024aa}
{Hashimoto}, J., {Dong}, R., {Muto}, T., {Liu}, H.~B., \& {Terada}, Y. 2024,
  \aj, 167, 75, \dodoi{10.3847/1538-3881/ad1b5e}

\bibitem[{{Hunziker} {et~al.}(2021){Hunziker}, {Schmid}, {Ma}, {Menard},
  {Avenhaus}, {Boccaletti}, {Beuzit}, {Chauvin}, {Dohlen}, {Dominik}, {Engler},
  {Ginski}, {Gratton}, {Henning}, {Langlois}, {Milli}, {Mouillet}, {Tschudi},
  {van Holstein}, \& {Vigan}}]{HunzikerSchmid+:2021aa}
{Hunziker}, S., {Schmid}, H.~M., {Ma}, J., {et~al.} 2021, \aap, 648, A110,
  \dodoi{10.1051/0004-6361/202040166}

\bibitem[{{Kataoka} {et~al.}(2016){Kataoka}, {Tsukagoshi}, {Momose}, {Nagai},
  {Muto}, {Dullemond}, {Pohl}, {Fukagawa}, {Shibai}, {Hanawa}, \&
  {Murakawa}}]{KataokaTsukagoshi+:2016ae}
{Kataoka}, A., {Tsukagoshi}, T., {Momose}, M., {et~al.} 2016, \apjl, 831, L12,
  \dodoi{10.3847/2041-8205/831/2/L12}

\bibitem[{{Kothe} {et~al.}(2013){Kothe}, {Blum}, {Weidling}, \&
  {G^^c3^^bcttler}}]{KotheBlum+:2013aa}
{Kothe}, S., {Blum}, J., {Weidling}, R., \& {G^^c3^^bcttler}, C. 2013, \icarus,
  225, 75, \dodoi{10.1016/j.icarus.2013.02.034}

\bibitem[{{Kratter} \& {Lodato}(2016)}]{KratterLodato:2016aa}
{Kratter}, K., \& {Lodato}, G. 2016, \araa, 54, 271,
  \dodoi{10.1146/annurev-astro-081915-023307}

\bibitem[{{Kuffmeier} {et~al.}(2021){Kuffmeier}, {Dullemond}, {Reissl}, \&
  {Goicovic}}]{KuffmeierDullemond+:2021aa}
{Kuffmeier}, M., {Dullemond}, C.~P., {Reissl}, S., \& {Goicovic}, F.~G. 2021,
  \aap, 656, A161, \dodoi{10.1051/0004-6361/202039614}

\bibitem[{{Leedham} {et~al.}(2025){Leedham}, {Booth}, \&
  {Clarke}}]{LeedhamBooth+:2025aa}
{Leedham}, C.~S., {Booth}, R.~A., \& {Clarke}, C.~J. 2025, \mnras, 539, 2780,
  \dodoi{10.1093/mnras/staf644}

\bibitem[{{Les} \& {Lin}(2015)}]{LesLin:2015aa}
{Les}, R., \& {Lin}, M.-K. 2015, \mnras, 450, 1503,
  \dodoi{10.1093/mnras/stv712}

\bibitem[{{Lesur} {et~al.}(2025){Lesur}, {Latter}, \&
  {Ogilvie}}]{LesurLatter+:2025aa}
{Lesur}, G., {Latter}, H., \& {Ogilvie}, G.~I. 2025, \aap, 703, A225,
  \dodoi{10.1051/0004-6361/202555944}

\bibitem[{{Lesur} {et~al.}(2023){Lesur}, {Flock}, {Ercolano}, {Lin}, {Yang},
  {Barranco}, {Benitez-Llambay}, {Goodman}, {Johansen}, {Klahr}, {Laibe},
  {Lyra}, {Marcus}, {Nelson}, {Squire}, {Simon}, {Turner}, {Umurhan}, \&
  {Youdin}}]{LesurFlock+:2023aa}
{Lesur}, G., {Flock}, M., {Ercolano}, B., {et~al.} 2023, in Astronomical
  Society of the Pacific Conference Series, Vol. 534, Protostars and Planets
  VII, ed. S.~{Inutsuka}, Y.~{Aikawa}, T.~{Muto}, K.~{Tomida}, \& M.~{Tamura},
  465

\bibitem[{{Lim} {et~al.}(2026){Lim}, {Simon}, {Li}, {Brouillette}, {Rea}, \&
  {Lyra}}]{LimSimon+:2026aa}
{Lim}, J., {Simon}, J.~B., {Li}, R., {et~al.} 2026, \apj, 1000, 156,
  \dodoi{10.3847/1538-4357/ae47e9}

\bibitem[{{Lim} {et~al.}(2025){Lim}, {Simon}, {Li}, {Carrera}, {Baronett},
  {Youdin}, {Lyra}, \& {Yang}}]{LimSimon+:2025aa}
---. 2025, \apj, 981, 160, \dodoi{10.3847/1538-4357/adb311}

\bibitem[{{Lin} \& {Youdin}(2015)}]{LinYoudin2015}
{Lin}, M.-K., \& {Youdin}, A.~N. 2015, \apj, 811, 17,
  \dodoi{10.1088/0004-637X/811/1/17}

\bibitem[{{Longarini} {et~al.}(2021){Longarini}, {Lodato}, {Toci}, {Veronesi},
  {Hall}, {Dong}, \& {Patrick Terry}}]{LongariniLodato+:2021aa}
{Longarini}, C., {Lodato}, G., {Toci}, C., {et~al.} 2021, \apjl, 920, L41,
  \dodoi{10.3847/2041-8213/ac2df6}

\bibitem[{{Malygin} {et~al.}(2017){Malygin}, {Klahr}, {Semenov}, {Henning}, \&
  {Dullemond}}]{Malygin+2017}
{Malygin}, M.~G., {Klahr}, H., {Semenov}, D., {Henning}, T., \& {Dullemond},
  C.~P. 2017, \aap, 605, A30, \dodoi{10.1051/0004-6361/201629933}

\bibitem[{{Manger} {et~al.}(2021){Manger}, {Pfeil}, \&
  {Klahr}}]{MangerPfeil+:2021cm}
{Manger}, N., {Pfeil}, T., \& {Klahr}, H. 2021, \mnras, 508, 5402,
  \dodoi{10.1093/mnras/stab2599}

\bibitem[{{Marino} {et~al.}(2015){Marino}, {Perez}, \&
  {Casassus}}]{MarinoPerez+:2015aa}
{Marino}, S., {Perez}, S., \& {Casassus}, S. 2015, \apjl, 798, L44,
  \dodoi{10.1088/2041-8205/798/2/L44}

\bibitem[{{Matsakos} \& {K^^c3^^b6nigl}(2017)}]{MatsakosKonigl:2017aa}
{Matsakos}, T., \& {K^^c3^^b6nigl}, A. 2017, \aj, 153, 60,
  \dodoi{10.3847/1538-3881/153/2/60}

\bibitem[{{Mayama} {et~al.}(2012){Mayama}, {Hashimoto}, {Muto}, {Tsukagoshi},
  {Kusakabe}, {Kuzuhara}, {Takahashi}, {Kudo}, {Dong}, {Fukagawa}, {Takami},
  {Momose}, {Wisniewski}, {Follette}, {Abe}, {Akiyama}, {Brandner}, {Brandt},
  {Carson}, {Egner}, {Feldt}, {Goto}, {Grady}, {Guyon}, {Hayano}, {Hayashi},
  {Hayashi}, {Henning}, {Hodapp}, {Ishii}, {Iye}, {Janson}, {Kandori}, {Kwon},
  {Knapp}, {Matsuo}, {McElwain}, {Miyama}, {Morino}, {Moro-Martin},
  {Nishimura}, {Pyo}, {Serabyn}, {Suto}, {Suzuki}, {Takato}, {Terada},
  {Thalmann}, {Tomono}, {Turner}, {Watanabe}, {Yamada}, {Takami}, {Usuda}, \&
  {Tamura}}]{MayamaHashimoto+:2012aa}
{Mayama}, S., {Hashimoto}, J., {Muto}, T., {et~al.} 2012, \apjl, 760, L26,
  \dodoi{10.1088/2041-8205/760/2/L26}

\bibitem[{{Mayama} {et~al.}(2018){Mayama}, {Akiyama}, {Pani^^c4^^87}, {Miley},
  {Tsukagoshi}, {Muto}, {Dong}, {de Leon}, {Mizuki}, {Oh}, {Hashimoto}, {Sai},
  {Currie}, {Takami}, {Grady}, {Hayashi}, {Tamura}, \&
  {Inutsuka}}]{MayamaAkiyama+:2018aa}
{Mayama}, S., {Akiyama}, E., {Pani^^c4^^87}, O., {et~al.} 2018, \apjl, 868, L3,
  \dodoi{10.3847/2041-8213/aae88b}

\bibitem[{{Mej^^c3^^ada} {et~al.}(2005){Mej^^c3^^ada}, {Durisen}, {Pickett}, \&
  {Cai}}]{MejiaDurisen+:2005aa}
{Mej^^c3^^ada}, A.~C., {Durisen}, R.~H., {Pickett}, M.~K., \& {Cai}, K. 2005,
  \apj, 619, 1098, \dodoi{10.1086/426707}

\bibitem[{{Miranda} \& {Rafikov}(2020{\natexlab{a}})}]{MirandaRafikov:2020aa}
{Miranda}, R., \& {Rafikov}, R.~R. 2020{\natexlab{a}}, \apj, 892, 65,
  \dodoi{10.3847/1538-4357/ab791a}

\bibitem[{{Miranda} \& {Rafikov}(2020{\natexlab{b}})}]{MirandaRafikov:2020ab}
---. 2020{\natexlab{b}}, \apj, 904, 121, \dodoi{10.3847/1538-4357/abbee7}

\bibitem[{{Montesinos} {et~al.}(2016){Montesinos}, {Perez}, {Casassus},
  {Marino}, {Cuadra}, \& {Christiaens}}]{MontesinosPerez+:2016aa}
{Montesinos}, M., {Perez}, S., {Casassus}, S., {et~al.} 2016, \apjl, 823, L8,
  \dodoi{10.3847/2041-8205/823/1/L8}

\bibitem[{{Mullin} {et~al.}(2026){Mullin}, {Lucas}, {Dong}, {Hashimoto},
  {Jiang}, {Johnstone}, {Lawson}, {Brittain}, {Guyon}, {Kudo}, {Lozi},
  {Najita}, {Sun}, {Tamura}, \& {Wagner}}]{MullinLucas+:2026aa}
{Mullin}, C., {Lucas}, M., {Dong}, R., {et~al.} 2026, \aj, 171, 241,
  \dodoi{10.3847/1538-3881/ae473d}

\bibitem[{{Musiolik} \& {Wurm}(2019)}]{MusiolikWurm:2019aa}
{Musiolik}, G., \& {Wurm}, G. 2019, \apj, 873, 58,
  \dodoi{10.3847/1538-4357/ab0428}

\bibitem[{{Nealon} {et~al.}(2020){Nealon}, {Cuello}, \&
  {Alexander}}]{NealonCuello+:2020aa}
{Nealon}, R., {Cuello}, N., \& {Alexander}, R. 2020, \mnras, 491, 4108,
  \dodoi{10.1093/mnras/stz3186}

\bibitem[{{Nealon} {et~al.}(2018){Nealon}, {Dipierro}, {Alexander}, {Martin},
  \& {Nixon}}]{NealonDipierro+:2018aa}
{Nealon}, R., {Dipierro}, G., {Alexander}, R., {Martin}, R.~G., \& {Nixon}, C.
  2018, \mnras, 481, 20, \dodoi{10.1093/mnras/sty2267}

\bibitem[{{Nelson} {et~al.}(2013){Nelson}, {Gressel}, \&
  {Umurhan}}]{NelsonGresselUmurhan2013}
{Nelson}, R.~P., {Gressel}, O., \& {Umurhan}, O.~M. 2013, \mnras, 435, 2610,
  \dodoi{10.1093/mnras/stt1475}

\bibitem[{{Nowak} {et~al.}(2024){Nowak}, {Rowther}, {Lacour}, {Meru}, {Nealon},
  \& {Price}}]{NowakRowther+:2024aa}
{Nowak}, M., {Rowther}, S., {Lacour}, S., {et~al.} 2024, \aap, 683, A6,
  \dodoi{10.1051/0004-6361/202347748}

\bibitem[{{Ohashi} {et~al.}(2025){Ohashi}, {Muto}, {Tsukamoto}, {Kataoka},
  {Tsukagoshi}, {Momose}, {Fukagawa}, \& {Sakai}}]{OhashiMuto+:2025aa}
{Ohashi}, S., {Muto}, T., {Tsukamoto}, Y., {et~al.} 2025, Nature Astronomy,
  \dodoi{10.1038/s41550-024-02454-x}

\bibitem[{{Ohashi} {et~al.}(2018){Ohashi}, {Kataoka}, {Nagai}, {Momose},
  {Muto}, {Hanawa}, {Fukagawa}, {Tsukagoshi}, {Murakawa}, \&
  {Shibai}}]{OhashiKataoka+:2018aa}
{Ohashi}, S., {Kataoka}, A., {Nagai}, H., {et~al.} 2018, \apj, 864, 81,
  \dodoi{10.3847/1538-4357/aad632}

\bibitem[{{Orihara} \& {Momose}(2025)}]{OriharaMomose:2025aa}
{Orihara}, R., \& {Momose}, M. 2025, \apj, 986, 215,
  \dodoi{10.3847/1538-4357/add890}

\bibitem[{{Ormel} \& {Cuzzi}(2007)}]{OrmelCuzzi2007}
{Ormel}, C.~W., \& {Cuzzi}, J.~N. 2007, \aap, 466, 413,
  \dodoi{10.1051/0004-6361:20066899}

\bibitem[{{Oshiro} {et~al.}(2025){Oshiro}, {Tatsuuma}, {Okuzumi}, \&
  {Tanaka}}]{OshiroTatsuuma+:2025aa}
{Oshiro}, H., {Tatsuuma}, M., {Okuzumi}, S., \& {Tanaka}, H. 2025, \apj, 983,
  75, \dodoi{10.3847/1538-4357/adbf04}

\bibitem[{{Owen} \& {Lai}(2017)}]{OwenLai:2017aa}
{Owen}, J.~E., \& {Lai}, D. 2017, \mnras, 469, 2834,
  \dodoi{10.1093/mnras/stx1033}

\bibitem[{{Perez} {et~al.}(2015){Perez}, {Casassus}, {M^^c3^^a9nard}, {Roman},
  {van der Plas}, {Cieza}, {Pinte}, {Christiaens}, \&
  {Hales}}]{PerezCasassus+:2015aa}
{Perez}, S., {Casassus}, S., {M^^c3^^a9nard}, F., {et~al.} 2015, \apj, 798, 85,
  \dodoi{10.1088/0004-637X/798/2/85}

\bibitem[{{Pfeil} {et~al.}(2023){Pfeil}, {Birnstiel}, \&
  {Klahr}}]{PfeilBirnstiel+:2023aa}
{Pfeil}, T., {Birnstiel}, T., \& {Klahr}, H. 2023, \apj, 959, 121,
  \dodoi{10.3847/1538-4357/ad00af}

\bibitem[{{Pfeil} {et~al.}(2024){Pfeil}, {Birnstiel}, \&
  {Klahr}}]{PfeilBirnstiel+:2024aa}
---. 2024, \aap, 687, L5, \dodoi{10.1051/0004-6361/202449323}

\bibitem[{{Pfeil} \& {Klahr}(2019)}]{PfeilKlahr2019}
{Pfeil}, T., \& {Klahr}, H. 2019, \apj, 871, 150,
  \dodoi{10.3847/1538-4357/aaf962}

\bibitem[{{Pinilla} {et~al.}(2012){Pinilla}, {Birnstiel}, {Ricci}, {Dullemond},
  {Uribe}, {Testi}, \& {Natta}}]{PinillaBirnstiel+:2012vz}
{Pinilla}, P., {Birnstiel}, T., {Ricci}, L., {et~al.} 2012, \aap, 538, A114,
  \dodoi{10.1051/0004-6361/201118204}

\bibitem[{{Pinilla} {et~al.}(2018){Pinilla}, {Benisty}, {de Boer}, {Manara},
  {Bouvier}, {Dominik}, {Ginski}, {Loomis}, \& {Sicilia
  Aguilar}}]{PinillaBenisty+:2018aa}
{Pinilla}, P., {Benisty}, M., {de Boer}, J., {et~al.} 2018, \apj, 868, 85,
  \dodoi{10.3847/1538-4357/aae824}

\bibitem[{{Raettig} {et~al.}(2013){Raettig}, {Lyra}, \&
  {Klahr}}]{RaettigLyra+:2013aa}
{Raettig}, N., {Lyra}, W., \& {Klahr}, H. 2013, \apj, 765, 115,
  \dodoi{10.1088/0004-637X/765/2/115}

\bibitem[{{Ricci} {et~al.}(2010){Ricci}, {Testi}, {Natta}, {Neri}, {Cabrit}, \&
  {Herczeg}}]{RicciTesti+:2010aa}
{Ricci}, L., {Testi}, L., {Natta}, A., {et~al.} 2010, \aap, 512, A15,
  \dodoi{10.1051/0004-6361/200913403}

\bibitem[{{Rometsch} {et~al.}(2021){Rometsch}, {Ziampras}, {Kley}, \&
  {B^^c3^^a9thune}}]{RometschZiampras+:2021aa}
{Rometsch}, T., {Ziampras}, A., {Kley}, W., \& {B^^c3^^a9thune}, W. 2021, \aap,
  656, A130, \dodoi{10.1051/0004-6361/202142105}

\bibitem[{{Shariff} \& {Umurhan}(2024)}]{ShariffUmurhan:2024aa}
{Shariff}, K., \& {Umurhan}, O.~M. 2024, \apj, 977, 272,
  \dodoi{10.3847/1538-4357/ad90a5}

\bibitem[{{Shi} \& {Chiang}(2014)}]{ShiChiang:2014aa}
{Shi}, J.-M., \& {Chiang}, E. 2014, \apj, 789, 34,
  \dodoi{10.1088/0004-637X/789/1/34}

\bibitem[{{Shi} {et~al.}(2016){Shi}, {Zhu}, {Stone}, \&
  {Chiang}}]{ShiZhu+:2016aa}
{Shi}, J.-M., {Zhu}, Z., {Stone}, J.~M., \& {Chiang}, E. 2016, \mnras, 459,
  982, \dodoi{10.1093/mnras/stw692}

\bibitem[{{Sierra} \& {Lizano}(2020)}]{SierraLizano:2020aa}
{Sierra}, A., \& {Lizano}, S. 2020, \apj, 892, 136,
  \dodoi{10.3847/1538-4357/ab7d32}

\bibitem[{{Sierra} {et~al.}(2024){Sierra}, {P^^c3^^a9rez}, {Sotomayor},
  {Benisty}, {Chandler}, {Andrews}, {Carpenter}, {Henning}, {Testi}, {Ricci},
  \& {Wilner}}]{SierraPerez+:2024ab}
{Sierra}, A., {P^^c3^^a9rez}, L.~M., {Sotomayor}, B., {et~al.} 2024, \apj, 974,
  306, \dodoi{10.3847/1538-4357/ad7460}

\bibitem[{{Soon} {et~al.}(2019){Soon}, {Momose}, {Muto}, {Tsukagoshi},
  {Kataoka}, {Hanawa}, {Fukagawa}, {Saigo}, \& {Shibai}}]{SoonMomose+:2019aa}
{Soon}, K.-L., {Momose}, M., {Muto}, T., {et~al.} 2019, \pasj, 71, 124,
  \dodoi{10.1093/pasj/psz112}

\bibitem[{{Stoll} \& {Kley}(2014)}]{StollKley2014}
{Stoll}, M. H.~R., \& {Kley}, W. 2014, \aap, 572, A77,
  \dodoi{10.1051/0004-6361/201424114}

\bibitem[{{Stoll} \& {Kley}(2016)}]{StollKley:2016vp}
---. 2016, \aap, 594, A57, \dodoi{10.1051/0004-6361/201527716}

\bibitem[{{Su} \& {Bai}(2024)}]{SuBai:2024aa}
{Su}, Z., \& {Bai}, X.-N. 2024, \apj, 975, 126,
  \dodoi{10.3847/1538-4357/ad7581}

\bibitem[{{Su} \& {Wei}(2025)}]{SuWei:2025aa}
{Su}, Z., \& {Wei}, X. 2025, \apj, 983, 89, \dodoi{10.3847/1538-4357/adc0ff}

\bibitem[{{Takahashi} {et~al.}(2016){Takahashi}, {Tsukamoto}, \&
  {Inutsuka}}]{TakahashiTsukamoto+:2016aa}
{Takahashi}, S.~Z., {Tsukamoto}, Y., \& {Inutsuka}, S. 2016, \mnras, 458, 3597,
  \dodoi{10.1093/mnras/stw557}

\bibitem[{{Takeuchi} \& {Lin}(2002)}]{TakeuchiLin2002}
{Takeuchi}, T., \& {Lin}, D.~N.~C. 2002, \apj, 581, 1344,
  \dodoi{10.1086/344437}

\bibitem[{{Tarczay-Neh^^c3^^a9z} {et~al.}(2020){Tarczay-Neh^^c3^^a9z},
  {Reg^^c3^^a1ly}, \& {Vorobyov}}]{Tarczay-NehezRegaly+:2020aa}
{Tarczay-Neh^^c3^^a9z}, D., {Reg^^c3^^a1ly}, Z., \& {Vorobyov}, E. 2020,
  \mnras, 493, 3014, \dodoi{10.1093/mnras/staa364}

\bibitem[{{Tazaki} {et~al.}(2021){Tazaki}, {Murakawa}, {Muto}, {Honda}, \&
  {Inoue}}]{TazakiMurakawa+:2021aa}
{Tazaki}, R., {Murakawa}, K., {Muto}, T., {Honda}, M., \& {Inoue}, A.~K. 2021,
  \apj, 921, 173, \dodoi{10.3847/1538-4357/ac1f8c}

\bibitem[{{Teague} {et~al.}(2022){Teague}, {Bae}, {Benisty}, {Andrews},
  {Facchini}, {Huang}, \& {Wilner}}]{TeagueBae+:2022aa}
{Teague}, R., {Bae}, J., {Benisty}, M., {et~al.} 2022, \apj, 930, 144,
  \dodoi{10.3847/1538-4357/ac67a3}

\bibitem[{{Temmink} {et~al.}(2023){Temmink}, {Booth}, {van der Marel}, \& {van
  Dishoeck}}]{TemminkBooth+:2023aa}
{Temmink}, M., {Booth}, A.~S., {van der Marel}, N., \& {van Dishoeck}, E.~F.
  2023, \aap, 675, A131, \dodoi{10.1051/0004-6361/202346272}

\bibitem[{{Urpin}(2003)}]{Urpin2003}
{Urpin}, V. 2003, \aap, 404, 397, \dodoi{10.1051/0004-6361:20030513}

\bibitem[{{Uyama} {et~al.}(2020){Uyama}, {Muto}, {Mawet}, {Christiaens},
  {Hashimoto}, {Kudo}, {Kuzuhara}, {Ruane}, {Beichman}, {Absil}, {Akiyama},
  {Bae}, {Bottom}, {Choquet}, {Currie}, {Dong}, {Follette}, {Fukagawa},
  {Guidi}, {Huby}, {Kwon}, {Mayama}, {Meshkat}, {Reggiani}, {Ricci}, {Serabyn},
  {Tamura}, {Testi}, {Wallack}, {Williams}, \& {Zhu}}]{UyamaMuto+:2020aa}
{Uyama}, T., {Muto}, T., {Mawet}, D., {et~al.} 2020, \aj, 159, 118,
  \dodoi{10.3847/1538-3881/ab7006}

\bibitem[{{van der Marel}(2023)}]{van-der-Marel:2023aa}
{van der Marel}, N. 2023, European Physical Journal Plus, 138, 225,
  \dodoi{10.1140/epjp/s13360-022-03628-0}

\bibitem[{{Warren} \& {Brandt}(2008)}]{WarrenBrandt:2008aa}
{Warren}, S.~G., \& {Brandt}, R.~E. 2008, Journal of Geophysical Research
  (Atmospheres), 113, D14220, \dodoi{10.1029/2007JD009744}

\bibitem[{{W^^c3^^b6lfer} {et~al.}(2021){W^^c3^^b6lfer}, {Facchini},
  {Kurtovic}, {Teague}, {van Dishoeck}, {Benisty}, {Ercolano}, {Lodato},
  {Miotello}, {Rosotti}, {Testi}, \& {Ubeira
  Gabellini}}]{WolferFacchini+:2021aa}
{W^^c3^^b6lfer}, L., {Facchini}, S., {Kurtovic}, N.~T., {et~al.} 2021, \aap,
  648, A19, \dodoi{10.1051/0004-6361/202039469}

\bibitem[{{Whipple}(1972)}]{Whipple:1972vv}
{Whipple}, F.~L. 1972, in From Plasma to Planet, ed. A.~{Elvius}, 211

\bibitem[{{Woitke} {et~al.}(2016){Woitke}, {Min}, {Pinte}, {Thi}, {Kamp},
  {Rab}, {Anthonioz}, {Antonellini}, {Baldovin-Saavedra}, {Carmona}, {Dominik},
  {Dionatos}, {Greaves}, {G^^c3^^bcdel}, {Ilee}, {Liebhart}, {M^^c3^^a9nard},
  {Rigon}, {Waters}, {Aresu}, {Meijerink}, \& {Spaans}}]{WoitkeMin+:2016aa}
{Woitke}, P., {Min}, M., {Pinte}, C., {et~al.} 2016, \aap, 586, A103,
  \dodoi{10.1051/0004-6361/201526538}

\bibitem[{{Yen} \& {Gu}(2020)}]{YenGu:2020aa}
{Yen}, H.-W., \& {Gu}, P.-G. 2020, \apj, 905, 89,
  \dodoi{10.3847/1538-4357/abc55a}

\bibitem[{{Youdin} \& {Goodman}(2005)}]{YoudinGoodman:2005aa}
{Youdin}, A.~N., \& {Goodman}, J. 2005, \apj, 620, 459, \dodoi{10.1086/426895}

\bibitem[{{Youdin} \& {Lithwick}(2007)}]{YoudinLithwick2007}
{Youdin}, A.~N., \& {Lithwick}, Y. 2007, \icarus, 192, 588,
  \dodoi{10.1016/j.icarus.2007.07.012}

\bibitem[{{Zhang} {et~al.}(2024){Zhang}, {Huang}, \&
  {Dong}}]{ZhangHuang+:2024aa}
{Zhang}, M., {Huang}, P., \& {Dong}, R. 2024, \apj, 961, 86,
  \dodoi{10.3847/1538-4357/ad055c}

\bibitem[{{Zhang} \& {Zhu}(2020)}]{ZhangZhu:2020aa}
{Zhang}, S., \& {Zhu}, Z. 2020, \mnras, 493, 2287,
  \dodoi{10.1093/mnras/staa404}

\bibitem[{{Zhang} \& {Zhu}(2024)}]{ZhangZhu:2024aa}
---. 2024, \apjl, 974, L38, \dodoi{10.3847/2041-8213/ad815f}

\bibitem[{{Zhang} {et~al.}(2025){Zhang}, {Zhu}, \&
  {Fairbairn}}]{ZhangZhu+:2025aa}
{Zhang}, S., {Zhu}, Z., \& {Fairbairn}, C.~W. 2025, \apjl, 995, L33,
  \dodoi{10.3847/2041-8213/ae2023}

\bibitem[{{Zhu}(2019)}]{Zhu:2019aa}
{Zhu}, Z. 2019, \mnras, 483, 4221, \dodoi{10.1093/mnras/sty3358}

\bibitem[{{Zhu} {et~al.}(2025){Zhu}, {Zhang}, \& {Johnson}}]{ZhuZhang+:2025aa}
{Zhu}, Z., {Zhang}, S., \& {Johnson}, T.~M. 2025, \apj, 980, 259,
  \dodoi{10.3847/1538-4357/adae0d}

\bibitem[{{Ziampras} {et~al.}(2025){Ziampras}, {Dullemond}, {Birnstiel},
  {Benisty}, \& {Nelson}}]{ZiamprasDullemond+:2025aa}
{Ziampras}, A., {Dullemond}, C.~P., {Birnstiel}, T., {Benisty}, M., \&
  {Nelson}, R.~P. 2025, \mnras, 540, 1185, \dodoi{10.1093/mnras/staf785}

\bibitem[{{Zormpas} {et~al.}(2022){Zormpas}, {Birnstiel}, {Rosotti}, \&
  {Andrews}}]{ZormpasBirnstiel+:2022aa}
{Zormpas}, A., {Birnstiel}, T., {Rosotti}, G.~P., \& {Andrews}, S.~M. 2022,
  \aap, 661, A66, \dodoi{10.1051/0004-6361/202142046}

\bibitem[{{Zubko} {et~al.}(1996){Zubko}, {Mennella}, {Colangeli}, \&
  {Bussoletti}}]{ZubkoMennella+:1996aa}
{Zubko}, V.~G., {Mennella}, V., {Colangeli}, L., \& {Bussoletti}, E. 1996,
  \mnras, 282, 1321, \dodoi{10.1093/mnras/282.4.1321}

\end{thebibliography}

\appendix
\section{Input images and gradient maps for the shadow reconstruction}\label{appendix:images}

\begin{figure*}[t]
    \begin{center}
    \includegraphics[width=\hsize,bb = 0 0 382 290]{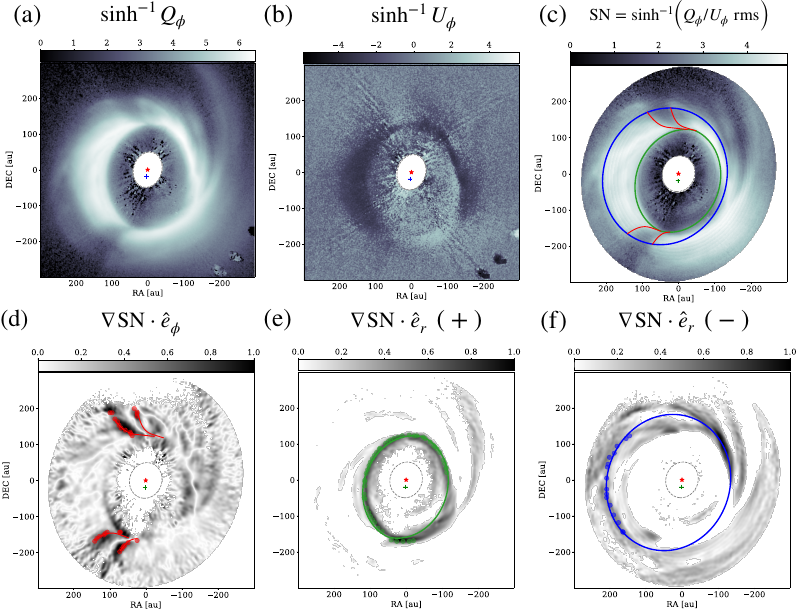}
    \end{center}
    \caption{
    Images and gradient maps used for the shadow-geometry reconstruction.
    (a) Inverse-hyperbolic-sine-stretched H-band $Q_\phi$ image.
    (b) Inverse-hyperbolic-sine-stretched H-band $U_\phi$ image.
    (c) Inverse-hyperbolic-sine-stretched S/N map, defined as $\sinh^{-1}(Q_\phi/U_{\phi,\mathrm{rms}})$, overlaid with the best-fit model curves.
    The green, red, and blue curves indicate the horizon curve, shadow boundaries, and apparent outer edge of the outer disk, respectively.
    The green cross and red star mark the positions of the outer-disk center and the central star, respectively.
    (d) Azimuthal gradient component of the S/N map, $\nabla {\rm S/N}\cdot \hat{\boldsymbol{e}}_\phi$, used to identify the shadow boundaries.
    (e) Positive radial gradient component, $\nabla {\rm S/N}\cdot \hat{\boldsymbol{e}}_r>0$, used to identify the horizon curve.
    (f) Negative radial gradient component, $\nabla {\rm S/N}\cdot \hat{\boldsymbol{e}}_r<0$, used to identify the apparent outer-edge curve.
    The colored circles in panels (d)--(f) show the sampled data points used in the simultaneous fitting.
    }
    \label{fig:images}
\end{figure*}

Figure~\ref{fig:images} shows the images and gradient maps used to define the fitting data for the shadow-geometry reconstruction. The H-band $Q_\phi$ image traces the polarized scattered light from the disk surface, while the corresponding $U_\phi$ image is used to assess residual noise and possible non-azimuthal polarization signals. We construct the S/N map as $Q_\phi/U_{\phi,\mathrm{rms}}(r)$, where $U_{\phi,\mathrm{rms}}(r)$ is the radial rms profile measured from the $U_\phi$ image. An inverse hyperbolic sine stretch is applied to these images for visualization and, in the case of the S/N map, for the subsequent gradient-based feature extraction. 

The geometric features used in the fitting are identified from the gradients of the asinh-stretched S/N map. The shadow-boundary points are sampled from pixels with large azimuthal gradients, $\nabla {\rm S/N}\cdot \hat{\boldsymbol{e}}_\phi$, because the shadows appear as localized changes in brightness along the azimuthal direction. The horizon-curve points are sampled from pixels with large positive radial gradients, $\nabla {\rm S/N}\cdot \hat{\boldsymbol{e}}_r>0$, whereas the apparent outer-edge points are sampled from pixels with large negative radial gradients, $\nabla {\rm S/N}\cdot \hat{\boldsymbol{e}}_r<0$. The sampled points shown in figure~\ref{fig:images} are used as the fitting data for the simultaneous modeling of the horizon curve, shadow boundaries, and apparent outer-edge curve.

\section{Posterior distributions of the shadow-geometry model}\label{appendix:corner_plots}

\begin{figure*}[t]
    \begin{center}
    \includegraphics[width=\hsize,bb = 0 0 1466 1489]{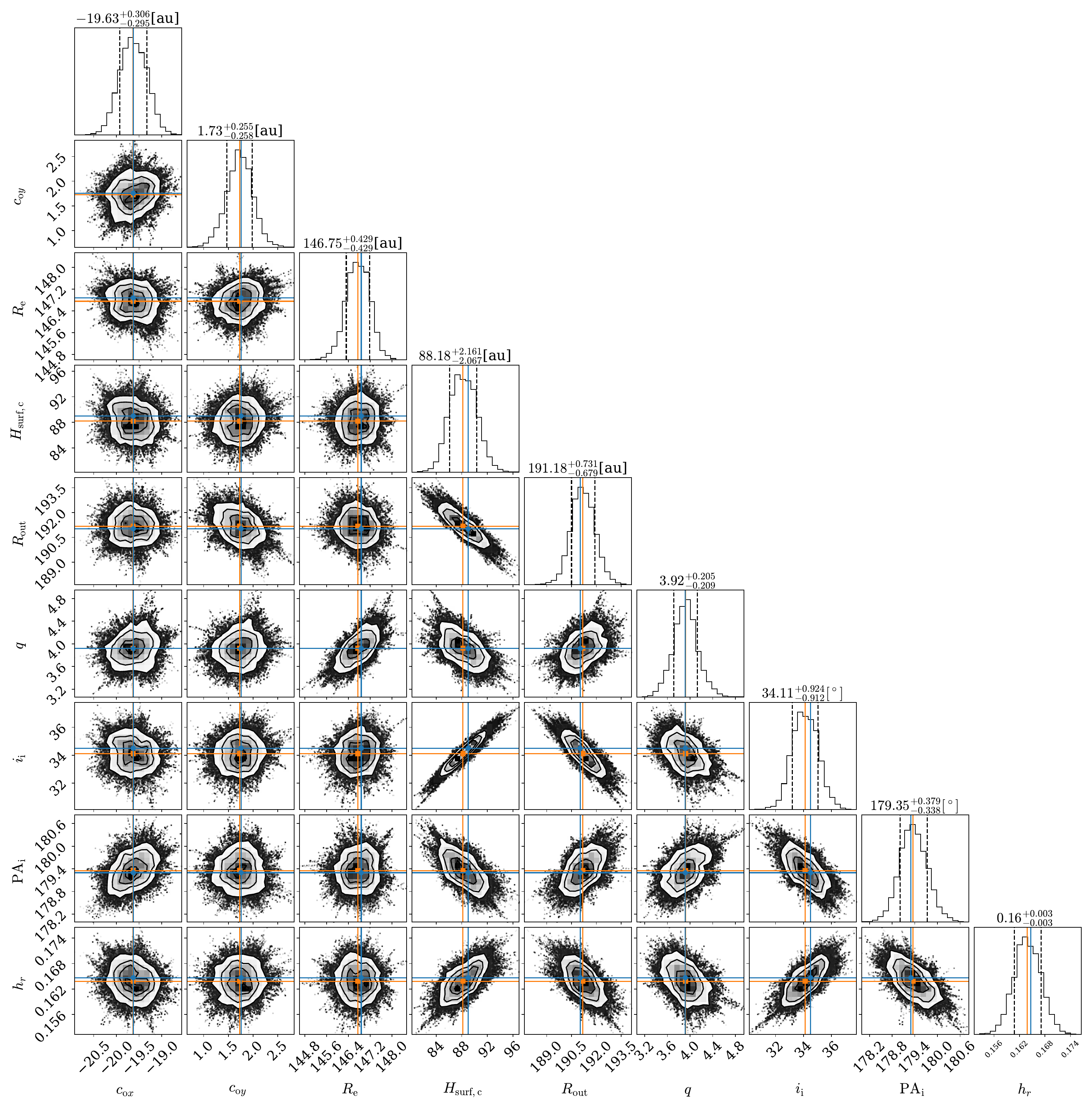}
    \end{center}
    \caption{
    Posterior distributions of the disk model parameters listed in Table~\ref{tab:disk_params}.
    The distributions were obtained from the MCMC fitting of the horizon curve, shadow boundaries, and apparent outer-edge curve for the HD~142527 disk.
    The diagonal panels show the marginalized one-dimensional posterior distributions, while the off-diagonal panels show the two-dimensional posterior distributions and parameter correlations.
    The orange solid lines indicate the median values of the posterior distributions, whereas the blue solid lines indicate the maximum-likelihood values adopted for the best-fit model.
    The dashed lines indicate the 16th and 84th percentiles.
    The parameter values and uncertainties quoted above the one-dimensional histograms correspond to the median and the 16th--84th percentile ranges, respectively.
    }
    \label{fig:corner_plots}
\end{figure*}

Figure~\ref{fig:corner_plots} shows the posterior distributions of the disk model parameters obtained from the MCMC fitting of the shadow-geometry model. We explored the posterior distributions using the affine-invariant ensemble sampler implemented in \texttt{emcee}, with 36 walkers evolved for 20,000 steps. The first 10,000 steps were discarded as burn-in, and the posterior distributions shown here were constructed from the remaining samples from all walkers. The fitted data points were sampled from high-gradient pixels corresponding to the horizon curve, shadow boundaries, and apparent outer-edge curve, as described in subsection~\ref{subsubsec:shadow_reconstruction}. The diagonal panels show the marginalized one-dimensional posterior distributions, while the off-diagonal panels show the two-dimensional posterior distributions and correlations between parameters.

The orange solid lines indicate the median values of the posterior distributions, and the dashed lines indicate the 16th and 84th percentiles. These percentile ranges are adopted as the statistical uncertainties listed in table~\ref{tab:disk_params}. The blue solid lines indicate the best-fit parameter values used to draw the model curves in figure~\ref{fig:images}. For most parameters, the best-fit values are consistent with the posterior medians within the quoted uncertainties, indicating that the adopted best-fit model lies within a high-probability region of the posterior distribution.

Some parameter correlations are visible in the two-dimensional posterior distributions. In particular, parameters related to the vertical structure and radial extent of the scattering surface are partially correlated, reflecting the fact that the observed horizon, shadow boundaries, and apparent outer edge jointly constrain the projected shape of the disk surface. Nevertheless, the posterior distributions are well localized within the adopted search ranges, and the resulting uncertainties are sufficiently small for the subsequent temperature analysis.

\section{Dust absorption and scattering opacities}\label{appendix:opacity}

\begin{figure*}[t]
    \begin{center}
    \includegraphics[width=\hsize,bb = 0 0 1371 396]{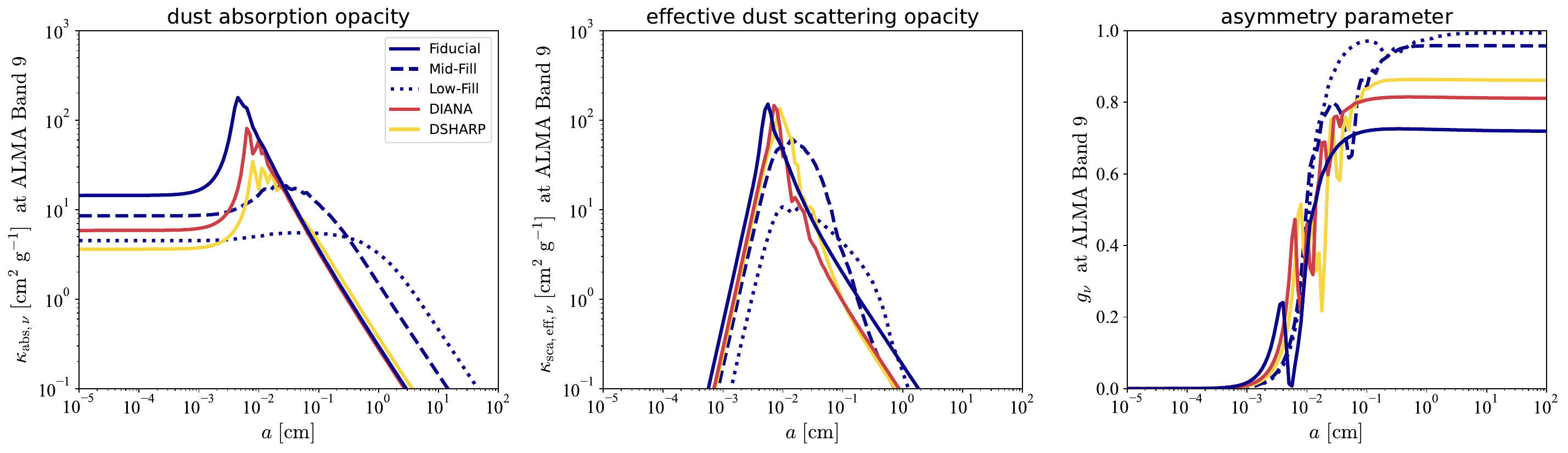}
    \end{center}
    \caption{Dust absorption opacity (left panel), effective dust scattering opacity (middle panel), and asymmetry parameter (right panel) for different opacity models and porosity models as a function of the dust grain size with the frequency of $\nu = 697~{\rm GHz}$. }
    \label{fig:opacity_Ricci_DIANA_DSHARP}
\end{figure*}

In sections \ref{sec:method} and \ref{sec:results}, we use the opacity model based on \citet{RicciTesti+:2010aa} as the fiducial model to evaluate the disk temperature [equation \eqref{eq:Iobs}], the emission height [equations \eqref{eq:tau_z} and \eqref{eq:z_tau_1}], and the cooling timescale (subsection \ref{subsubsec:method_cooling}).
To calculate the frequency-dependent effective dust extinction opacity per dust mass $\kappa_{\rm d,\nu}$ for a given dust grain size, we use the absorption opacity $\kappa_{\rm abs,\nu}$, the scattering opacity $\kappa_{\rm sca,\nu}$, and the asymmetry parameter $g_\nu$, as described in subsection \ref{subsubsec:method_opacity}.
Figure \ref{fig:opacity_Ricci_DIANA_DSHARP} shows $\kappa_{\rm abs,\nu}$, $\kappa_{\rm sca,eff,\nu}( = [1-g_\nu]\kappa_{\rm sca,\nu})$, and $g_\nu$ as a function of the dust grain size $a$ with $\nu = 697~{\rm GHz}$ for the fiducial model.
In this frequency, both $\kappa_{\rm abs,\nu}$ and $\kappa_{\rm sca,eff,\nu}$ have a peak at the dust grain with $\approx 100~{\rm \mu m}$, where the size parameter ($=2\pi a/\lambda$) takes the value of almost unity.

These dust opacities can also depend on the structure of the dust particles.
In section \ref{subsec:AnalyticModel_results}, we treat the dust filling factor $f_{\rm dust}$ as a parameter on which the dust opacity depends.
Figure \ref{fig:opacity_Ricci_DIANA_DSHARP} also plots $\kappa_{\rm abs,\nu}$, $\kappa_{\rm sca,eff,\nu}$, and $g_\nu$ for different values of $f_{\rm dust}$.
As $f_{\rm dust}$ decreases, the peak value of $\kappa_{\rm abs,\nu}$ at $a \sim 100~{\rm \mu m}$ decreases, while $\kappa_{\rm abs,\nu}$ at smaller and larger dust than $100~{\rm \mu m}$ decreases and increases, respectively.
For $\kappa_{\rm sca,eff,\nu}$, the porous dust leads to a peak value lower than that of the compact dust.

Furthermore, the composition of dust particles can affect the dust extinction opacity and thereby the estimate of the dust grain size.
For instance, the DIANA \citep{WoitkeMin+:2016aa} and DSHARP \citep{BirnstielDullemond+:2018aa} opacity models assume a different composition from the model based on \citet{RicciTesti+:2010aa}, leading to a different value of the opacities.
Figure \ref{fig:opacity_Ricci_DIANA_DSHARP} plots $\kappa_{\rm abs,\nu}$, $\kappa_{\rm sca,eff,\nu}$, and $g_\nu$ for the Ricci, DIANA, and DSHARP opacity models as a function of the dust grain size.
For the DIANA and DSHARP models, $\kappa_{\rm abs,\nu}$ at the dust of $a \lesssim 100~{\rm \mu m}$ becomes smaller compared to the Ricci opacity, while $\kappa_{\rm sca,eff,\nu}$ takes almost the same values as the Ricci opacity.
We present the results of the grain size estimation for the DIANA and DSHARP models in appendix \ref{appendix:result_opacity}.

\section{Disk temperature and gas scale height}\label{appendix:temperature}

\begin{figure*}[t]
    \begin{center}
    \includegraphics[width=\hsize,bb = 0 0 1395 405]{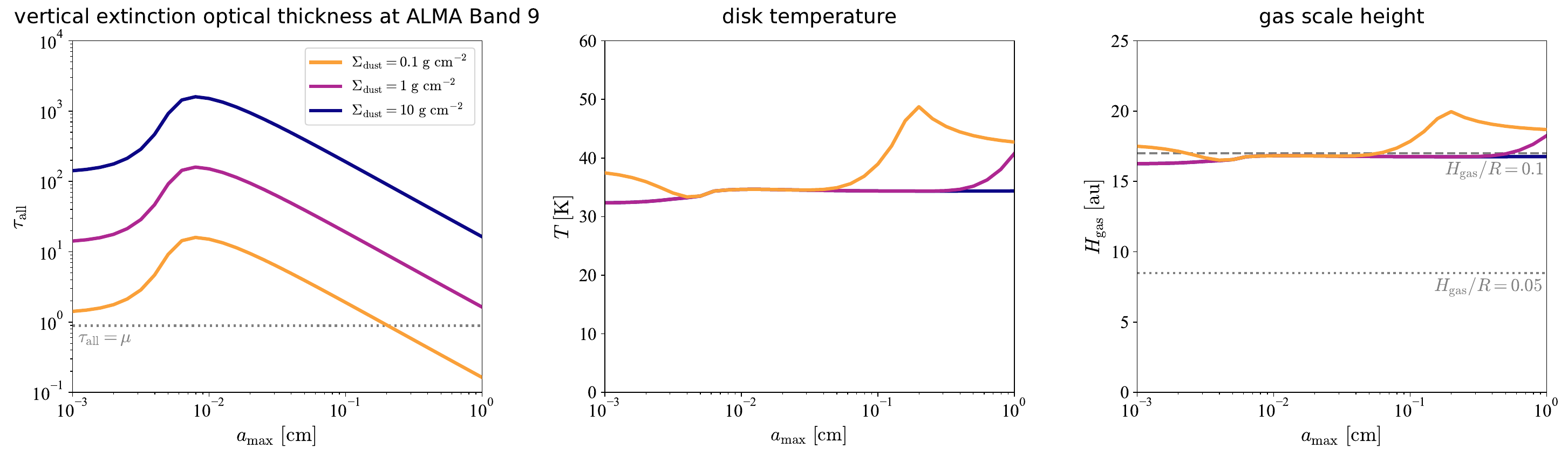}
    \end{center}
    \caption{Calculated disk's vertical extinction optical thickness at ALMA Band 9 (left panel), disk temperature (middle panel), and gas scale height (right panel) as a function of the maximum dust grain size for different values of $\Sigma_{\rm dust}$ in the fiducial model. The horizontal dotted line in the left panel shows $\tau_{\rm all}=\mu$. The horizontal dashed and dotted lines in the right panel show $H_{\rm gas}/R=0.1$ and $0.05$, respectively. }
    \label{fig:tauall_T_Hgas_amax}
\end{figure*}

In section \ref{subsec:AnalyticModel_results}, we derive the disk temperature from the vertical extinction optical thickness at ALMA Band 9 with a given dust grain size and the observed intensity of the dust thermal emission.
The left and middle panels of figure \ref{fig:tauall_T_Hgas_amax} show the vertical extinction optical thickness at $\nu = 697~{\rm GHz}$, $\tau_{\rm all}$, and disk temperature $T$ as a function of the dust maximum grain size $a_{\rm max}$.
This figure indicates that in the cases of $\Sigma_{\rm dust}=1~{\rm g~cm^{-2}}$ and $10~{\rm g~cm^{-2}}$, $\tau_{\rm all}$ for all maximum grain sizes exceed unity.
This shows that the north region of the HD~142527 disk is almost optically thick at ALMA Band 9 for the fiducial model.
For the case of $\Sigma_{\rm dust}=0.1~{\rm g~cm^{-2}}$, the dust with $a_{\rm max}\lesssim 2~{\rm mm}$ leads to larger $\tau_{\rm all}$ than unity.

From $\tau_{\rm all}$ and the observed dust thermal emission intensity $I_{\nu,\rm obs}$, we obtain $T$ using equation \eqref{eq:Iobs}.
The middle panel of this figure shows that for the case of $\Sigma_{\rm dust}=10~{\rm g~cm^{-2}}$, the temperature rises slightly as $a_{\rm max}$ increases up to $\approx 100~{\rm \mu m}$, and remains nearly constant at $T\approx 35~{\rm K}$ for sizes larger than that.
This is due to the scattering effect, which corresponds to the third term on the right-hand side of equation \eqref{eq:Iobs}.
For the cases of the low dust surface density, $T$ increases sharply to $50~{\rm K}$ at most as $\tau_{\rm all}$ approaches unity.
This temperature range is also consistent with an estimate from molecular line emission observations of $^{13}$C$^{18}$O and DCO$^+$ at the midplane of the north point \citep{TemminkBooth+:2023aa}.

The dependence of $T$ on the dust grain size determines that on the gas scale height.
We present $H_{\rm gas}$ in the right panel of figure \ref{fig:tauall_T_Hgas_amax} as a function of maximum dust grain size $a_{\rm max}$.
The gas scale height takes approximately $17~{\rm au}$ for almost all cases and varies from $16~{\rm au}$ to $20~{\rm au}$ depending on changes in $T$.

The disk temperature also determines the occurrence of the gravitational instability.
The gravitational instability operates when Toomre's $Q_{\rm T}$ parameter, defined by $Q_{\rm T}=c_{\rm s}\Omega_{\rm K}/\pi G \Sigma_{\rm gas}$, is smaller than approximately two.
The middle panel of figure \ref{fig:tauall_T_Hgas_amax} shows that the minimum value of the disk temperature approximates to $32.5~{\rm K}$ with $\Sigma_{\rm gas}=10~{\rm g~cm^{-2}}$, yielding $Q_{\rm T}\approx2.3$.
Therefore, our model predicts that the outer disk of HD~142527 is stable for the gravitational instability.

\begin{figure*}[t]
    \begin{center}
    \includegraphics[width=\hsize,bb = 0 0 1143 783]{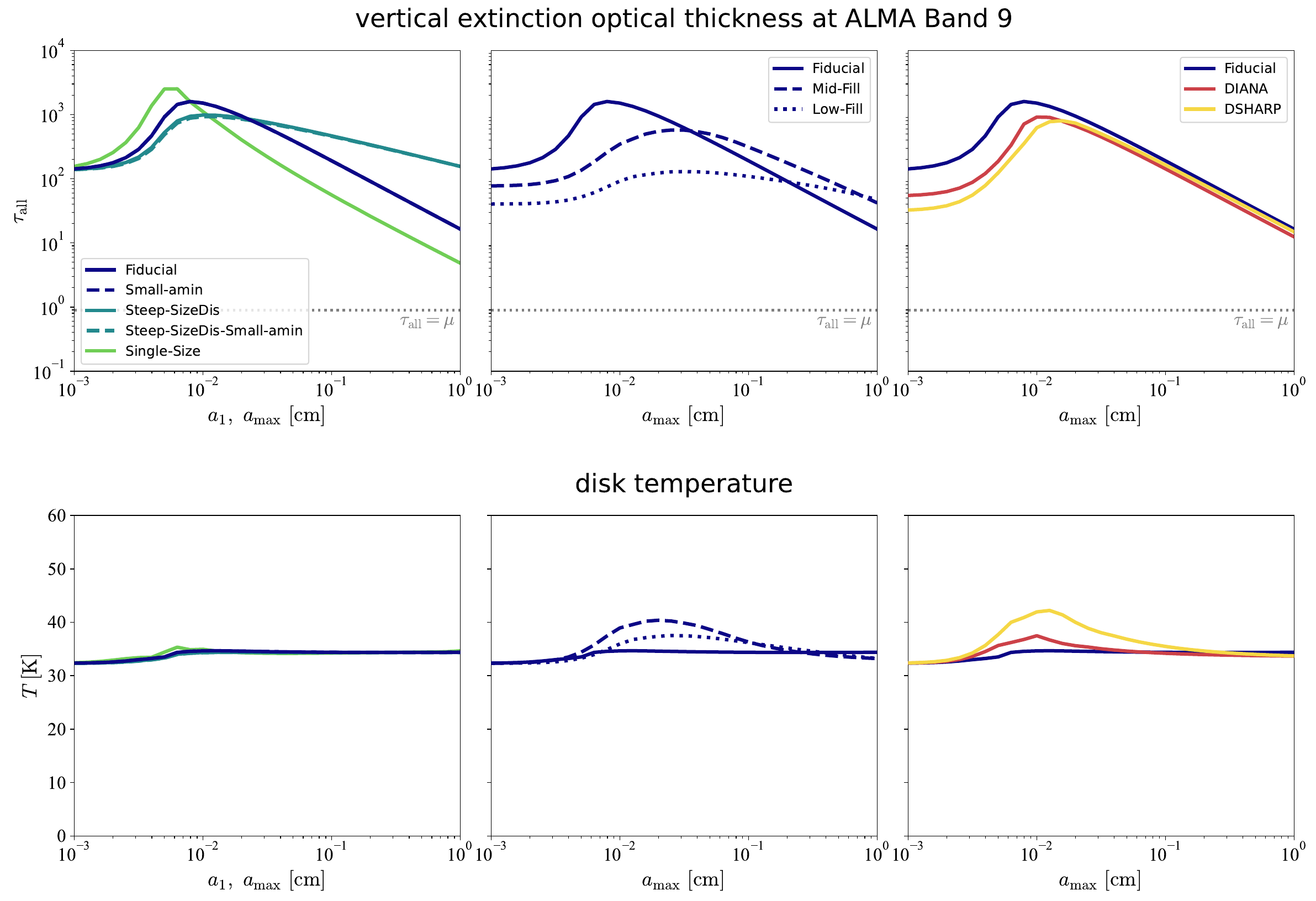}
    \end{center}
    \caption{Upper row: Same as the left of figure \ref{fig:tauall_T_Hgas_amax}, but for the different models of the size distribution (left panel), porosity (middle panel), and opacity (right panel), with $\Sigma_{\rm dust}=10~{\rm g~cm^{-2}}$. The horizontal dotted lines show $\tau_{\rm all}=\mu$. Lower row: Same as the middle panel of figure \ref{fig:tauall_T_Hgas_amax}, but for the different models of the size distribution (left panel), porosity (middle panel), and opacity (right panel), with $\Sigma_{\rm dust}=10~{\rm g~cm^{-2}}$. }
    \label{fig:tauall_T_Hgas_amax_parameter}
\end{figure*}

Furthermore, $\tau_{\rm all}$ and $T$ depend on the size distribution, dust filling factor, and the opacity models.
Figure \ref{fig:tauall_T_Hgas_amax_parameter} shows $\tau_{\rm all}$ (upper row) and $T$ (lower row) for different models as a function of dust grain size.
Changes in these parameters affect $\tau_{\rm all}$, but the magnitude of those changes remains within one order of magnitude of the fiducial model for each dust grain size.
Therefore, $T$ is insensitive to fluctuations in these parameters, with the maximum variation limited to a factor of 1.3.

\section{Parameter dependence on calculated ALMA Band 9 emission height}\label{appendix:observed_height}

\begin{figure*}[t]
    \begin{center}
    \includegraphics[width=\hsize,bb = 0 0 1360 355]{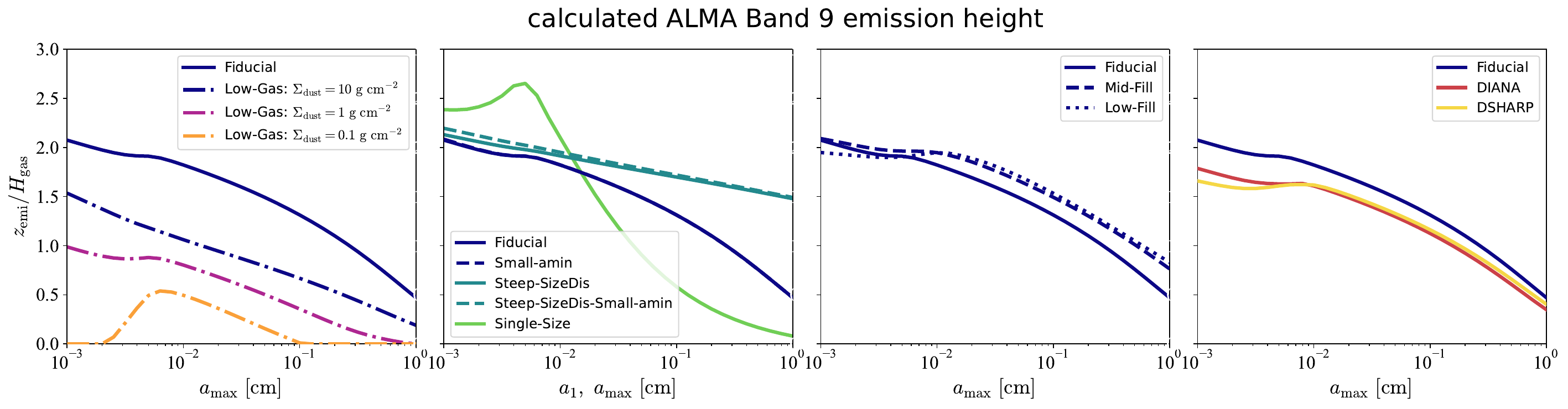}
    \end{center}
    \caption{Same as figure \ref{fig:zobs_amax_fiducial}, but for the different models of the Low-Gas, size distribution, porosity, and opacity from left to right, with $\Sigma_{\rm dust}=10~{\rm g~cm^{-2}}$ except for the Low-Gas model. }
    \label{fig:zobs_parameter}
\end{figure*}

In section \ref{subsec:AnalyticModel_results}, we display the cooling timescale at the ALMA Band 9 emission height, $t_{\rm cool,cal}$, to see the dependence of the gas surface density (figure \ref{fig:map_color_Sigmag1e0}), size distribution (figure \ref{fig:map_color_SizeDis}), and dust filling factor (figure \ref{fig:map_color_Poro}).
These parameters affect the ALMA Band 9 emission height $z_{\rm emi}$ through the optical depth as well as the vertical distribution of the cooling timescale.

Figure \ref{fig:zobs_parameter} shows $z_{\rm emi}$ for all models presented in this study, as a function of dust grain size with $\Sigma_{\rm dust}=10~{\rm g~cm^{-2}}$.
The left panel of this figure shows that a decrease in the gas surface density $\Sigma_{\rm gas}$ results in an overall lower $z_{\rm emi}$ than that of the fiducial model presented in figure \ref{fig:zobs_amax_fiducial}.
The dust amount at high altitudes, as well as the dust scale height, decreases because a decrease in $\Sigma_{\rm gas}$ increases the Stokes number of the dust [see equation \eqref{eq:St}].
Therefore, the ALMA Band 9 optical depth measured from infinity down, $\tau_\nu(z)$, in equation \eqref{eq:tau_z} becomes small at high altitudes, yielding the low $z_{\rm emi}$.  

For the difference in the size distribution, $z_{\rm emi}$ is also changed through the dust number density and optical depth at ALMA Band 9.
The variation in $z_{\rm emi}$ with respect to the size distribution models reflects the dependence of the effective dust extinction opacity on the dust grain size.
For the Single-Size model, the profile of $z_{\rm emi}$ has a peak at $a_1\sim100~{\rm \mu m}$, which corresponds to a peak of the dust absorption and scattering opacities (see figure \ref{fig:opacity_Ricci_DIANA_DSHARP} in appendix \ref{appendix:opacity}).
As the dust grain size increases, $z_{\rm emi}$ for the Single-Size model decreases to values smaller than those for the fiducial model.
For the Steep-SizDis model, $z_{\rm emi}$ is higher than that for the fiducial model due to an increased abundance of small dust particles with high opacity.
However, a change in the minimum dust grain size has only a minor effect on $z_{\rm emi}$.
This is because the dust particles smaller than $1~{\rm \mu m}$ possess a constant opacity.

Furthermore, for porous dust, $z_{\rm emi}$ is almost the same as that of compact dust, especially in the case of $a_{\rm max}\lesssim 100~{\rm \mu m}$ with $\Sigma_{\rm dust}=10~{\rm g~cm^{-2}}$.
For a case of $a_{\rm max}\gtrsim 100~{\rm \mu m}$, $z_{\rm emi}$ is slightly higher than that of the fiducial model due to the large opacity.
We note, however, for a case of $a_{\rm max}\lesssim 100~{\rm \mu m}$ with low $\Sigma_{\rm dust}$, $z_{\rm emi}$ decreases as the filling factor decreases because the absorption opacity decreases (see figure \ref{fig:opacity_Ricci_DIANA_DSHARP} in appendix \ref{appendix:opacity}). 

Moreover, different opacity models cause a change in $z_{\rm emi}$.
The right panel of figure \ref{fig:zobs_parameter} shows $z_{\rm emi}$ for different opacity models (for details of the opacity model, see appendix \ref{appendix:opacity}).
For the DIANA and DSHARP models, $z_{\rm emi}$ is slightly lower than that of the fiducial model, which uses the Ricci opacity model.
This is due to the low absorption opacity at $a\lesssim 100~{\rm \mu m}$, as shown in figure \ref{fig:opacity_Ricci_DIANA_DSHARP} in appendix \ref{appendix:opacity}.
These slightly low $z_{\rm emi}$ have a minor impact on $t_{\rm cool,cal}$ (see appendix \ref{appendix:result_opacity}).

\section{Dependence on the opacity model}\label{appendix:result_opacity}
As mentioned in appendix \ref{appendix:opacity}, the dust extinction opacity controls disk temperature, the emission height at ALMA Band 9, and the cooling timescale.
In this appendix, we explain the results for cases of different opacity models, the DIANA \citep{WoitkeMin+:2016aa} and DSHARP \citep{BirnstielDullemond+:2018aa} opacity models.
The absorption opacity $\kappa_{\rm abs,\nu}$, the effective scattering opacity $\kappa_{\rm sca,\nu}$, and the asymmetry parameter $g_\nu$ as functions of the dust grain size for both models are shown in figure \ref{fig:opacity_Ricci_DIANA_DSHARP} in appendix \ref{appendix:opacity}.
For the DIANA and DSHARP models, we reset the dust material density to $\rho_{\rm m}=2.08{\rm ~g~cm^{-3}}$ and $1.68{\rm ~g~cm^{-3}}$, respectively.

\begin{figure}[t]
    \begin{center}
    \includegraphics[width=\hsize,bb = 0 0 558 843]{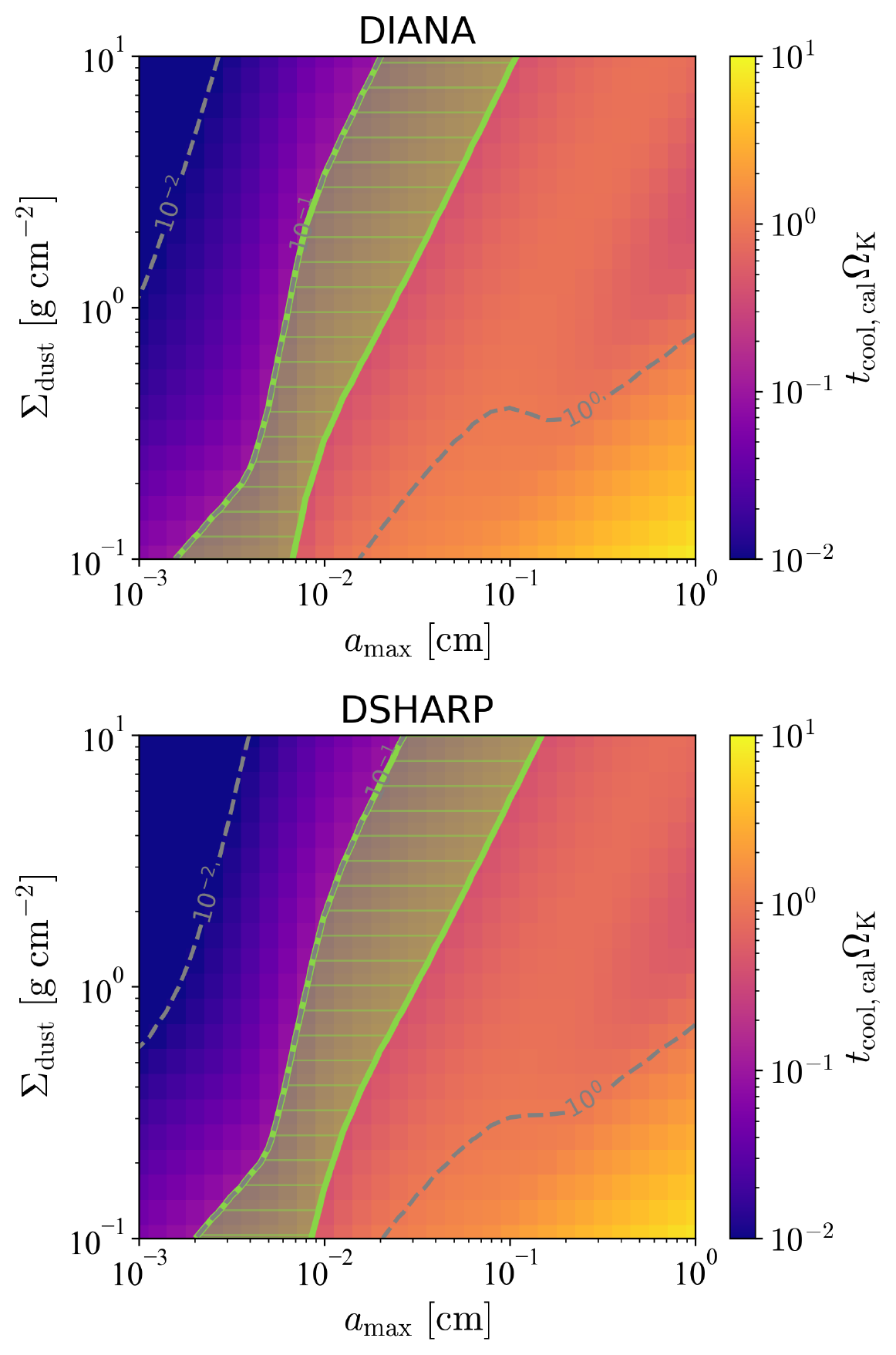}
    \end{center}
    \caption{Same as figure \ref{fig:map_color_fiducial}, but for the DIANA model (upper panel) and DSHARP model (lower panel). }
    \label{fig:map_color_opacity}
\end{figure}

We show in figure \ref{fig:map_color_opacity} the cooling timescale at the ALMA Band 9 emission height, $t_{\rm cool,cal}$, as a function of $a_{\rm max}$ and $\Sigma_{\rm dust}$ for the DIANA (upper panel) and DSHARP (lower panel) models.
For both models, $t_{\rm cool,cal}$ increases slightly compared to that for the fiducial model, as shown in figure \ref{fig:map_color_fiducial}.
This is because a low absorption opacity with $a\lesssim100~{\rm \mu m}$ (see the left panel of figure \ref{fig:opacity_Ricci_DIANA_DSHARP}) leads to a high emission height at ALMA Band 9 (see the right panel of figure \ref{fig:zobs_parameter}).
The constrained maximum dust grain size with $\Sigma_{\rm dust}=10~{\rm g~cm^{-2}}$ becomes $200~{\rm \mu m}\lesssim a_{\rm max}\lesssim1.0~{\rm mm}$ and $250~{\rm \mu m}\lesssim a_{\rm max}\lesssim1.5~{\rm mm}$ for the DIANA and DSHARP models, respectively.

\begin{figure}[t]
    \begin{center}
    \includegraphics[width=\hsize,bb = 0 0 422 720]{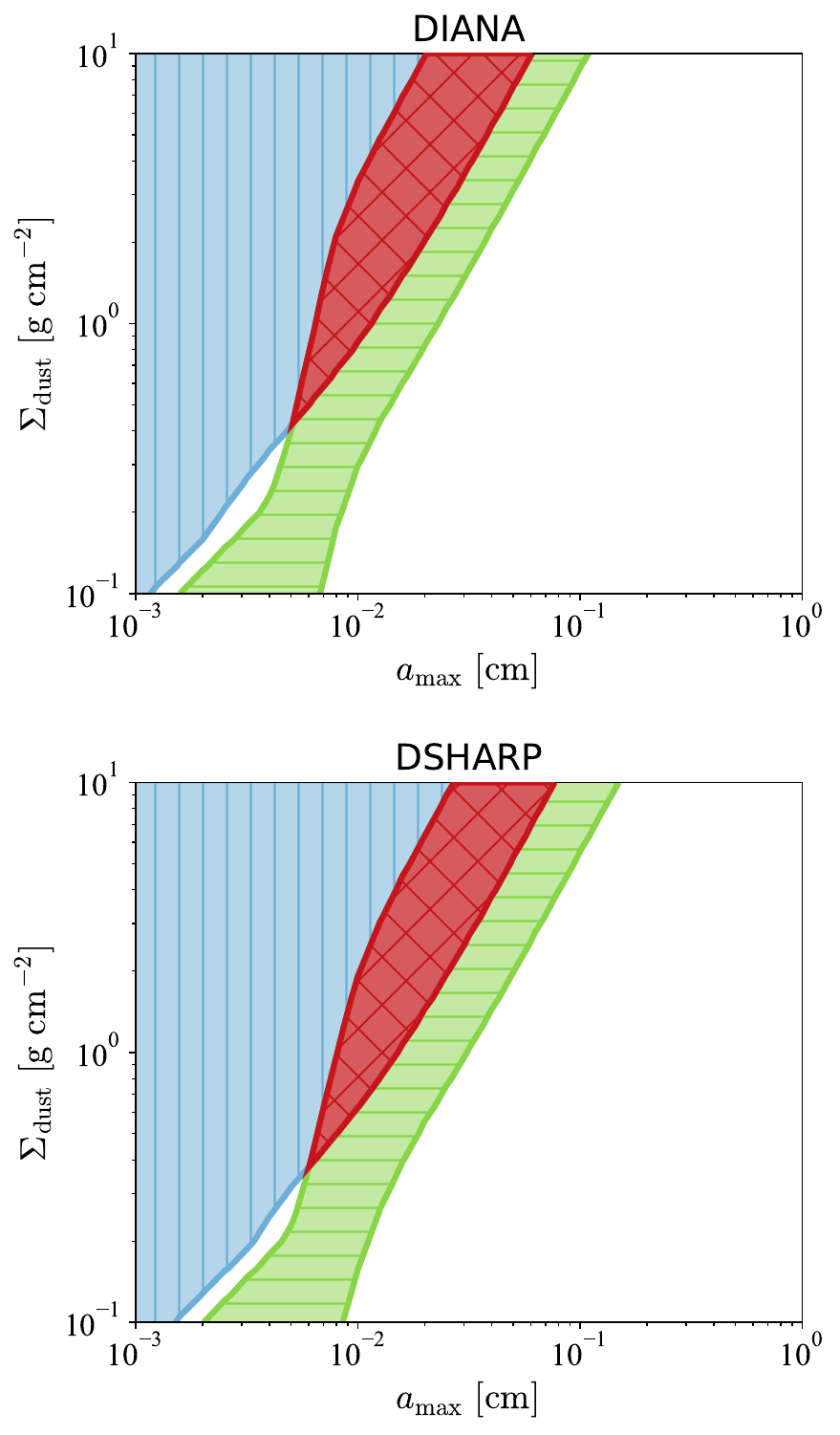}
    \end{center}
    \caption{Same as figure \ref{fig:map_VSI_models}, but for the DIANA model (upper panel) and DSHARP model (lower panel).}
    \label{fig:map_VSI_opacity}
\end{figure}

According to section \ref{sec:VSI_application}, we also compute the thickness of the VSI-active layers and search the parameter space satisfying the cooling and global VSI conditions.
Figure \ref{fig:map_VSI_opacity} maps the $a_{\rm max}$--$\Sigma_{\rm dust}$ parameter space that satisfies the cooling and global VSI conditions for the DIANA (upper panel) and DSHARP (lower panel) models.
As with $t_{\rm cool,cal}$, variations in the opacity models slightly alter the $a_{\rm max}$--$\Sigma_{\rm dust}$ parameter space that satisfies the cooling and global VSI conditions.
The maximum dust grain size reproducing $0.1<t_{\rm cool,cal}\Omega_{\rm K}<0.4$ and $\alpha_z=2\times 10^{-3}$ with $\Sigma_{\rm dust}=10~{\rm g~cm^{-2}}$ becomes $200~{\rm \mu m}\lesssim a_{\rm max}\lesssim 600~{\rm \mu m}$ and $250~{\rm \mu m}\lesssim a_{\rm max}\lesssim 730~{\rm \mu m}$ for the DIANA and DSHARP models, respectively.
For both models, there is also no maximum grain size satisfying the cooling and global VSI condition at $\Sigma_{\rm dust}\lesssim 0.3~{\rm g~cm^{-2}}$.

As a result, we conclude that differences in the opacity models or dust material components have little effect on the results of the grain-size limitation.

\end{document}